\documentclass[preprint,12pt,authoryear]{elsarticle}

\usepackage{amssymb}

\usepackage{lineno}

\usepackage[left=2.cm,top=2.cm,right=2.cm,bottom=1.9cm]{geometry}  
\usepackage{amsmath}
\usepackage{graphicx, pgfplots} 
\newcommand{\stkout}[1]{\ifmmode\text{\sout{\ensuremath{#1}}}\else\sout{#1}\fi}
\usepackage{cancel, comment}
\usepackage{url, doi, cleveref}
\usepackage[normalem]{ulem}

\makeatletter
\let\jnl@style=\rm
\def\ref@jnl#1{{\jnl@style#1}}
\def\aap{\ref@jnl{A\&A}}                
\def\aj{\ref@jnl{AJ}}                   
\def\grl{\ref@jnl{Geophys.~Res.~Lett.}} 
\def\icarus{\ref@jnl{Icarus}}           
\def\mnras{\ref@jnl{MNRAS}}             
\def\nat{\ref@jnl{Nature}}              
\def\planss{\ref@jnl{Planet.~Space~Sci.}}   
\pgfplotsset{compat=1.18}
\makeatother

\journal{Icarus}

\begin{document}

\begin{frontmatter}



\title{Librations and obliquity of the largest moons of Uranus}


\author[inst1]{Rose-Marie Baland }
\ead{rose-marie.baland@observatoire.be}
\author[inst1,inst2]{Valerio Filice}
\author[inst1,inst2]{Sébastien Le Maistre}
\author[inst1]{Antony Trinh}
\author[inst1]{Marie Yseboodt}
\author[inst1,inst3]{Tim Van Hoolst}

\affiliation[inst1]{organization={Royal Observatory of Belgium},
            addressline={Ringlaan 3}, 
            city={Brussels},
            postcode={1180}, 
            country={Belgium}}

\affiliation[inst2]{organization={Earth and Life Institute, Université catholique de Louvain},
            addressline={Place Louis Pasteur 3}, 
            city={Louvain‐la‐Neuve},
            postcode={1348}, 
            country={Belgium}}
            
\affiliation[inst3]{organization={Instituut voor Sterrenkunde, KU Leuven},
            addressline={Celestijnenlaan 200D}, 
            city={Leuven},
            postcode={3001}, 
            country={Belgium}}

\begin{abstract}

Following the discovery of several ocean worlds in the solar system, and the selection of Uranus as the highest priority objective by the Planetary Science and Astrobiology Decadal Survey 2023-2032, the five largest moons of Uranus (Miranda, Ariel, Umbriel, Titania and Oberon) have been receiving renewed attention as they may also harbor a subsurface ocean.
We assess how rotation measurements could help confirm the internal differentiation of the bodies and detect internal oceans if any. Because of the time-varying gravitational torque of Uranus on the flattened shape of its synchronous satellites, the latter librate with respect to their mean rotation and precess with a non zero obliquity. 
For a range of interior models with a rocky core surrounded by a hydrosphere, either solid or divided into an outer ice shell with a liquid ocean underneath, we compute their diurnal libration amplitude and obliquity.
We find that if the Uranian satellites were two-layer solid bodies, libration measurement accuracies from around $0.25$ m for Oberon to around $6$ m for Miranda would rule out the possibility of homogeneous interiors. In combination with independent estimates of the mean moment of inertia (MOI), libration measurements could also be used to detect the presence of an ocean, the measurement precision required for this depending on the actual value of the libration amplitude. 
To compute the obliquity, we first build series for the orbital precession of all five satellites with a secular perturbations model. With the exception of Miranda, we show that due to the mutual gravitational interactions between the satellites, the obliquity of the large Uranian moons exhibits relatively large periodic variations around the mean value. We find that an obliquity measurement accuracy from around $1$ m for Ariel to around $400$ m for Oberon can rule out the homogeneous case. The presence of an internal global ocean could allow a resonant amplification of the obliquity, facilitating its detection. If no such resonance occurs, the obliquity would be almost indistinguishable from that expected for a solid body. 
The effect of tidal deformations on the rotation of the small to medium-sized Uranian moons is showed to be limited. Librations would be reduced by up to $10\%$ and obliquity increased by up to $15\%$ for Titania and Oberon, the effects being negligible for Miranda.
\end{abstract}


\end{frontmatter}


\newpage
\tableofcontents 

\newpage
\section{Introduction}

Named after characters of the English literature, Miranda, Ariel, Umbriel, Titania, and Oberon are the five largest moons of Uranus. They are small or medium-sized icy satellites about which we still have almost everything to learn. Discovered by William Herschel (Titania and Oberon, 1787), William Lassell (Ariel and Umbriel, 1851), and Gerard Kuiper (Miranda, 1948), they have only been visited by one space mission, during the flyby of Uranus by Voyager 2 in 1986.

The last few decades have shown that water can exist in liquid state in the outer Solar System, beneath the surface of icy moons of Jupiter and Saturn \citep{Nimmo2016}.
The Galileo mission extensively studied the satellite system of Jupiter, discovering induced magnetic fields in the vicinity of Europa, Ganymede, and Callisto, indicating the existence of subsurface oceans \citep{Khu98,Kiv02}.
Magnetic induction is just one of the multiple techniques for detecting these oceans. Observing the surface can be evocative, as in the case of Europa and its ‘icebergs’ \citep{Car98}. More recent Hubble Space Telescope (HST) observations of the low-amplitude oscillations of auroral ovals on Ganymede indicate the presence of an underground ocean beneath a thick layer of ice \citep{Sau15}. 

The Cassini mission has extended the range of techniques used to detect the subsurface oceans of icy satellites. The small Enceladus showed extraordinary activity at its anomalously warm south pole, with plumes of water vapor and ice particles interpreted as evidence of liquid water beneath the surface, in contact with an underlying rocky core \citep{Por06,Spe06,Spe09,Pos11}. The shape, gravity field, and diurnal librations of Enceladus are consistent with a global water reservoir beneath an ice shell around $20$ or $30$ km thick (\cite{Park24} and references therein). The gravity and shape of Dione can also be explained in terms of an ice shell overlying a global water ocean \citep{Beu16,Zan20}.  
The observed obliquity of Titan is about three times larger than expected for an entirely solid and rigid Titan (\cite{Bal19} and references therein). The significant diurnal tidal deformations experienced by the satellite also testify to the presence of a global ocean inside Titan \citep{Ies12,Durante2019,Goossens2024}, whereas a thick ice shell over a global ocean explains the electric field measurements by the Huygens probe \citep{Beg12}. The latest to join the club of icy satellites with an ocean may be the highly cratered Mimas. Its libration may be compatible with the presence of a liquid layer at depth \citep{Taj2014} and the evolution of its orbit suggests the presence of a very young ocean at a depth of $20-30$ km below the surface \citep{Lai24}. 

Voyager, Galileo, and Cassini-Huygens were Flagship-class missions. Neptune, Uranus and their moons are still awaiting ambitious exploration missions. The Uranus Orbiter and Probe (UOP) has been prioritized as the next Flagship-class mission by the 2023-2033 Planetary Science and Astrobiology Decadal Survey \citep{NAP26522}. Among other things, the mission, during its 4-year tour, would aim to address important questions regarding the five largest Moons. What are their rock-to-ice mass ratios and internal structures? Which moons have internal oceans? The answers to these questions will also help constraining their formation and evolution.

In this paper, we aim to define ranges for the rotational observables (libration and obliquity) of Uranus's largest satellites that could be estimated, among other techniques, by landmark tracking from images obtained with the Narrow and Wide Angle Cameras (NAC and WAC) that will be part of the future UOP spacecraft's payload. Optical data could be used jointly with the radiometric data from the radioscience experiment aboard the spacecraft to infer the satellites' interior \citep{Filice2024}. Although not considered in this study, magnetic induction has also been proposed for investigating subsurface oceans within the largest moons of Uranus, e.g. \cite{Cochrane2021}.

The paper is organized as follows. In the Section \ref{Sec2}, based on existing literature, we define a range of three-layer interiors (shell, ocean and core) for each satellite, assumed to be in hydrostatic equilibrium. The interiors are constrained by the mass and radius and are characterized by the total mean moment of inertia (MOI) and degree-2 gravity coefficients, as well as by the density, dimension and flattenings of each layer. Previous studies did not explore the influence of a difference in density between the ocean and the ice shell. We briefly discuss how this might affect the interpretation of an estimated MOI and gravity coefficients in terms of hydrosphere thickness and/or global differentiation. In Section \ref{Sec3}, we compute the amplitude of diurnal libration and the obliquity of the solid layers (shell and core) of the three-layer interiors defined in Section \ref{Sec2}, assuming that the satellites are locked in a Cassini state. We discuss the precision required on rotational measurements so that these measurements can be interpreted in terms of interior properties and/or global differentiation. Concluding remarks are presented in Section \ref{Sec4}, including a discussion about the similarities and differences between our libration results and the recent study by \cite{Hem24}.

\section{Satellites' interior and gravity field}
\label{Sec2}

\subsection{Mass, radius, and mean density}
The main observational constraints on the interior of the five largest Uranian satellites are their mass and radius (see Table \ref{data}). We consider the determination of their orbits and masses from Earth-based astrometry, Voyager 2 observations and ring occultations by \cite{Jac14}. The shape and radius estimations for the five satellites, from limb coordinates, date back to the Voyager 2 era \citep{Tho88}. 
An update of the mean radius of Umbriel, based on a stellar occultation analysis, is reported in \cite{Ass23}, but is not statistically different from the previous estimation.
The density thus obtained from the mass and radius, combined with observations of the surface, give the first indications of the composition and internal structure of the satellites. Ariel, Umbriel, Titania, and Oberon are medium-sized satellites with low density indicating interior mainly composed of water ice, whereas Miranda is a small satellite, like Enceladus and Mimas, but with a lower density and therefore a lower rock-to-ice ratio\citep{Hus06}. 

\begin{table}[h]
\begin{center}
\begin{tabular}{lccccccc}
\hline
& $a$ [km] & $b$ [km] & $c$ [km] & $R$ [km] & $G\!M$ [km$^3$s$^{-2}$] & $\bar\rho$ [kg\, m$^{-3}$] \\
\hline
Miranda 	 & $240.4\pm0.6$ & $234.2\pm0.9$ & $232.9\pm1.2$ & $235.8\pm0.7$ & $4.3\pm0.2$  & $1173.12$ \\
Ariel 	     & $581.1\pm0.9$ & $577.9\pm0.6$ & $577.7\pm1.0$ & $578.9\pm0.6$ & $83.5\pm1.4$ & $1539.51$ \\
Umbriel 	 & $-$ & $-$ & $-$ & $584.7\pm2.8$ & $85.1\pm1.9$  & $1522.78$ \\
Titania 	 & $-$ & $-$ & $-$ & $788.9\pm1.8$ & $226.9\pm4.1$ & $1653.01$ \\
Oberon 	     & $-$ & $-$ & $-$ & $761.4\pm2.6$ & $205.3\pm5.8$ & $1663.63$ \\
\hline
\end{tabular}
\end{center}
\caption{Equatorial and polar radii ($a>b>c$), mean radius $R$, gravitational parameter $G\!M$, and mean density $\bar\rho$ for the five large Uranian satellites, after \cite{Tho88} and \cite{Jac14}.} 
\label{data}
\end{table}

\subsection{Moments of inertia and gravity field}
\label{Sec21}

The normalized mean moment of inertia $MOI=I/M\!R^2$ is an indicator of interior differentiation and has yet to be determined for the Uranian satellites. It can be considered to vary between around $0.3$ and $0.4$ (e.g.~$0.31$ and $0.36$ for Ganymede and Callisto, respectively, two end members in terms of differentiation, \citep{Sch04}).

\begin{figure}[!htb]
      \begin{center}
        \hspace{0cm}
\includegraphics[height=4.2 cm]{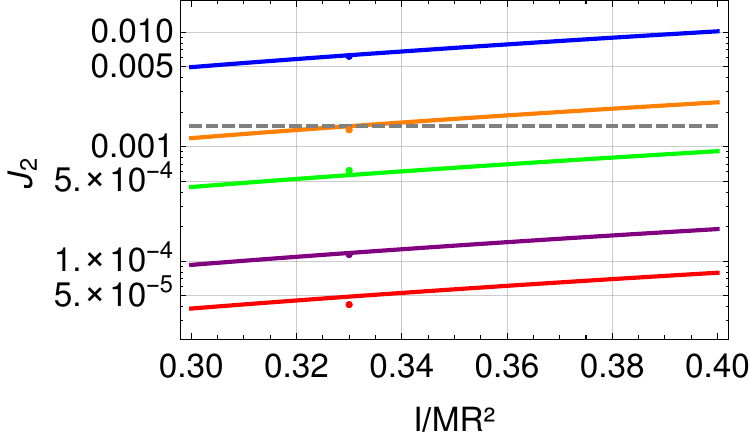}
\includegraphics[height=4.3 cm]{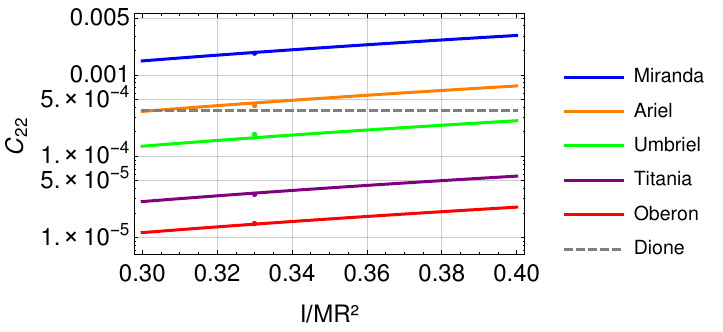}
\caption{Unormalized gravity coefficients $J_{2}$ and $C_{22}$ of the Uranian satellites as a function of the normalized mean moment of inertia $I/M\!R^2$, computed from Eq.~(\ref{Eq1}) assuming hydrostatic equilibrium. The dots indicate the values of \cite{Che14} for $C/M\!R^2$ (taken here as a proxy for $I/M\!R^2)$ $=0.33$. The gray dashed lines indicate the estimated values for Dione \citep{Zan20}. \label{FigC20C22}}
      \end{center}  
\end{figure}

Assuming hydrostatic equilibrium, the values of the coefficients $J_{2}$ and $C_{22}$ of the gravity field can be obtained, for any value of $I/M\!R^2$, from Radau's equation and the hydrostatic relation between $J_{2}$ and $C_{22}$ (e.g.~\cite{VH2008}):
\begin{subequations}
\label{Eq1}
\begin{eqnarray}
MOI&=&\frac{I}{M R^2}=\frac{2}{3}\left[1-\frac{2}{5}\left(\frac{25}{4}\frac{q_r}{\frac{3}{2}J_{2}+\frac{5}{4}q_r}-1\right)^{1/2}\right],\\
 C_{22}&=&\frac{3}{10}J_{2},
\end{eqnarray}
\end{subequations}
where $q_r=n^2 R^3 /G\!M$ is the ratio of the centrifugal acceleration to the gravitational acceleration with $n$ the mean motion, $R$ the mean radius, and $G\!M$ the gravitational parameter. Note that Radau's equation relates degree-2 gravity coefficients to the normalized mean moment of inertia ($I/M\!R^2$), not to the polar moment of inertia ($C/M\!R^2=I/M\!R^2+2J_{2}/3$) as often found in the literature (e.g.~Eq.~7 of \cite{Cas23}), see discussion after Eq.~(47) in \cite{VH2008}. Fig.~\ref{FigC20C22} shows the $J_{2}$ and $C_{22}$ ranges as a function of $I/M\!R^2$ for the five satellites. We here obtain values consistent with the previous estimates \citep{Che14, Cas23}, based on the same hypothesis. These calculations are primarily intended to identify orders of magnitude. They do not require any assumptions about the composition of the moons' interiors or the number of layers into which they are divided.  

\begin{figure}[!htb]
      \begin{center}
        \hspace{0cm}
\includegraphics[height=5 cm]{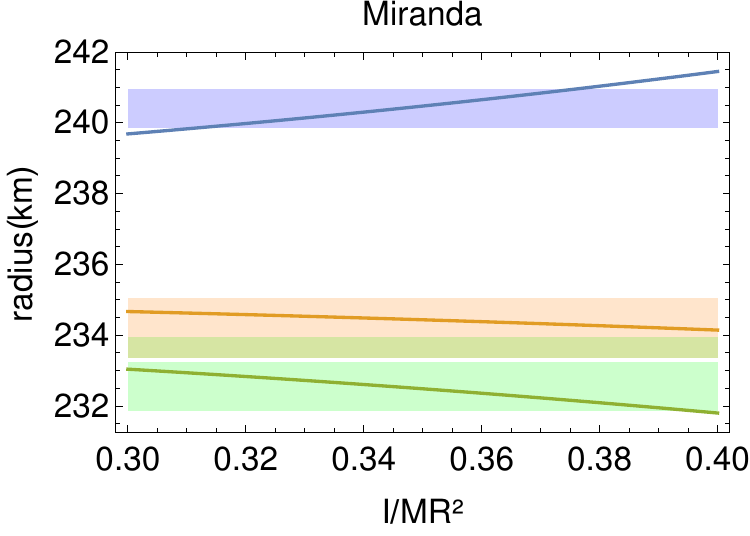}\quad
\includegraphics[height=5 cm]{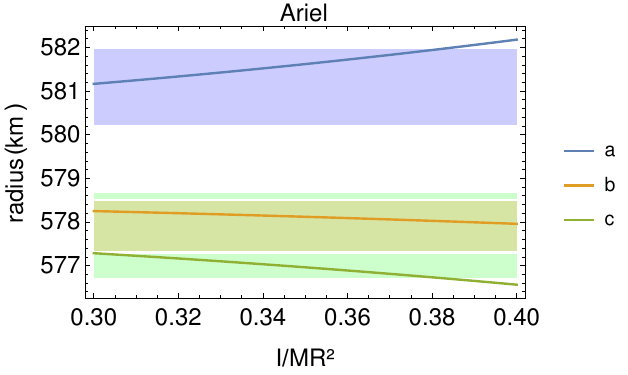}
\caption{Equatorial and polar radii ($a>b>c$) of Miranda and Ariel as a function of the MOI ($I/M\!R^2$), defined as in Eq.~(35) of \cite{VH2008} and computed from Eq.~(\ref{Eq2}) and the mean radius $R=(a+b+c)/3$, assuming hydrostatic equilibrium (solid lines). The shaded areas correspond to the estimates by \cite{Tho88} to 1$\sigma$. }
\label{Figabc} 
\end{center}
\end{figure}

The orders of magnitude thus obtained are compatible with those of the gravitational coefficients of Dione, a satellite of Saturn of approximately the same size as Ariel and Umbriel that is known not to satisfy hydrostatic equilibrium \citep{Zan20}. The ellipsoidal shape ($a>b>c$) of Miranda and Ariel was estimated by \cite{Tho88}. It is difficult to conclude with certainty whether or not these shapes are compatible with the hydrostatic equilibrium hypothesis, given Voyager 2's partial coverage of the satellites, which probably makes the small formal uncertainties on $a,b,$ and $c$ somewhat optimistic. Like \cite{Cas23}, we find that the individual values of $a$, $b$ and $c$ would be compatible with the values expected for hydrostatic equilibrium (see Fig.~\ref{Figabc}). Unlike \cite{Cas23}, we find that the combination $(b-c)/(a-c)$ ($0.17\pm0.17$ and $0.06\pm0.33$ for Miranda and Ariel, respectively) is also compatible with the expected hydrostatic value of $0.25$. 
Until there is clear evidence to the contrary, it remains reasonable to assume that the large Uranian satellites are in hydrostatic equilibrium when modeling their rotation, as we do in the following.\\

\subsection{Three-layer interiors}
\label{Secinteriors}

Following \cite{Hus06,Bie22,Cas23}, we here assume that the satellites are all differentiated into a rocky core and an external hydrosphere. The hydrosphere is either an entirely ice-I mantle, or divided into an ice-I shell and an ocean of liquid water in contact with the rocky core. \cite{Hus06,Bie22,Cas23} have come to different conclusions about the possibility of a liquid ocean. According to \cite{Hus06} and \cite{Bie22}, a water ocean is only possible for Titania and Oberon, below an ice shell at least $100$ km thick. \cite{Cas23} predict residual deep oceans less than $30$ km thick in Ariel and Umbriel and less than $50$ km in Titania and Oberon below a thick ($> 200$ km) ice shell. It should be noted that \cite{Bie22}, in contrast to \cite{Cas23}, have adopted a conservative approach by neglecting tidal heating, which could help maintain an ocean in the other satellites. All exclude the possibility of a subsurface ocean in Miranda, since that moon is so small that outside of tidal resonance, an ocean would freeze in just a few tens of My.

In this exploratory study, we consider relatively conservative ranges for the densities and dimensions of the two or three layers, with the aim of describing the influence of these parameters on rotation. In that spirit, we also consider that Miranda could harbor a liquid ocean. We use a wide density range for ice and water (between $700$ and $1050$ kg~m$^{-3}$, the lowest density implies a high porosity while the largest implies the presence of impurities like ammonia and/or salts) following what \cite{Cas23} used for the hydrosphere in their Section~6.1. Unlike these authors, here we explore the consequences of a variable density contrast between water and ice (with $\rho_o\geq\rho_s$). Our range of rock densities goes from $2400$ kg~m$^{-3}$ (the lowest value of \cite{Cas23}) to $3500$ kg~m$^{-3}$, as in \cite{Hus06} and \cite{Bie22}). We consider a maximum of $50$ km for the thickness of the ocean, following \cite{Cas23}.

To build our set of three-layer interiors, we first set the ocean thickness $h_o$ to a value between $2.5$ and $50$~km, and then derive the ice shell thickness $h_s$ and rock core radius $R_c= R-h_s-h_o$ from ice, water and rock densities and constraints on total radius and mass. Since the ocean is chosen to have a limited thickness ($h_o\leq 50$ km), the ice shell is relatively thick for all five satellites ($h_s\gtrsim 30$ km for Miranda, and $h_s\gtrsim 70$ km for the others, see Table \ref{Tab1}). 
The MOI of a three-layer interior is given by (e.g.~\citealt{VH2008})
\begin{equation}
    MOI=\frac{8\pi}{15 \,MR^2}\left(\rho_c R_c^5+\rho_o (R_o^5-R_c^5)+\rho_s(R^5-R_o^5)\right)
\end{equation}
with $R_o=R-h_s$, the mean ocean radius.
The ranges for MOI and gravity coefficients of our set of differentiated interiors are smaller than the values for homogeneous bodies. In this work, we neglect the effects of pressure on the density of materials, so that a body that is completely homogeneous in terms of its composition is of uniform density and has a MOI equal to $0.4$.

\begin{table}[h]
\small
\begin{center}
\footnotesize
\begin{tabular}{llccccc}
\hline
&& Miranda & Ariel & Umbriel & Titania & Oberon  \\
\hline
\textit{Three-layer interiors} &&&&&&\\
&&&&&&\\
Ice-I shell density& $\rho_s$ (kg~m$^{-3}$) & 700 - 1050& 700 - 1050& 700 - 1050& 700 - 1050& 700 - 1050\\
Water ocean density &$\rho_o$ (kg~m$^{-3}$) &700 - 1050& 700 - 1050& 700 - 1050& 700 - 1050& 700 - 1050\\
Rocky core density &$\rho_c$ (kg~m$^{-3}$) &2400 - 3500& 2400 - 3500&2400 - 3500& 2400 - 3500& 2400 - 3500\\
Ice-I shell thickness &$h_s$ (km) & 31.8 - 146.3 & 71.3 - 238.0 & 75.6 - 244.3  & 88.4 - 292.0 & 81.3 - 279.0\\
Water ocean thickness &$h_o$  (km)&2.5 - 50& 2.5 - 50&2.5 - 50& 2.5 - 50& 2.5 - 50\\
Rocky core radius &$R_c$ (km) & 87.0 - 154.0 & 338.4 - 457.6 & 337.9 - 459.1 & 494.4 - 650.5 & 479.9 - 630.1\\
&&&&&&\\
MOI & &  0.288 - 0.367&0.280 - 0.337&0.279 - 0.338&0.282 - 0.339&0.282 - 0.340\\
$J_{2}$ ($\times 10^5$) & & 442.0 - 810.6& 98.2 - 158.1&36.7 - 59.1&7.8 - 12.5&3.3 - 5.2\\
$C_{22}$ ($\times 10^5$) & & 132.6 - 243.2 & 29.5 - 47.4 & 11.0 - 17.7 & 2.3 - 3.8 & 1.0 - 1.6 \\
&&&&&&\\
\textit{Homogeneous interiors} &&&&&&\\
&&&&&&\\
MOI & & 0.4 & 0.4 & 0.4 & 0.4 & 0.4 \\
$J_{2}$ ($\times 10^5$) & & 1008.4 & 241.7 & 90.4 & 18.9 & 7.8\\
$C_{22}$ ($\times 10^5$) & & 302.5 & 72.5  & 27.1 &  5.7 & 2.4 \\
\\\hline
\end{tabular}
\end{center}
\caption{Density and dimension ranges of the three layers of the Uranian satellite interiors considered in this study to compute the rotational observables. The ranges of total MOI and degree-2 gravity coefficients $J_2$ and $C_{22}$ are also given, and compared with their values for homogeneous satellites of the same mass and radius. }
\label{Tab1} 
\end{table}

Since our three-layer interior ranges are essentially conservative extensions of \cite{Cas23}'s two-layer interior ranges, we here continue the discussion initiated in their Section 6.1. about the MOI and $C_{22}$ as observables for a future mission. $C_{22}$ is a classical result of radioscience experiments carried of space missions, while the MOI can be derived from $C_{22}$ using the Radau equation if the hydrostatic assumption applies, see Section \ref{Sec21}. 

\begin{figure}[!htb]
\begin{center}
\hspace{0cm}
\includegraphics[height=4.8 cm]{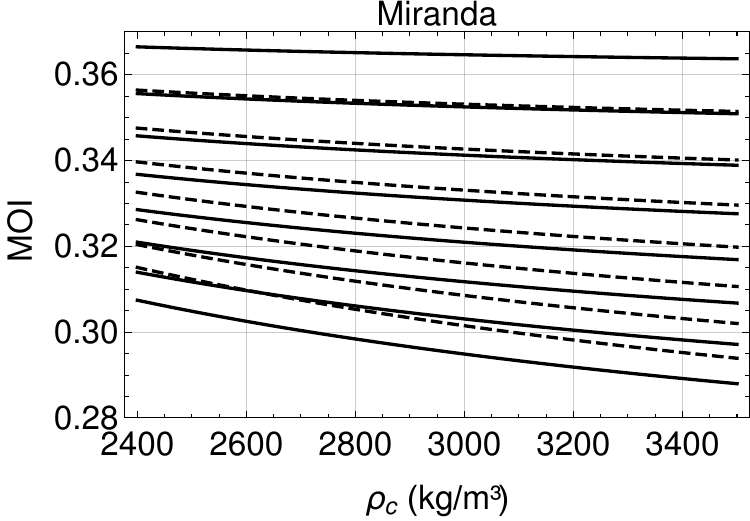}\quad
\includegraphics[height=4.8 cm]{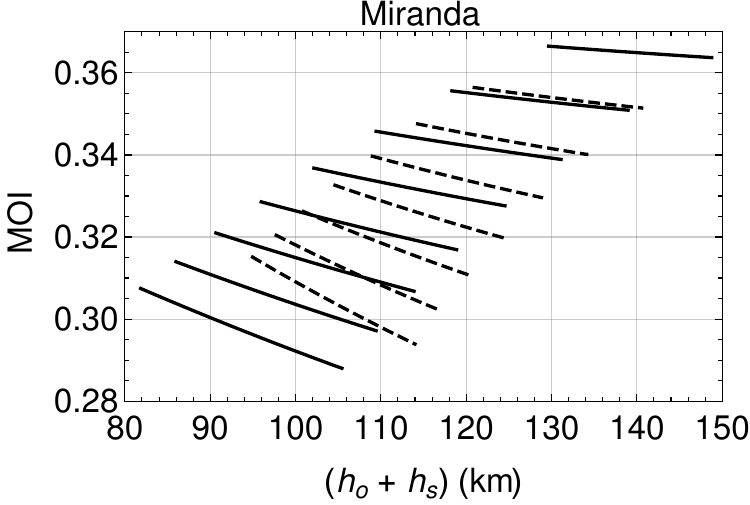}\\
\vspace{0.5cm}
\includegraphics[height=5 cm]{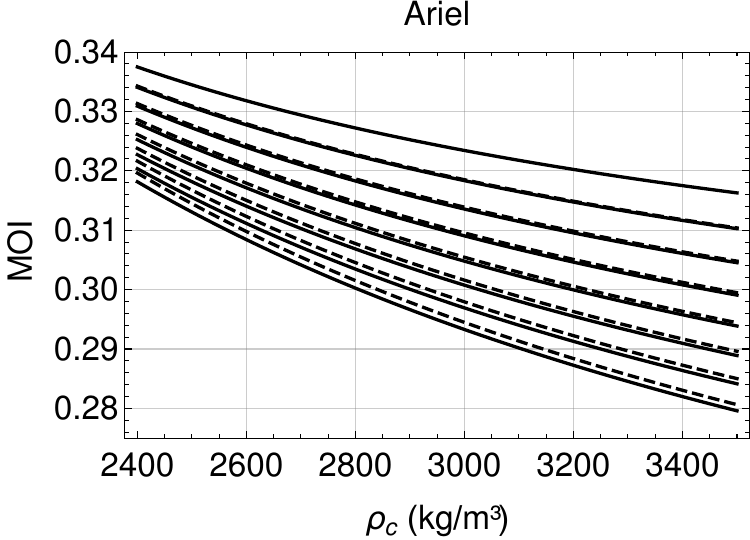}\quad
\includegraphics[height=5 cm]{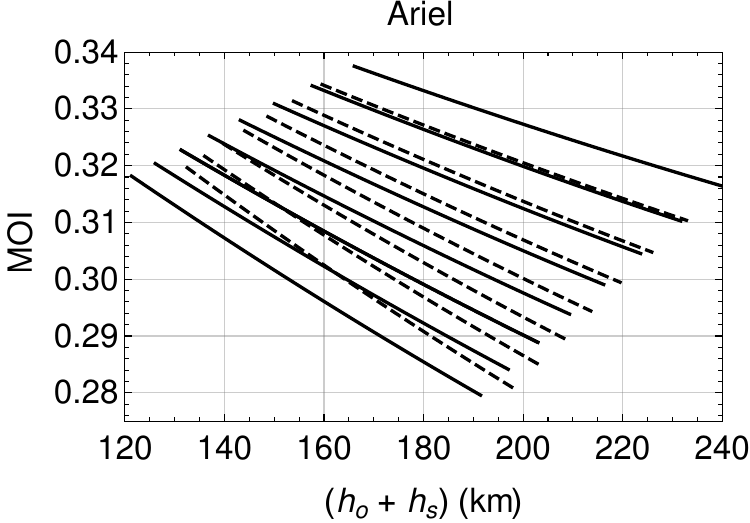}\\
\vspace{0.5cm}
\includegraphics[height=5 cm]{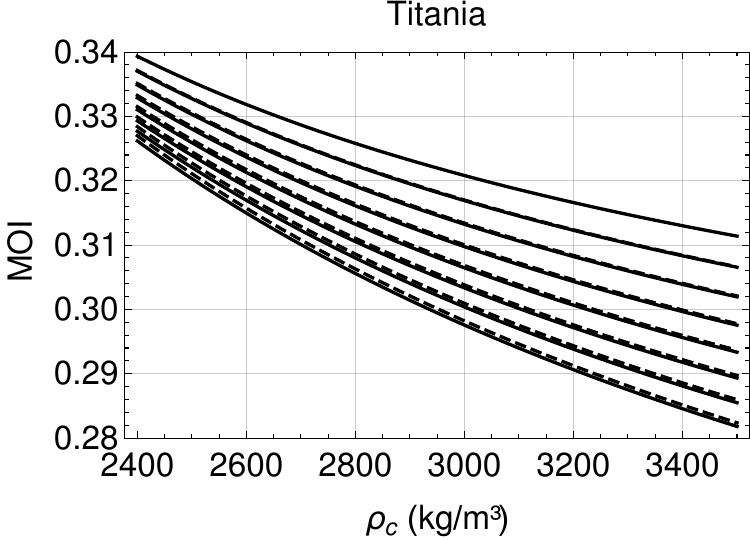}\quad
\includegraphics[height=5 cm]{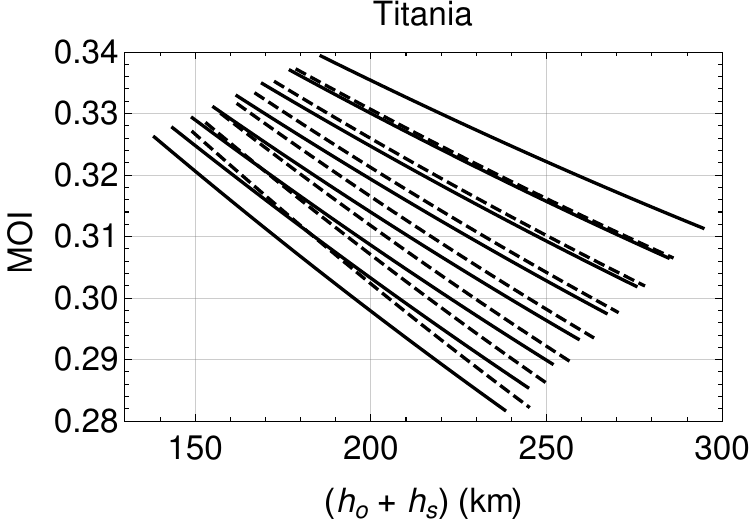}
\caption{MOI as a function of the rocky core density $\rho_c$ (left panels) and hydrosphere thickness $h_s+h_o$ (right panels) for three-layer Miranda, Ariel (proxy for Umbriel), and Titania (proxy for Oberon), as defined in Table \ref{Tab1}. This figure is based on Fig.~11 of \cite{Cas23}. 
The lines correspond, from bottom to top, to $\rho_s = 700$ kg/m$^3$ to $1050$ kg/m$^3$, in steps of $50$ kg/m$^3$ ($\rho_o=\rho_s$ and $\rho_o=1050$ kg/m$^3$ for the solid and dashed lines, respectively). The dashed lines also correspond to a fixed ocean thickness $h_o=50$ km, demonstrating the maximum impact that a density contrast can have on MOI.}
 \label{Fig3}
\end{center}
\end{figure}

The relationship between MOI and rocky core density and hydrosphere thickness shown in Fig.~11 of \cite{Cas23} concerns two-layer models with a rocky core and a hydrosphere. These two-layer models are identical to our three-layer models with equal density for shell and ocean in terms of mass distribution. In Fig.~\ref{Fig3}, we show the relationship between MOI and $\rho_c$ and $h_s+h_o$ for our three-layer interiors which also include a density difference between the shell and the ocean.

The prospects for constraining the interior from the MOI are questionable, due to the limited constraints on ice and water densities. 
We consider here, as \cite{Cas23} do, a precision of $\pm0.005$ on a future MOI determination, corresponding to an error of around $1.5\%$, which seems reasonable given the actual precision on gravity field determination of small icy satellites like Enceladus and Dione \citep{Ies12,Zan20}.
Here, with the conservative ranges considered for rock density and for the contrast between shell and ocean density, an accuracy of $\pm0.005$ on the MOI corresponds to an uncertainty of around $30$ km on the hydrosphere thickness of Miranda (versus the $15$ km reported in \cite{Cas23}) and even if the shell density was determined with a precision better than $50$ kg/m$^3$, the ocean density would remain essentially unconstrained for small $\rho_s$. 
For the medium-sized satellites Ariel, Umbriel, Titania and Oberon, an accuracy of $\pm0.005$ on MOI corresponds to no better than $50$ km  on the hydrosphere thickness, as already indicated by \cite{Cas23}. We note that, as with Miranda, considering a higher maximal core density, as we do here, tends to extend the range of hydrosphere thicknesses allowed for a given MOI. 

If we can't determine the interior parameters, and therefore the ice-water/rock ratio, can we at least determine whether a body is homogeneous or differentiated? This is the question that \cite{Cas23} addresses via their Fig.~12 for $C_{22}$ as a function of the MOI. First, we would like to point out that, assuming hydrostatic equilibrium, the MOI information is fully equivalent to the $C_{22}$ information. Fig.~12 of \cite{Cas23} illustrates Radau's relationship, which does not depend on the detailed interior structure, as shown here in Fig.~\ref{FigC20C22}. We thus focus the discussion here on the MOI, for which the precision required depends on the actual value of the parameter relative to its homogeneous counterpart of $0.40$. With the exception of Miranda, all the MOI values are below 0.34, even for low rock density. Lower MOI values would make it easier to distinguish between a homogeneous and a differentiated body, because the closer the value is to 0.4, the more accurate the measurement should be. For instance, if Oberon has an MOI of $0.34$, a precision of $\le0.06$ would be required to distinguish between differentiated and homogeneous interiors. If Miranda were to have an MOI of $0.29$, a precision of less than $0.11$ would be sufficient. In terms of $C_{22}$, this translates in precisions of $0.8\times10^{-5}$ and  $170\times10^{-5}$, respectively.

\subsection{Flattenings}

To model the rotation of a three-layer synchronous satellite in hydrostatic equilibrium, we need to calculate the moments of inertia of its layers, which depend on the polar ($\alpha$) and equatorial ($\beta$) flattenings of the surface and of the shell-ocean and ocean-core ellipsoidal interfaces. In the rest of the paper, the subscripts $s$, $o$, and $c$ refer to the flattenings at the surface, ocean radius, and core radius, respectively. Synchronous satellites are flattened by both the rotation and the static tides. The polar ($\alpha_s$) and equatorial ($\beta_s$) flattenings of the surface are defined as follows \citep{VH2008}
\begin{subequations}
\label{Eq2}
\begin{eqnarray}
    \alpha_s&=&\frac{3}{2}J_2+\frac{5}{4}q_r\\
    \beta_s&=&6C_{22}+\frac{3}{2}q_r=\frac{6}{5}\alpha_s
\end{eqnarray}
    \end{subequations}
and are therefore univocally related to the MOI through the gravity coefficients. If polar and/or equatorial geometric flattenings could be measured more precisely than the $J_2$ and $C_{22}$ gravity coefficients, then they could be used to derive MOI using the Radau equation. The flattenings at ocean radius $R_o$ ($\alpha_o$, $\beta_o$) and core radius $R_c$ ($\alpha_c$, $\beta_c$) are obtained by integration of Clairaut's differential equation. The relation $\beta=6 \alpha/5$ is valid at any depth (see details in \cite{VH2008}). 

\begin{figure}[!htb]
      \begin{center}
        \hspace{0cm}
\includegraphics[height=5 cm]{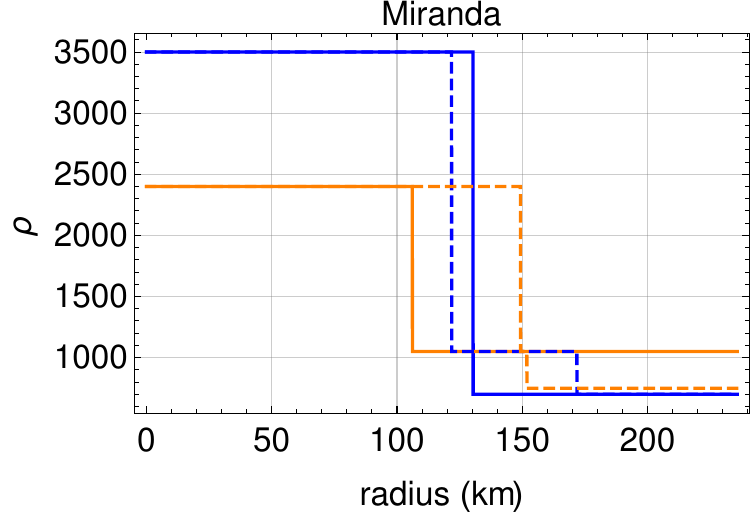}\quad 
\includegraphics[height=5 cm]{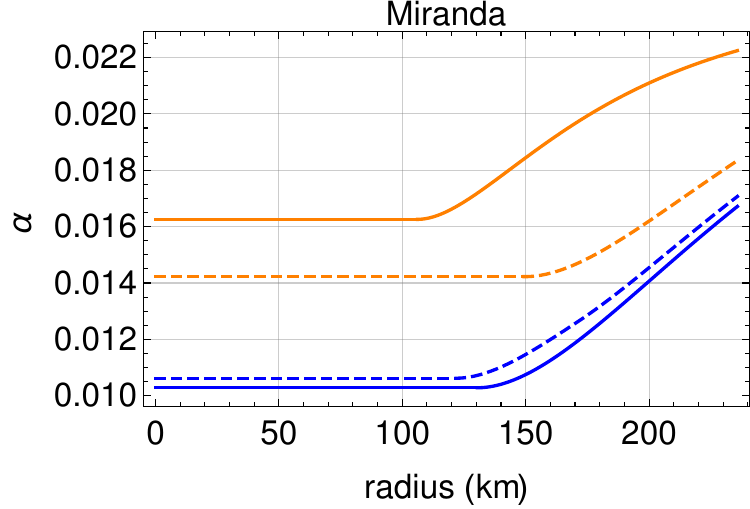}
 \caption{Density (left) and polar flattening (right) of Miranda as a function of the radius for interiors with minimal (solid blue line: $\rho_s=\rho_o=700$ kg/m$^3$, $\rho_c=3500$ kg/m$^3$, and $h_o=50$ km) and maximal (solid orange line: $\rho_s=\rho_o=1050$ kg/m$^3$, $\rho_c=2400$ kg/m$^3$, and $h_o=2.5$ km) MOI. The dashed blue and orange lines correspond to similar interiors, but with $\rho_o=1050$ kg/m$^3$ and $\rho_s=700$ kg/m$^3$, respectively.}
 \label{Fig4}
\end{center}
\end{figure}

Fig.~\ref{Fig4} right panel illustrates the relationship between polar flattening and depth, in the case of Miranda, for end-members three-layer interiors whose density profiles are given in the left panel. It can be seen that $\alpha$ increases from the center to the surface. $\alpha_s$ decreases with decreasing MOI and therefore tends to decrease with decreasing shell density $\rho_s$. A shell-ocean density contrast can also influence the flattening of the shell (especially for large MOI), but it will always remain within the range allowed by the minimum and maximum MOI considered.

\subsection{Tidal Love number \texorpdfstring{$k_2$}{k2}}
\label{Seck2}

Uranus's gravitational field is responsible for periodic deformations of its satellites, resulting in periodic variations in the satellites' gravitational potential that are proportional to the tidal Love number $k_2$. A small $k_2$ value indicates a rather rigid body with little deformation.
The rotation of synchronous satellites depends on tidal deformations, as periodic deformation bulges influence the couplings experienced by the different layers. Although the influence of the periodic tidal deformations on the rotation of large satellites can be very important (see \cite{VH2013} for libration), it is small for small satellites like Enceladus \citep{VH2016,Bal16}. We therefore expect it to be small for Miranda too, and to be somewhat more significant for the mid-sized Uranian satellites. We will quantify the elastic effect on libration in Section \ref{Sec3}. We consider ranges of $k_2$ that include the values reported in \cite{Hus06} for two-layer bodies along with results we obtain for three-layer models to discuss the possible effect of tidal deformations, in the presence or not of an ocean, on rotation. We leave the discussion of the effect of viscosity aside, as the detailed computation of the Love numbers of the Uranian satellites is not our objective.\\

For a homogeneous body of rigidity $\mu$, the tidal Love number can be expressed as \cite{Mel73}
\begin{equation}
\label{k2hom}
k_2=\frac{3}{2}\frac{1}{1+\frac{19\mu}{2 g \rho R}}.
\end{equation}
The values for the satellites of Uranus are given in the second column of table \ref{Tabk2} for $\mu=4\times 10^9$ Pa (a value that is best representative of pure ice), and are in agreement with the values of \cite{Che14}. For a two-layer solid body, the dependence of $k_2$ on the core radius (or equivalently the ice mantle thickness) is small in absolute value. We therefore only consider the unique values provided by \cite{Hus06} and reported in the third column of our Tab.~\ref{Tabk2}. For a three-layer body with a liquid ocean, we use a standard propagator matrix technique (e.g.~\cite{Sab16}). Following \cite{Hus06}, we take $\mu_s=3.3\times 10^9$ Pa and $\mu_c=50\times 10^9$ Pa for the shell and core rigidity, respectively. For the case where $\rho_s=\rho_o$, the \textit{Homogeneous crust model} of \cite{Beu15} (see its Appendix~A) is accurate to about $1\%$, compared with the standard model. As the Homogeneous crust model does not depend explicitly on the radius and density of the core, it sheds light on the results obtained with the standard model which do not vary significantly in relation to these parameters.  The ranges obtained with the general model are given in column 4 of table \ref{Tabk2}. The three-layer $k_2$ are $2$ to $20$ times larger than the two-layer $k_2$ due to the inclusion of a subsurface global ocean in the former models, while the value for the homogeneous case is similar to the lowest values of the 3-layer case.

\begin{table}[h]
\begin{center}
\begin{tabular}{lccc}
\hline
& Eq.~(\ref{k2hom})  & \cite{Hus06} & This paper \\
 & homogeneous & 2-layer & 3-layer \\
\hline
Miranda 	 & $0.00084$   & $0.00049$ & 0.00082 - 0.00422  \\
Ariel 	     & $0.00871$   & $0.00172$ & 0.00790 - 0.03659  \\
Umbriel 	 & $0.00870$   & $0.00225$ & 0.00788 - 0.0353  \\
Titania 	 & $0.01854$   & $0.00310$ & 0.01642 - 0.07308  \\
Oberon 	     & $0.01750$   & $0.00305$ & 0.01554 - 0.07206  \\
\hline
\end{tabular}
\end{center}
\caption{Love number $k_2$ for each satellite, for the solid homogeneous case (second column), the solid two-layer case (third column - values from \cite{Hus06}), and in the three-layer case with a liquid ocean (last column).}
\label{Tabk2}
\end{table}

\begin{figure}[!htb]
      \begin{center}
        \hspace{0cm}
\includegraphics[height=5 cm]{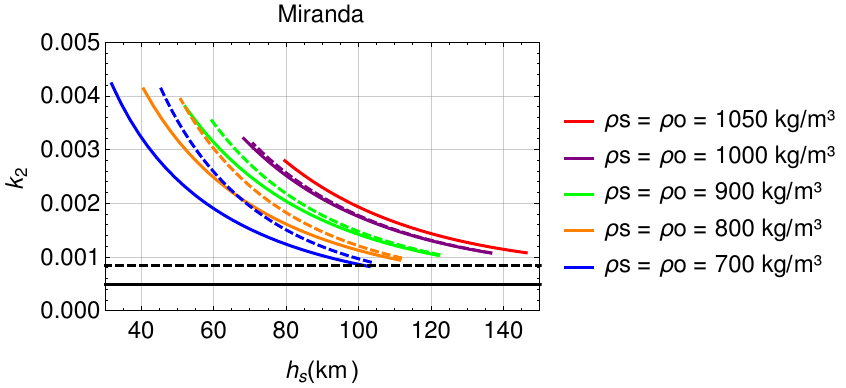}\\
\vspace{0.25cm}
\includegraphics[height=5 cm]{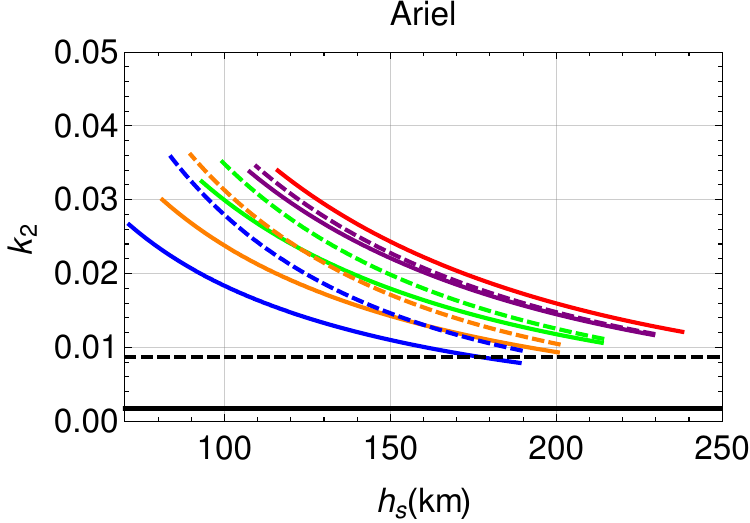}\quad
\includegraphics[height=5 cm]{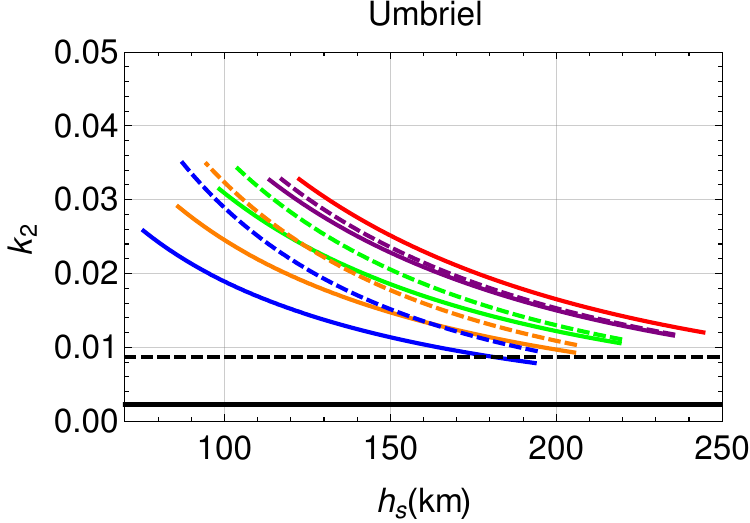}\\
\vspace{0.25cm}
\includegraphics[height=5 cm]{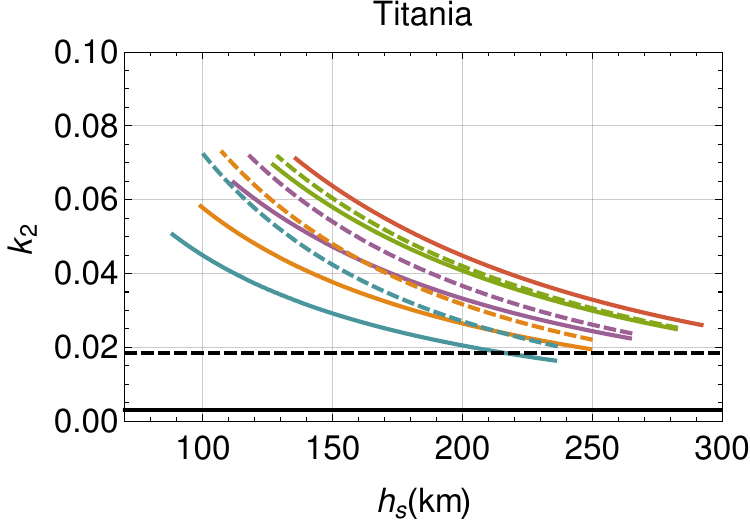}\quad
\includegraphics[height=5 cm]{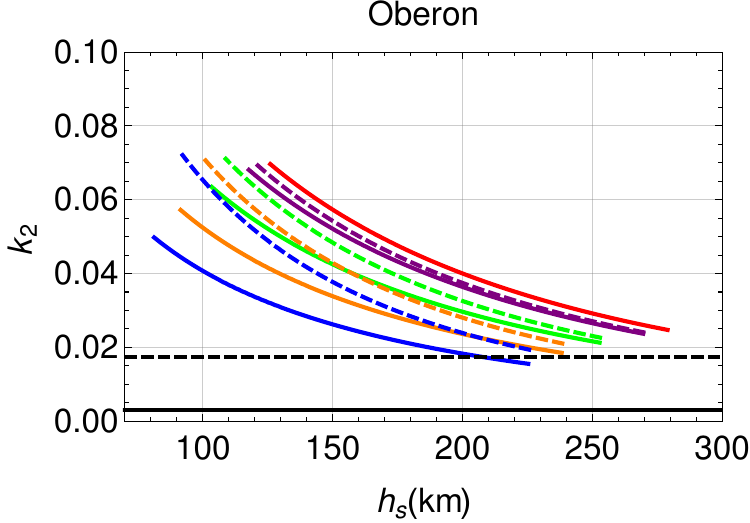}
 \caption{Love number $k_2$ as a function of the ice shell thickness $h_s$, for three-layer interiors as defined in Table~\ref{Tab1}. Colored solid lines correspond to cases where $\rho_o=\rho_s$ (values are given by the color code). Colored dashed lines correspond to the same shell densities and $\rho_o=1050$ kg/m$^3$. Solid and dashed black lines correspond to the 2-layer and homogeneous solid cases as reported in Table \ref{Tabk2}. }
 \label{Figk2}
\end{center}
\end{figure}

In the three-layer case, $k_2$ increases with decreasing shell thickness $h_s$, as illustrated on Fig.~\ref{Figk2}. $k_2$ can be as large as about $0.004$ for Miranda, $0.035$ for Ariel and Umbriel, and $0.07$ for Titania and Oberon. The lowest values (about $0.0008$ for Miranda, $0.008$ for Ariel and Umbriel, $0.016$ for Titania and Oberon) are close to those of the homogeneous case, as a result of the rigidity considered in the latter case. $k_2$ depends, but to a lesser extent, on ice rigidity and ice and ocean densities. It increases as rigidity decreases (not shown here). 
Although we expect that the denser the shell and ocean, the larger the tidal Love numbers, it should be noted that dense shell and ocean imply a thicker shell in order to satisfy the mass constraint, which mitigates the increase in $k_2$ in our set of interior models.

\newpage
\section{Rotation}
\label{Sec3}

The equator of Uranus is nearly at a right angle to its orbital plane. The orbits of the five large Uranian satellites are almost coplanar with Uranus's equator and have relatively small eccentricities. The inclinations are below $0.30$ degrees (with the exception of Miranda which has a $4.4^\circ$ inclination). Images taken by Voyager 2 have confirmed theoretical expectations that these satellites, like the Moon, are in synchronous rotation with their axis of rotation roughly parallel to Uranus's axis of rotation \citep{Smi86}. They are also most likely trapped in a low-obliquity Cassini State, see \cite{Gomes2024}.
In the following subsections, we study their forced diurnal librations and spin precession. 
We rely for this on previously published rotation models applied to e.g.~the Galilean satellites, Titan and Enceladus. 

\subsection{Orbital model}

We use orbital parameters derived from the URA111 ephemeris \citep{Jac14} as input of the rotation models. The level of detail required to describe a satellite's orbit varies according to what we aim to model. 

Regarding librations, we model only the diurnal component, which corresponds to variations in the rotation of a satellite in a Keplerian orbit. Other librations at other periods exist, due to orbit perturbations. Because of its short period, the diurnal libration could be identified as a signal distinct from other variations in space mission data (e.g.~\cite{Tho16}). It is also the rotation variation that depends most on the satellite's internal structure (long-period librations depend little on internal structure unless resonances would exist with internal modes). To model the diurnal libration, only the mean orbital eccentricity $e$ and mean motion $n$, as given in Table 2 of \cite{Jac14} and reproduced here in Table \ref{dataorb}, are relevant.

\begin{table}[h]
\begin{center}
\begin{tabular}{lccc}
\hline
& $e$ & $n$ (deg day$^{-1}$) & $2\pi/n$ (days) \\
\hline
Miranda 	 & $0.00135$  & $254.6357883$  & 1.41 \\
Ariel 	     & $0.00122$  & $142.8185916$  & 2.52 \\
Umbriel 	 & $0.00394$  & $86.8610921$   & 4.14 \\
Titania 	 & $0.00123$  & $41.3486828$   & 8.71 \\
Oberon 	     & $0.00140$  & $26.7387498$   & 13.46 \\
\hline
\end{tabular}
\end{center}
\caption{Mean orbital eccentricity $e$, mean motion $n$, and revolution period \citep{Jac14}.}
\label{dataorb}
\end{table}

Regarding spin precession and obliquity, the description of the satellite's orbit calls for greater caution. A satellite's obliquity and spin precession are driven by its inclination and orbital precession. The mean orbital elements with time-independent inclinations and precession rates of Table~2 of \cite{Jac14} are not a relevant representation in this context. The orbital precession is not necessarily uniform, and also inclination varies over periods much longer than the orbital period. Consequently, also the obliquity varies over long periods. These variations need to be modeled to predict obliquities from future spacecraft observations. We note that only information on the instantaneous (or averaged over a relatively short period) value of obliquity could be obtained from a visiting spacecraft (e.g.~\cite{Sti08}). To model spin precession and obliquity, we chose to use series for the orbital precession built with a secular perturbations model, see Appendix \ref{AppA}.

\subsection{Diurnal librations}

As a synchronous satellite revolves along its eccentric orbit, since its long axis does not quite point exactly towards the center of the planet, the planet exerts a time-varying gravitational torque on the satellite, causing diurnal longitudinal librations (e.g.~\cite{Com03}). In the presence of an internal ocean or any other global fluid layer, the response to the external gravitational torque is altered by the coupling between the satellite's different layers.
The amplitude of diurnal librations is also influenced by the periodic deformations of the body and its layers.

\subsubsection{Solid case}
\label{Sec321}

The amplitude of the forced diurnal libration of an entirely solid and rigid synchronous satellite is given by (e.g.~\cite{VH2008})
\begin{subequations}
\begin{eqnarray}
g &=& -6\frac{(B-A)}{C}\frac{n^2 e}{n^2 - \sigma_f^2},\\
\mathrm{where}\ \sigma_f &=& n\sqrt{\frac{3(B-A)}{C}}.
\end{eqnarray}
\end{subequations}
The amplitude $g$ is proportional to the orbital eccentricity $e$ and to the difference between the equatorial moments of inertia ($A<B$) of the satellite, and inversely proportional to the polar moment of inertia $C$. $(B-A)$ and $e$ represent the intensity of the gravitational torque exerted by the parent planet, while $C$ represents the resistance of the satellite to the applied torque. $\sigma_f$ is the natural frequency of the free libration. The libration amplitude of a solid and rigid satellite does not depend directly on the density and size of the different layers characterizing its interior structure, but only on its principal moments of inertia. Assuming hydrostatic equilibrium, it therefore follows from Radau's equation that the libration amplitude of a given satellite depends only on the MOI. Two interiors with different density profiles but with the same MOI will therefore have the same libration amplitude.

If Uranus's satellites were rigid, homogeneous bodies, their libration amplitude would range from around $1$ m for Oberon to $62$ m for Miranda at the surface, see Table \ref{TabLibObl}. The solid libration amplitudes show little or no resonant effect, as the free periods are about $4$ (Miranda) to $45$ (Oberon) times longer than forcing periods and $g\simeq -6 e (B-A)/C$. 
As a general rule, the closer a synchronous satellite is to its planet, the faster it rotates, the larger the difference $(B-A)/M\!R^2$ is and the larger the amplitude of libration. Satellites with particularly large eccentricities can contradict this rule (for example, Umbriel's eccentricity is almost 4 times larger than Ariel's, so its libration amplitude is also larger).
For a body differentiated into two layers, the reduction in $C$ compared with the homogeneous case does not compensate for the reduction in ($B-A$) due to differentiation. Consequently, a smaller MOI means a smaller (absolute) libration amplitude. For the two-layer interiors of \cite{Hus06}, the decrease in libration amplitude relative to the homogeneous case ranges from around $0.5$ m (Oberon) to around $13$ m (Miranda), see Table \ref{TabLibObl} (last column). These differences give an idea of the order of magnitude needed on a libration measurement to distinguish between a homogeneous and a differentiated solid body. This is mostly relevant for Miranda, which is unlikely to have an internal ocean \citep{Hus06,Bie22,Cas23}.

\begin{table}[h]
\begin{center}
\begin{tabular}{lcccccccc}
\hline
& $2\pi/ \sigma_f$ & $-g$  & $-g_{el}$ & Diff & $-g$ [m] & $-g_{el}$ [m] & Diff  & Differentiation\\
& [days]& [m]  &  [m] &   &  [m] &  [m] &  & effect\\
& \multicolumn{1}{l}{[} &
\multicolumn{2}{c}{MOI $=0.4$}    & \multicolumn{1}{r}{]}& \multicolumn{1}{l}{[} & \multicolumn{1}{c}{MOI $<0.4$} & \multicolumn{1}{r}{]}& [m]\\
\hline
Miranda 	 &4.73 &62.40 & 62.36 & -0.05$\%$ & 49.44  & 49.42&-0.04$\%$ &12.95\\
Ariel 	     &17.12 &31.28 & 31.13 & -0.50$\%$ & 20.79  & 20.75&-0.19$\%$ &10.49\\
Umbriel 	 &45.98 &37.74 & 37.55 & -0.49$\%$ & 26.82  & 26.77&-0.22$\%$&10.91\\
Titania 	 &211.30 &3.30  & 3.27  & -1.04$\%$ & 2.21   & 2.2  &-0.34$\%$&1.09\\
Oberon 	     &506.86 &1.51  & 1.49  & -0.98$\%$ & 1.01   & 1.01 &-0.33$\%$&0.49\\
\hline
\end{tabular}
\end{center}
\caption{Libration amplitude in the solid case. Columns 2 to 5 show the free libration period, the rigid and elastic amplitudes, and their relative difference for homogeneous bodies (MOI of $0.4$). Columns 6 to 8 show the rigid and elastic amplitudes and their difference, for the two-layer interiors as defined in \cite{Hus06} (MOI $<0.4$). The last column shows the difference between the solid amplitudes of the homogeneous and two-layer interiors.}
\label{TabLibObl}
\end{table}

When a solid satellite is elastic and deformable, the gravitational torque is exerted not only on its static bulge, but also on its periodic tidal bulge. Moreover, the satellite's response to the torque is altered not only by the tides, which modify its resistance to changes in rotational speed, but also by the deformations caused by the changes in rotational speed themselves. The amplitude of the forced diurnal libration of an entirely solid and elastic synchronous satellite is given by (e.g.~\cite{VH2013})
\begin{subequations}
\begin{eqnarray}
g_{el} &=& -6 \frac{k_f-5k_2/6}{k_f}\frac{( B- A)}{\tilde C}\frac{n^2 e}{n^2 - \sigma_{el,f}^2},\\
\mathrm{where}\ \sigma_{el,f} &=& n\sqrt{\frac{3(B-A)}{\tilde C} \frac{k_f-k_2}{k_f}},\\
\mathrm{and}\ \tilde C&=&C+\frac{4k_2 n^2 R^5}{9G} .
\label{Ct}
\end{eqnarray}
\end{subequations}
Here $k_f$ is the fluid Love number obtained for instance from Eq.~(A3) of \cite{VH2008} as a function of the MOI ($k_f=3/2$ for a homogeneous body), and $k_2$ is the classical dynamical Love number (see Section~\ref{Seck2}). In general, elastic deformations tend to reduce the (absolute) amplitude of forced diurnal libration, compared with the rigid case. For large icy satellites considered as solid bodies, the decrease is limited (about $7\%$ for Ganymede, about $1\%$ for Europa, see \cite{VH2013}). For the small/medium-sized satellites of Uranus, the difference is less than about $1\%$ for both homogeneous and two-layer interiors, see Table \ref{TabLibObl}. 

\subsubsection{Ocean case }
\label{Sec322}

The amplitude of the diurnal librations of the shell and of the solid interior (the rocky core here) is given by Eqs.~(74-75) of \cite{Bal10} for rigid solid layers and by Eqs.~(62-63) of \cite{VH2013} for elastic layers. 
We will note $g_s$ and $g_c$ the amplitude of shell and core librations, respectively. Through an order-of-magnitude analysis, \cite{Bal10} have approximated their solution for $g_s$ and obtained an expression (their Eq.~81) that is valid for a relatively thin shell. Generalizing this approach for thicker shells, we obtain :
\begin{equation}
\label{gsapp}
 g_s \approx - 6 e\, \beta_s  \frac{ R^5+ R_o^5 (\rho_o-\rho_s)/\rho_s}{ R^5 - R_o^5}.
\end{equation}
This approximation is not accurate (differences up to $20\%$ for Oberon, mainly because the approximate expression does not include the effects of periodic deformations) but behaves in the same way as the exact solution in terms of interior parameters (layer's density and thickness). It will therefore greatly help us understand the numerical results obtained with the exact solution, and in particular their dependence on $R_o$ (or equivalently on $h_s=R-R_o$), on $\beta_s$ (or equivalently on the MOI), and on $(\rho_o-\rho_s)/\rho_s$. Note that \cite{Hem24} consider a fixed value of $100$ kg/m$^3$ for the density contrast between shell and ocean (\cite{Park2020} considered the same densities for shell and ocean). By not varying the density difference between shell and ocean, one could draw somewhat over-optimistic conclusions about the possibility of constraining shell thickness from a libration measurement.

The effect of periodic deformations on the diurnal librations of a satellite that harbors an internal liquid layer has been assessed in details by \cite{VH2013,VH2016,VH2020} (see also \cite{Jar14}). They showed that deformation reduces the effective gravitational torque on the satellite by a factor equal to the relative ratio between the fluid and tidal Love numbers $(k_f-k_2)/k_f$. Since this torque is the cause of the librations, the Love number ratio gives a first indication of the reducing effect of deformation of libration. 
As $k_2$ tends to increase with satellite size, the Love numbers' relative ratio of large satellites tends to be well below $1$ (e.g.~$(k_f-k_2)/k_f\simeq 0.75$ and $0.40$ for Europa and Ganymede, respectively, \cite{Wu2001,Wah06,VH2013}) and their librations are strongly reduced (up to one order of magnitude) by periodic deformations.  
The librations of Enceladus, one of the smallest quasi-spherical satellites in the solar system, have been shown to be almost unaffected (about $10\%$ decrease at most) by periodic deformations ($(k_f-k_2)/k_f\simeq 1$ for interior models with a shell between $30$ and $50$ km thick, and $\simeq 0.9$ for a $10$ km thick shell, \cite{VH2016,Bal16}). 

From Section \ref{Seck2}, we infer that for the interior defined in Section \ref{Secinteriors}, the ratio $(k_f-k_2)/k_f$ ranges from $0.92$ for Oberon and Titania to almost $1$ for Miranda. For Ariel and Umbriel, the ratio is somewhere in between. These large ratios are explained by the fact that we are considering relatively thick shells. We therefore anticipate that the effect of periodic deformations on the amplitude of shell libration is negligible for the small Miranda (a situation similar to Enceladus with a thick shell). As for the other (medium-sized) satellites, we expect a situation more similar to that of Enceladus with a thin shell and a reduction in the amplitude of shell libration by about $10\%$. 

\begin{figure}[!htb]
      \begin{center}
        \hspace{0cm}
\includegraphics[height=5 cm]{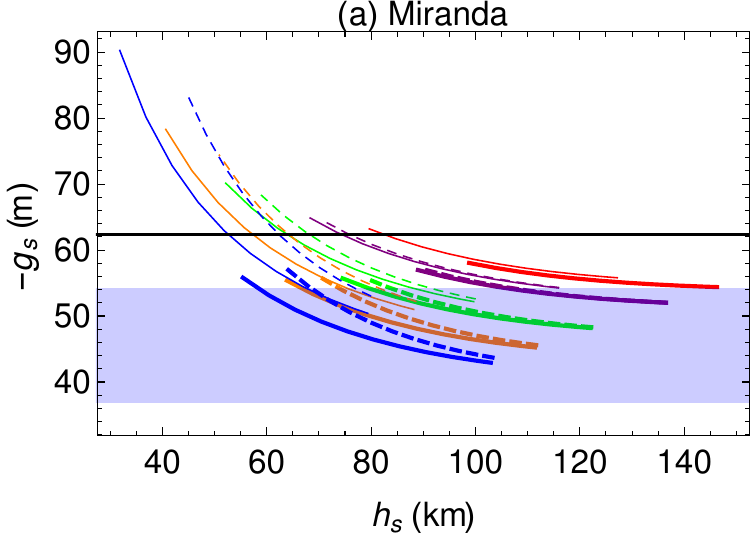}\quad
\includegraphics[height=5 cm]{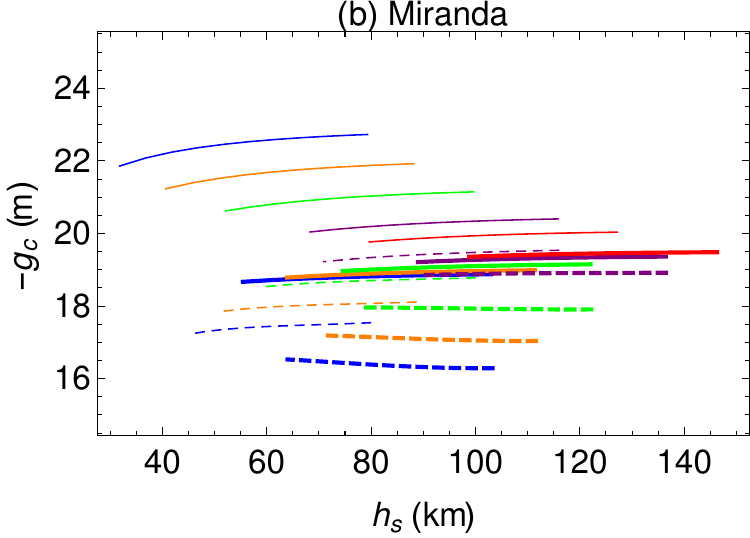}\\
\vspace{0.25cm}
\includegraphics[height=5 cm]{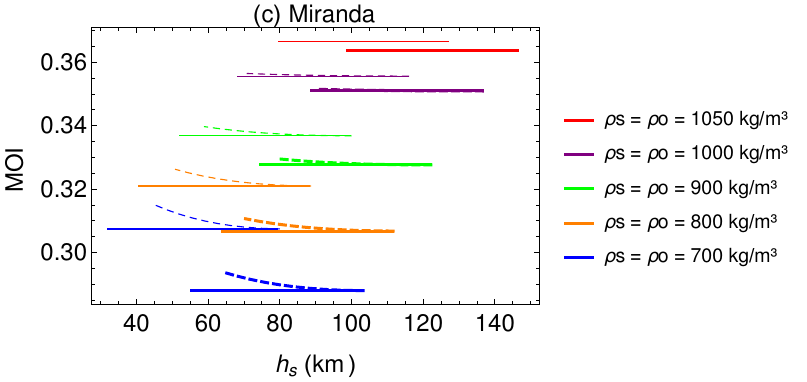}
\caption{Top: Miranda shell and core libration amplitude $g_s$ and $g_c$, both multiplied by the total radius, as a function of the shell thickness $h_s$ for a selection of interiors representative of the range of interiors defined in Section \ref{Secinteriors}. Solid lines correspond to cases where $\rho_o=\rho_s$ (values are given by the color code). Dashed lines correspond to the same shell densities as for solid lines, but with $\rho_o=1050$ kg/m$^3$. Thick (thin) lines correspond to interiors with $\rho_c=3500$ ($2400$) kg/m$^3$. The blue region in the first panel corresponds to the range of possible values for two-layer solid models with a MOI ranging from $0.288$ to $0.367$. The black line corresponds to the libration of a single-layer homogeneous body. The effects of periodic elastic deformations are considered here, for both solid and ocean cases.
Bottom: relationship between $h_s$ and MOI for the same three-layer interiors.}
\label{gsMiranda}     
\end{center}
\end{figure}

\subsubsection*{Small-sized Miranda}

For a given set of layer densities $\rho_s$, $\rho_o$ and $\rho_c$, the amplitude of libration of the shell $g_s$ increases with decreasing shell thickness $h_s$, see each indivudal curve on Fig.~\ref{gsMiranda} panel (a). This can be understood from the denominator in Eq.~(\ref{gsapp}) with $R_o=R-h_s$, which implies a $1/h_s$ dependence. 
For interiors with equal shell and ocean densities ($\rho_o=\rho_s$), since $g_s\propto \beta_s$ (see Eq.~\ref{gsapp}), and since $\beta_s$ increases with $\rho_s$, $g_s$ should increase with increasing shell density. However, this increase is counteracted by the effect of an increase in shell thickness that goes hand in hand with an increase in shell density in our set of interior models constrained by the total mass. Consequently, as for $k_2$ (see Fig.~\ref{Figk2}), an increase in shell density is characterised by a shift to the right of the curves, instead of a strictly upward shift. Similarly, a density contrast between the ocean and the shell results in a shift to the right of the curves, instead of the upward shift expected from the numerator of Eq.~(\ref{gsapp}).
Unlike $k_2$, $g_s$ is sensitive to the core density $\rho_c$. For each given set of layer densities, the MOI is almost constant (and so is $\beta_s$). As a result, when $\rho_c$ is decreased while the MOI is kept essentially unchanged, the shell is thinner and the libration is larger
(see solid thick orange and solid thin blue curves in Fig.~\ref{gsMiranda} panel (a), which correspond to interiors with the same MOI of about $0.3075$, see panel (c), but with different sets of layers densities).
The amplitude of libration of the core $g_c$ is about three times smaller than that of the shell, and depends very little on the thickness of the shell. It decreases with increasing core density and shell-ocean density contrast (panel b).

In the absence of any other constraint, such as the MOI of the satellite, it would be quite difficult to interpret a libration measurement. A measured amplitude of about $62$ m could correspond to a differentiated interior with ocean, but also to a homogeneous interior, as shown in Fig.~\ref{gsMiranda}. If we consider that a homogeneous interior is unlikely, an amplitude significantly larger than $54$ m would be necessary to detect the presence of an ocean. The larger the libration, the greater the possibility of constraining shell thickness, given the $1/h_s$ dependence of the libration amplitude.

\begin{figure}[!htb]
      \begin{center}
        \hspace{0cm}
\includegraphics[height=5 cm]{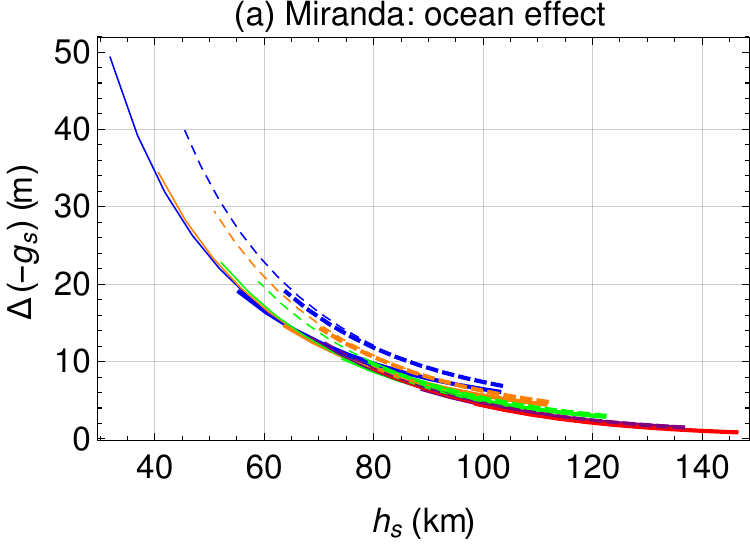}\quad
\includegraphics[height=5 cm]{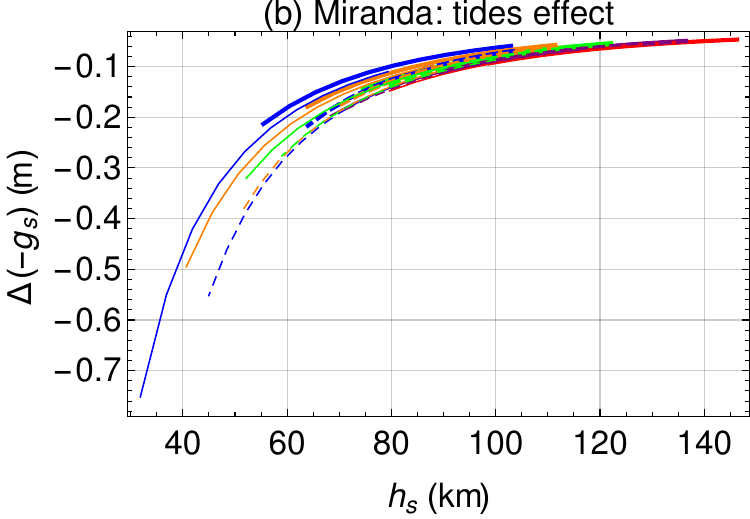}
 \caption{Left: Difference in libration at the surface of Miranda between the case with and without ocean (both with periodic deformations) as a function of shell thickness $h_s$. As each individual curve corresponds approximately to a given MOI (see bottom panel of Fig.~\ref{gsMiranda}), these differences correspond to the precision required to confirm the presence of the ocean from a surface libration measurement if the MOI is determined independently (from the gravity field, for example).
Right: Difference in libration at the surface of Miranda between ocean cases with and without periodic deformations as a function of shell thickness $h_s$. 
Solid and dashed lines defined as in Fig.~\ref{gsMiranda}.}
 \label{FigDeltagsMiranda} 
\end{center}
\end{figure}

If hydrostatic equilibrium applies, it should be possible to obtain independent estimates of the MOI from the gravity field measurement and Radau's equation. Assuming the MOI is known, the difference in libration between the case with and without the ocean 
increases with decreasing ice shell thickness, see Fig.~\ref{FigDeltagsMiranda}. If Miranda were to harbor a liquid ocean and its shell thickness were $40$ ($100$) km, then an accuracy of around $20$ ($2.5$) m would be required on the libration estimate to detect the ocean. Assuming the MOI is known, it would even be possible to constrain, for example, the shell thickness (especially for a relatively thin shell). For instance, if Miranda’s shell thickness were $60$ km, a precision of $10$ m on the libration measurement would result in a $20$ km precision on the shell thickness, see Fig.~\ref{FigDeltagsMiranda}, whereas only a lower bound at about $40$ km could be inferred if the MOI were unknown (see Fig.~\ref{gsMiranda}).

Although the presence of an ocean increases the amplitude of libration by about $50$ m for the thinnest shells considered here (Fig.~\ref{FigDeltagsMiranda}, panel a), it should be kept in mind that this increase is counteracted to some extent by the effect of periodic deformations. In the case of Miranda, the reduction is minimal (less than $1$ m, panel b), as anticipated above from the value of the $(k_f-k_2)/k_2$ ratio.

\begin{figure}[!htb]
      \begin{center}
        \hspace{0cm}
\includegraphics[height=5 cm]{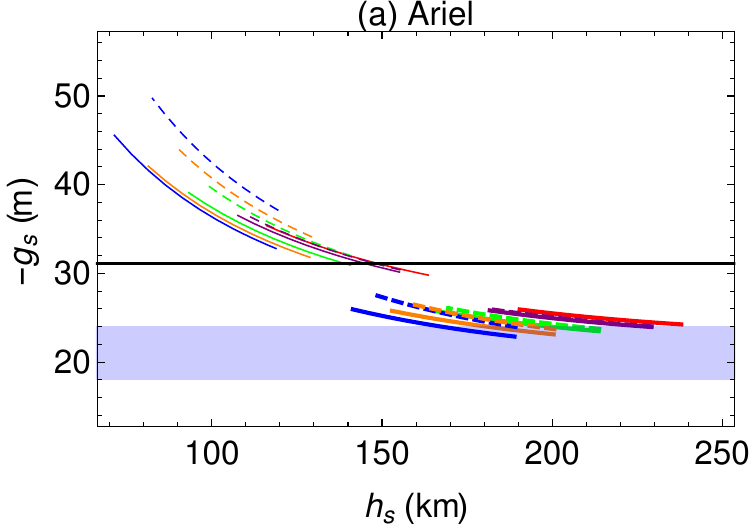}\quad
\includegraphics[height=5 cm]{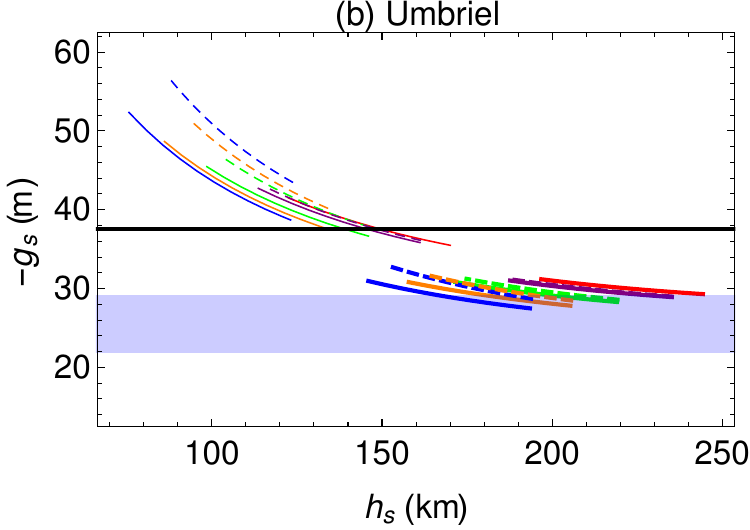}\\
\vspace{0.25cm}
\includegraphics[height=5 cm]{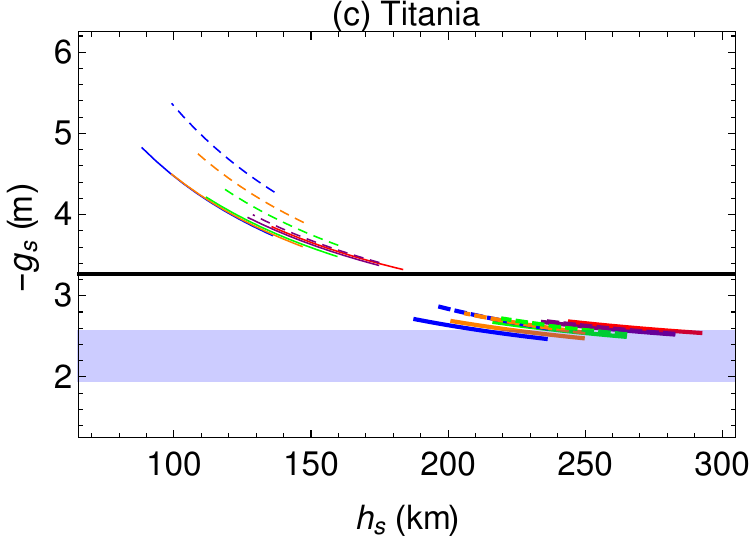}\quad
\includegraphics[height=5 cm]{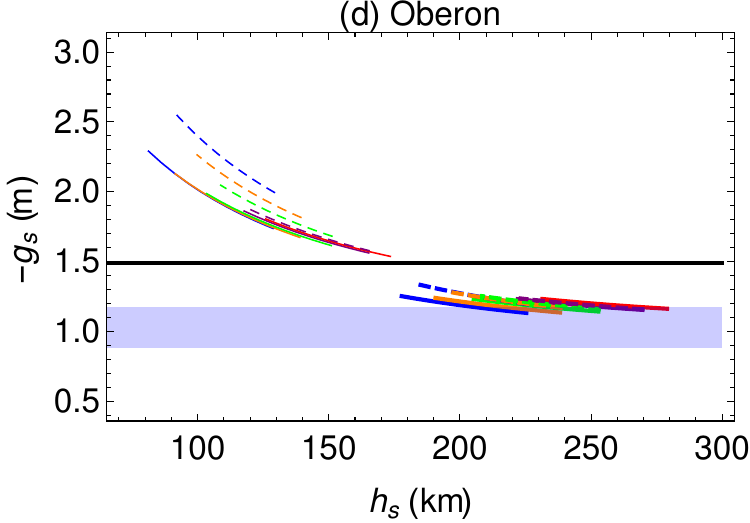}
 \caption{Ariel, Umbriel, Titania, and Oberon shell libration amplitude $g_s$ as a function of the shell thickness $h_s$ for a selection of interiors representative of the range of interiors defined in Section \ref{Secinteriors}. Solid and dashed lines defined as in Fig.~\ref{gsMiranda}. The blue region corresponds to the range of possible values for two-layer solid models with a MOI ranging from about $0.28$ to about $0.34$. The effects of periodic elastic deformations are considered here, for both solid and ocean cases.}
 \label{gsAriel} 
      \end{center}
\end{figure}

\newpage
\subsubsection*{Medium-sized satellites}

For Ariel, Umbriel, Titania and Oberon, as for Miranda, we observe that the shell libration amplitude depends mainly on the shell thickness and to a lesser extent on the density difference between the shell and the ocean. The dependence of $g_s$ on the MOI is less marked than for Miranda, see Fig.~\ref{gsAriel}, where the curves are more clustered than in Fig.~\ref{gsMiranda} panel (a). The possibility of interpreting a surface libration measurement in terms of shell thickness would therefore be less affected by the absence of an independent determination of MOI than for Miranda.
As for Miranda, the amplitude of libration of the core $g_c$ is about three times smaller than that of the shell, and depends very little on the thickness of the shell (not shown here). 

\begin{figure}[!htb]
      \begin{center}
        \hspace{0cm}
\includegraphics[height=5 cm]{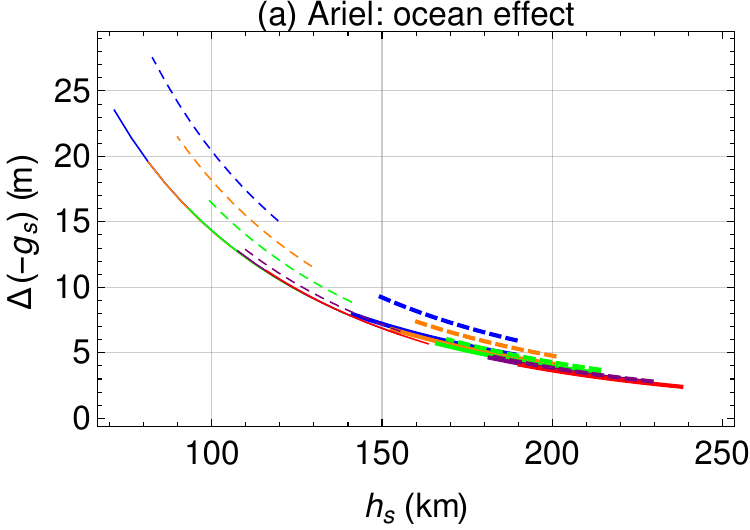}\quad
\includegraphics[height=5 cm]{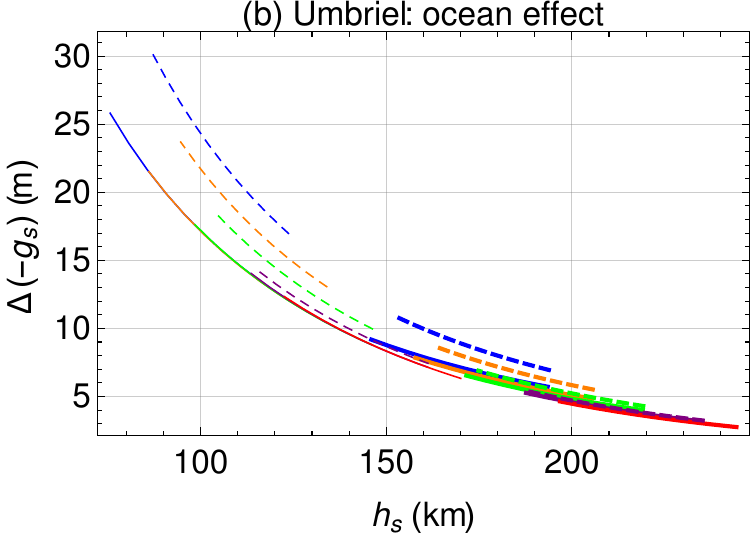}\\
\vspace{0.25cm}
\includegraphics[height=5 cm]{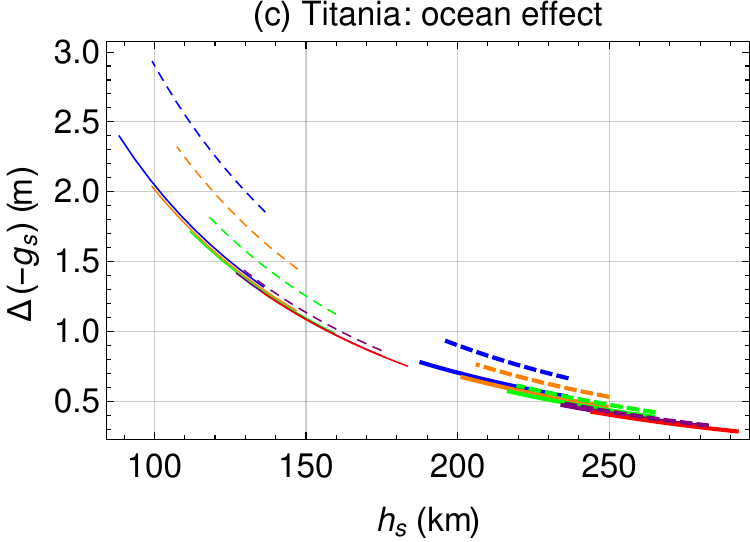}\quad
\includegraphics[height=5 cm]{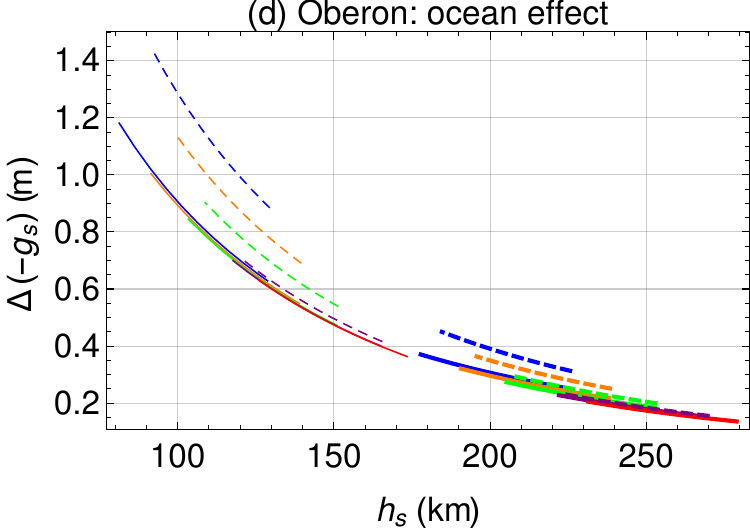}
\caption{Difference in libration at the surface of Ariel, Umbriel, Titania, and Oberon between the case with and without an ocean (both with periodic deformations) as a function of shell thickness $h_s$. Solid and dashed lines defined as in Fig.~\ref{gsMiranda}. As each individual curve corresponds approximately to a given MOI, these differences correspond to the precision required to confirm the presence of the ocean from a surface libration measurement if the MOI is determined independently (from the gravity field, for example).
Solid and dashed lines defined as in Fig.~\ref{gsMiranda}.}
\label{FigDeltagsAriel}
      \end{center}
\end{figure}

If the MOI is known, an accuracy of around $5$ m on a surface libration measurement should enable us to detect an ocean for Ariel and Umbriel, if the shell thickness is around $150$ km. For Titania and Oberon, one should aim for accuracies of around $0.65$ and $0.3$ m, respectively, see Fig.~\ref{FigDeltagsAriel}. More accurate measurements are required if the shell would be thicker.\

\begin{figure}[!htb]
      \begin{center}
        \hspace{0cm}
\includegraphics[height=5 cm]{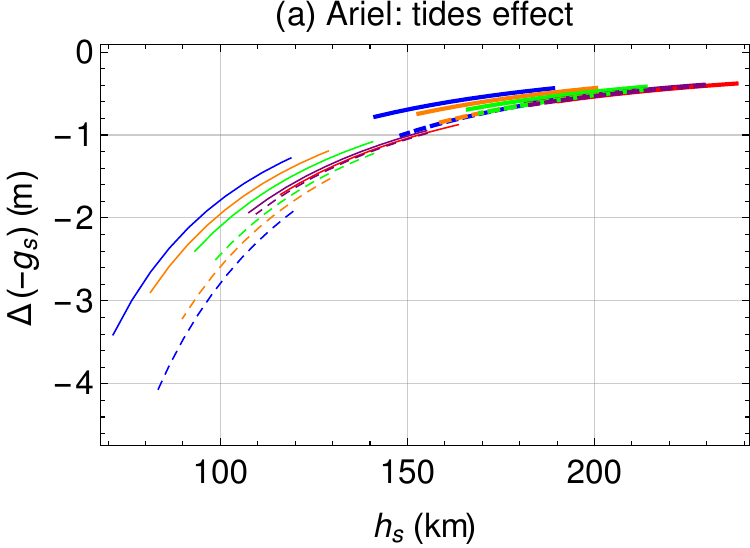}\quad
\includegraphics[height=5 cm]{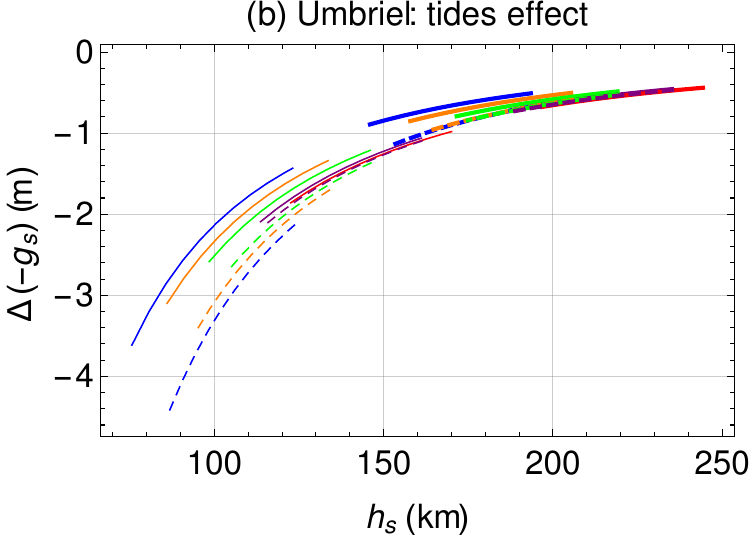}\\
\vspace{0.25cm}
\includegraphics[height=5 cm]{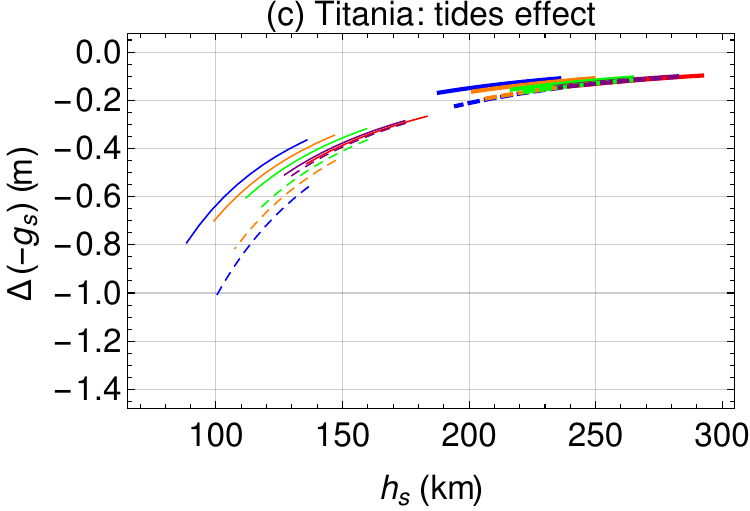}\quad
\includegraphics[height=5 cm]{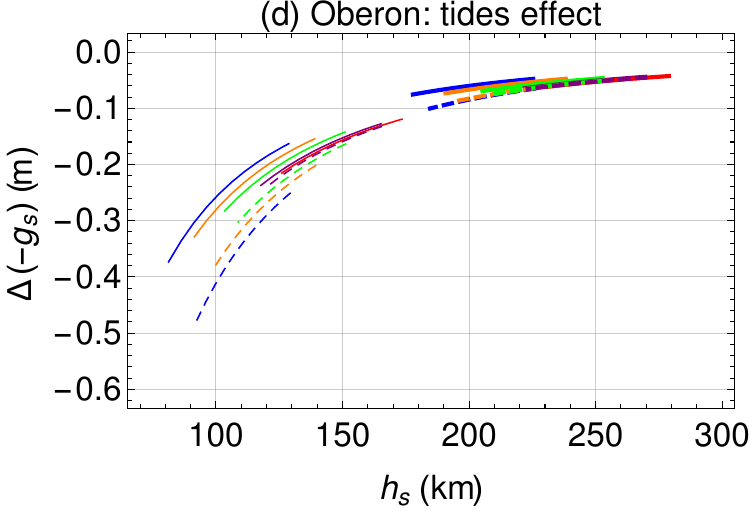}
\caption{Difference in libration at the surface of Ariel, Umbriel, Titania, and Oberon between the ocean cases with and without elastic deformations as a function of shell thickness $h_s$. 
Solid and dashed lines defined as in Fig.~\ref{gsMiranda}.}
\label{FigDeltagsk2Ariel}
      \end{center}
\end{figure}

Periodic deformations reduce shell librations by up to $8\%$ (Ariel, Umbriel) and $15\%$ (Titania, Oberon), see Fig.~\ref{FigDeltagsk2Ariel}.

\newpage

\subsection{Spin axis orientation}

As a synchronously rotating satellite revolves along its precessing and inclined orbit, its small axis - i.e.~the axis of rotation if we neglect polar motion - cannot remain parallel to the normal to the orbital plane. The planet exerts a gravitational torque on the satellite, causing its equatorial plane to precess around a mean plane called the Laplace plane, with an inclination that is usually referred to as the obliquity. Here, because of the large distance from the Sun, the equator of Uranus in J2000 (JD 2451545.0, 2000 January 1 12.0 h) is a good approximation of the Laplace plane.
When the orbital precession has a constant rate and a time-independent inclination, the spin precesses at the same rate as the orbit and the obliquity remains constant over time. In this configuration, known as a Cassini state, the rotation axis, the normal to the orbit, and the normal to the Laplace plane remain in the same plane (the Cassini plane), see e.g.~\cite{Hen04}. However, as discussed in \ref{AppA}, the orbits of the Uranian satellites do not precess at constant rates and with constant inclinations due to their mutual gravitational interactions. By expressing the orbital precession of a synchronous satellite as a series of superimposed uniform orbital precessions, it is possible to express the spin precession also as a series. \cite{Bil05} has shown that coplanarity is maintained on a mode-by-mode basis. This multi-frequency/non-uniform approach has already been applied to Titan, the Galilean satellites and Enceladus \citep{Bal11,Bal12,Bal16} with or without an internal ocean and more recently to the largest Uranian satellites considered as solid and rigid bodies by \cite{Gomes2024} in order to study the evolution of the system.

\begin{figure}[!htb]
      \begin{center}
        \hspace{0cm}
\includegraphics[width=12 cm]{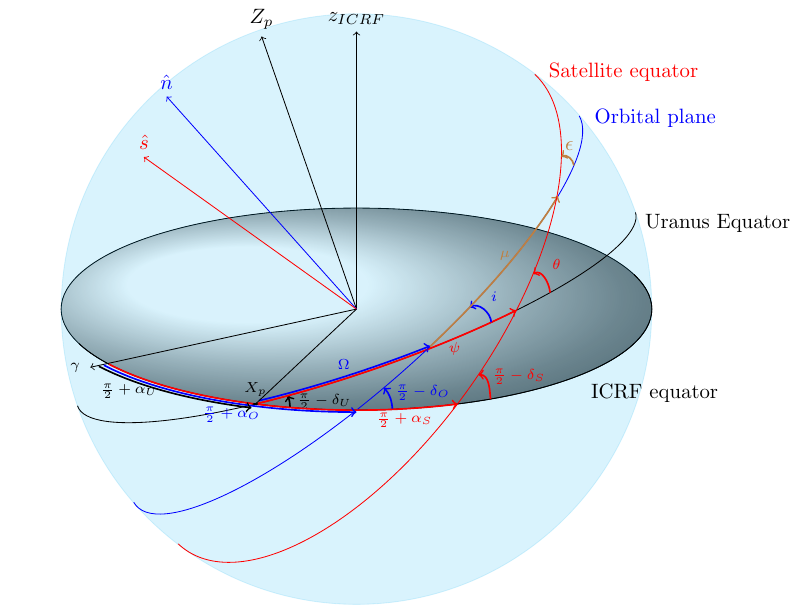}
\caption{The different planes and angles used to describe the precession of the Uranian satellites. The equator of Uranus, the orbital and equatorial planes of the satellite are oriented with respect to the ICRF equator with right ascension and declination angles ($\alpha$ and $\delta$). The orbital and equatorial planes of the satellite are oriented with respect to the Body Frame (BF) of Uranus with the Euler angles for node longitude and inclination $(\Omega,i)$ and $(\psi,\theta)$, respectively. The satellite equator orientation with respect to the orbital plane is defined by the Euler angles $\mu$ and $\varepsilon$. We refer to the latter as the obliquity, and to $\theta$ as the inertial obliquity. $X_p$ and $Z_p$ are the $x-$ and $z-$ axes of Uranus prograde BF, as defined in Fig.~\ref{FigBF}. $\hat s$ and $\hat n$ are the unit vectors along the spin axis and the orbital normal, respectively. $\gamma$ is in the direction of the vernal point. }
\label{FigAngles}
      \end{center}
\end{figure}

Similarly as for the variables $p(t)$ and $q(t)$ for the orbital precession, see Eq.~(\ref{pq}), we now express the spin precession of each satellite as a series and define the following variables
\begin{subequations}
\label{pqspin}
\begin{eqnarray}
    p_{spin}(t)&=&\sin \theta(t) \sin \psi(t)=\sum_{k=1}^{5} \sin{\theta_k} \sin (\dot\Omega_k t +\phi_k)\\
    q_{spin}(t)&=&\sin \theta(t) \cos\psi(t)=\sum_{k=1}^{5} \sin{\theta_k} \cos(\dot\Omega_k t +\phi_k),
\end{eqnarray}
\end{subequations}
where $\theta(t)$ is the time-variable obliquity of the spin axis, measured from the normal to the Laplace plane (called \textit{inertial obliquity} in the following) and $\psi(t)$ is its node longitude, see Fig.~\ref{FigAngles}.
Note that ($p_{spin},q_{spin})=(s_x,-s_y)$ and $(p,q)=(n_x,-n_y)$ with $(s_x,s_y)$ and $(n_x,n_y)$ the equatorial components of the unit vectors $\hat s$ and $\hat n$ along the spin axis and the orbit normal, respectively, in the coordinates of a frame associated with the Laplace plane. 

The inertial obliquity $\theta$ should not be confused with $\varepsilon$, the obliquity measured from the orbital plane ($\cos \varepsilon=\hat n . \hat s$, or for small inclination and obliquity, $\varepsilon\simeq\sqrt{(s_x-n_x)^2+(s_y-n_y)^2}$).
Although $\varepsilon$ is a more commonly used concept, $\theta$ is more convenient for describing the orientation of the axis of rotation in space. $\theta$ and $\psi$ are sufficient to orientate the spin axis with respect to Uranus Body Frame (BF) at J2000. To do that with $\varepsilon$ requires the use of three additional angles ($\mu$, $\Omega$ and $i$). $i$ and $\Omega$ are the orbital inclination and node longitude, respectively. $\mu$ is the angle from the node of the orbital plane on the Laplace plane to the node of the satellite equator on the orbital plane ($\sin\mu=\sin\theta \sin(\psi-\Omega)/\sin\varepsilon$). 
Only in the case of uniform orbital precession is $\varepsilon$ a convenient angle (in this case, $\theta=i+\varepsilon$, $\psi=\Omega$ and $\mu=0$). 
The angles $\theta$ and $\psi$ can also conveniently be transformed into equatorial coordinates $\alpha_S$ (right ascension) and $\delta_S$ (declination) expressed in the ICRF (International Celestial Reference Frame, see \citealt{Arc18}), in exactly the same way as $i$ and $\Omega$ are transformed to $\alpha_O$ and $\delta_O$ for the orbit (subscript $S$ refers to the spin axis and $O$ refers to the orbit).  

In our secular model, see Table~\ref{TabSeries}, the frequencies $\dot\Omega_k$ and phases $\phi_k$ in Eq.~(\ref{pqspin}) are common to all five satellites and are the same as those used to describe the orbital precession in Eq.~(\ref{pq}). The set of inertial obliquity amplitudes $\theta_k$ differ for each satellite. For each inclination amplitude $i_k$, there is a corresponding obliquity amplitude $\theta_k$, calculated using a Cassini state model chosen on the basis of the assumptions made (is the body rigid or elastic, with or without an internal ocean). On a mode-by-mode basis, we can define the obliquity amplitude as $\varepsilon_k=\theta_k-i_k$.

\subsubsection{Solid case}
\label{Sec331}

The obliquity amplitudes $\varepsilon_k$ and $\theta_k$ of an entirely solid and rigid synchronous satellite are given by (e.g.~\cite{Bal11})
\begin{subequations}
\label{Eq8}
\begin{eqnarray}
    \varepsilon_k&=&-\frac{ i_k \, \dot\Omega_k}{\omega_f+\dot\Omega_k}\\
    \theta_k &=&i_k+\varepsilon_k=\frac{ i_k \, \omega_f}{\omega_f+\dot\Omega_k}\\
    \mathrm{where}\ \omega_f&=&\frac{3}{2}n \, \frac{C-A}{C}.
\end{eqnarray}    
\end{subequations}
The relationship between the obliquity amplitudes and the principal moments of inertia is independent of whether or not the satellite is in hydrostatic equilibrium and therefore provides an alternative to Radau's equation for interpreting a degree-2 gravity field measurement in terms of interior differentiation (MOI $=C/M\!R^2-2J_2/3$ with $C$ as a function of the measured obliquity and of $(C-A)=MR^2(J_2+2C_{22})$). 

The torque applied on the satellite is proportional to the difference between the principal moments of inertia ($C-A$) and, as with libration, $C$ represents the satellite's resistance to the applied torque. The natural frequency of the free retrograde precession (written as a positive value) $\omega_f$, is proportional to $(C-A)$ but inversely proportional to $C$. As with the diurnal libration amplitude, the obliquity amplitudes of a solid and rigid satellite do not depend directly on the details of its interior (i.e.~density and size of the different layers), but only on its principal moments of inertia, and therefore only on the MOI, if hydrostatic equilibrium applies. 

When the satellite's response to the torque is altered by the periodic deformations, the obliquity amplitudes are given by (e.g.~\cite{Bal16})
\begin{subequations}
\begin{eqnarray}
\varepsilon_{el,k}&=&-\frac{C}{\tilde C}\frac{ i_k \, \dot\Omega_k}{\omega_{el,f}+\dot\Omega_k}\\
\theta_{el,k}&=&i_k+\varepsilon_{el,k}=\frac{i_k \, \omega_{el,f}}{(\omega_{el,f}+\dot\Omega_k)}+\frac{(\tilde C- C)}{\tilde C}\frac{i_k  \, \dot\Omega_k }{(\omega_{el,f}+\dot\Omega_k)}\\
\mathrm{where}\ \omega_{el,f}&=&\frac{3}{2}n \, \frac{(C-A)-k_2MR^2 q_r}{\tilde C}\\
\mathrm{and}\ \tilde C&=&C-k_2MR^2 q_r/2.
\end{eqnarray}
\end{subequations}
Note that $\tilde C$ does not have the same definition here as in the case of elastic librations, Eq.~(\ref{Ct}).

\begin{table}[h]
\begin{center}
\begin{tabular}{llccccc}
\hline
&&   $k=1$ & $k=2$ & $k=3$ &$k=4$ &  $k=5$\\
\hline
&&&&&&\\
\multicolumn{7}{l}{\textbf{MOI = 0.4}}\\
&&&&&&\\
Miranda &$\varepsilon_k$  (deg)  &  0.01622 & 0. & 0. & 0. & 0. \\
Ariel   &$\varepsilon_k$  (deg) & -0.00037 & 0.00013 & -0.00006 & 0.00002 & 0. \\
Umbriel &$\varepsilon_k$  (deg) & -0.00017 & -0.00012 & -0.00103 & 0.0004 & 0.00005 \\
Titania & $\varepsilon_k$ (deg) &  0.00102 & -0.00003 & 0.00149 & 0.00919 & 0.00193 \\
Oberon & $\varepsilon_k$  (deg) &  0.00006 & 0.00005 & -0.00285 & -0.04241 & 0.009 \\
&&&&&&\\
Miranda &$\varepsilon_{el,k}$  (deg)  &  0.01622 & 0. & 0. & 0. & 0. \\
Ariel   &$\varepsilon_{el,k}$  (deg) & -0.00037 & 0.00013 & -0.00006 & 0.00002 & 0. \\
Umbriel &$\varepsilon_{el,k}$  (deg) & -0.00017 & -0.00012 & -0.00103 & 0.0004 & 0.00005 \\
Titania & $\varepsilon_{el,k}$ (deg) &  0.00097 & -0.00003 & 0.00151 & 0.00929 & 0.00195 \\
Oberon & $\varepsilon_{el,k}$  (deg) &  0.00006 & 0.00005 & -0.00291 & -0.04304 & 0.00908 \\
&&&&&&\\
\multicolumn{7}{l}{\textbf{MOI after \cite{Hus06}}}\\
&&&&&&\\
Miranda &$\varepsilon_k$  (deg)  &  0.02010 & 0. & 0. & 0. & 0. \\
Ariel   &$\varepsilon_k$  (deg) & -0.00056 & 0.00019 & -0.00009 & 0.00004 & 0.00001 \\
Umbriel &$\varepsilon_k$  (deg) & -0.00025 & -0.00017 & -0.00145 & 0.00056 & 0.00007 \\
Titania & $\varepsilon_k$ (deg) &  0.00037 & -0.00005 & 0.00247 & 0.01463 & 0.00291 \\
Oberon & $\varepsilon_k$  (deg) &  0.00006 & 0.00003 & -0.01621 & -0.09412 & 0.0138 \\
&&&&&&\\
Miranda &$\varepsilon_{el,k}$  (deg)  & 0.02011 & 0. & 0. & 0. & 0. \\
Ariel   &$\varepsilon_{el,k}$  (deg) & -0.00056 & 0.00019 & -0.00009 & 0.00004 & 0.00001 \\
Umbriel &$\varepsilon_{el,k}$  (deg) & -0.00025 & -0.00017 & -0.00145 & 0.00056 & 0.00007 \\
Titania & $\varepsilon_{el,k}$ (deg) &  0.00037 & -0.00005 & 0.00248 & 0.01468 & 0.00292 \\
Oberon & $\varepsilon_{el,k}$  (deg) & 0.00006 & 0.00003 & -0.01669 & -0.09482 & 0.01384 \\
\hline
\end{tabular}
\end{center}
\caption{Rigid and elastic obliquity amplitudes of the Uranian satellites, seen as solid homogeneous (MOI $=0.4$) bodies or as two-layer bodies (MOI after \cite{Hus06}) for the 5 frequencies $k$. } 
\label{TabSerieseps}
\end{table}

\begin{figure}[!htb]
      \begin{center}
        \hspace{0cm}
\includegraphics[height=5cm]{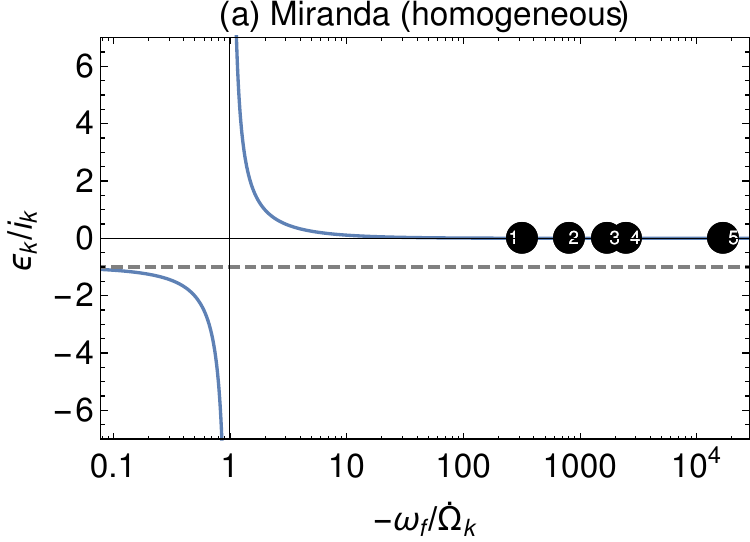}\quad
\includegraphics[height=5cm]{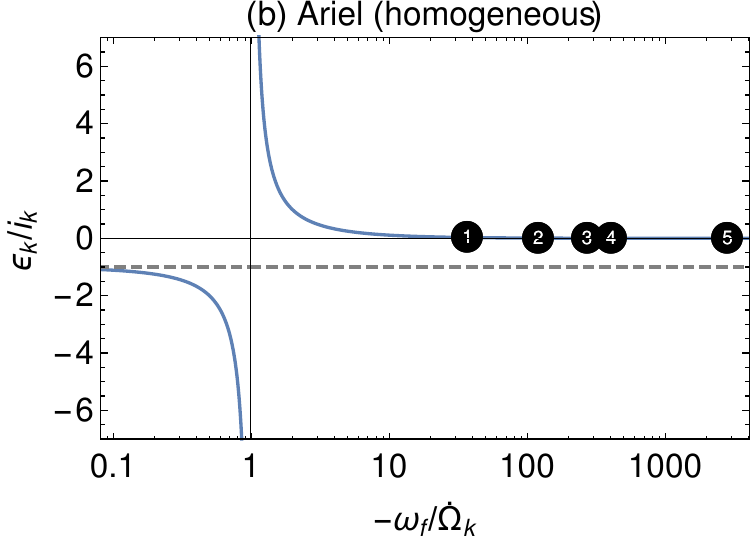}\\
\vspace{0.25cm}
\includegraphics[height=5cm]{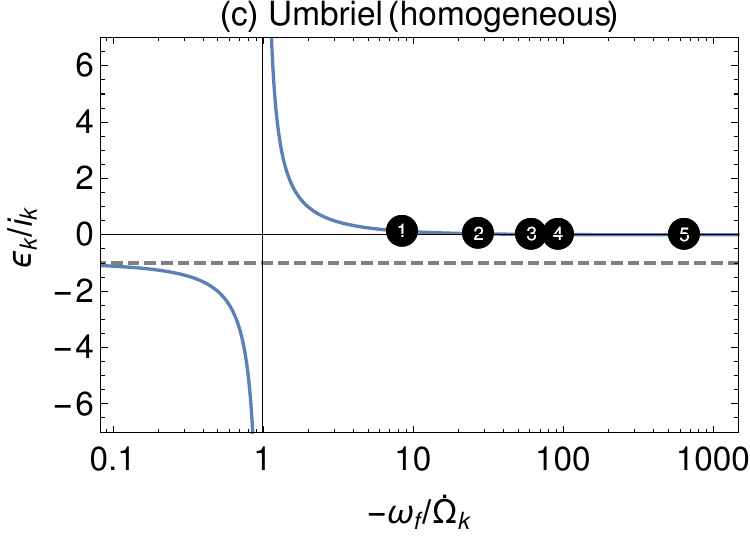}\quad
\includegraphics[height=5cm]{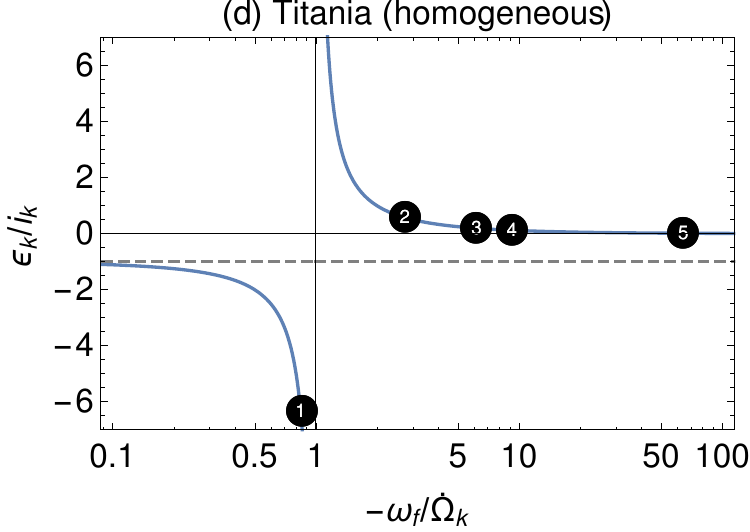}\\
\vspace{0.25cm}
\includegraphics[height=5cm]{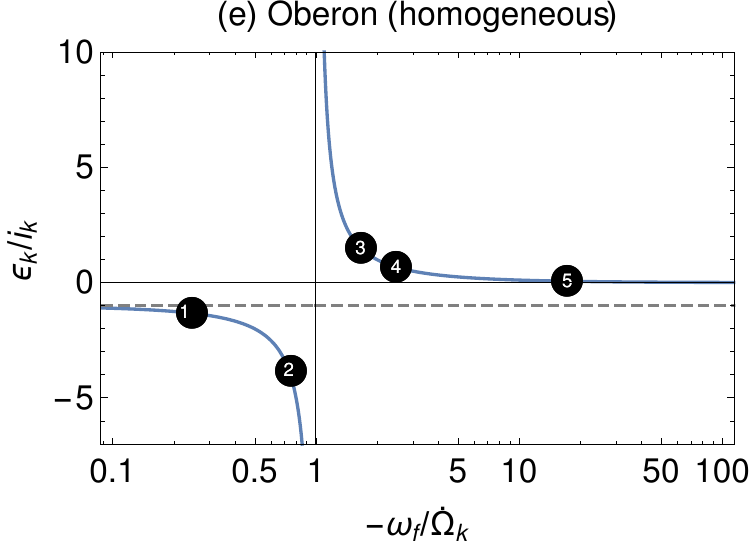}\quad
\includegraphics[height=5cm]{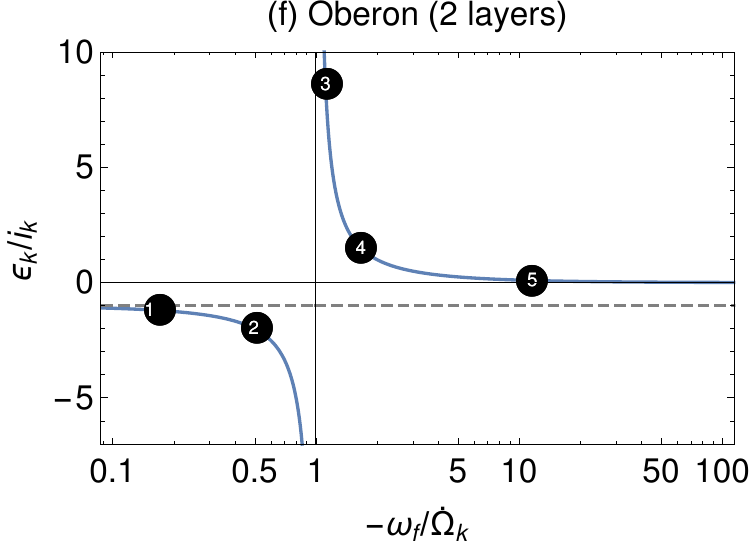}
 \caption{Ratio $\varepsilon_k/i_k$ as a function of the ratio $-\omega_f/\dot\Omega_k$ in the solid case, see Eq.~(\ref{Eq8}), in the homogeneous rigid case for the five satellites and also in the two-layer rigid case for Oberon. For the first four satellites, the plot would not differ much between the homogeneous and the two-layer cases. The labels $1$ to $5$ show the ratios at the forcing secular frequencies.}
\label{reson}
      \end{center}
\end{figure}

The rigid and elastic obliquity amplitudes of the five satellites are given in Table~\ref{TabSerieseps} for the case of homogeneous and two-layer solid bodies. 
The absolute values for the rigid two-layer solid bodies are in fairly good agreement with those of \cite{Gomes2024}.
To better understand the results, we also plotted in Fig.~\ref{reson} the ratio $\varepsilon_k/i_k$ as a function of the ratio of the free precession frequency to the orbital precession frequency $(-\omega_f/\dot\Omega_k)$. Since $C>A$ and $\dot\Omega_k<0$, the ratio $(-\omega_f/\dot\Omega_k)$ is positive. Depending on the value of the ratio, different regimes can be distinguished: 
\begin{description}
    \item[\textit{CSII regime:}] For $-\omega_f/\dot\Omega_k< 1$, $\varepsilon_k$ and $i_k$ are of  opposite signs, a situation similar to the Moon, which is said to be in Cassini State (CS) II. 
    In the limit case $-\omega_f/\dot\Omega_k\rightarrow 0$, the inertia  outweighs the torque, $\varepsilon_k$ tends towards $-i_k$ ($\theta_k\rightarrow 0 $), and, on a mode-by-mode basis, the axis of rotation tends to align with the normal to the Laplace plane. The free precession of the spin axis is then so slow that the spin axis cannot track the moving orbit normal.
    \item [\textit{CSI regime:}]  For $-\omega_f/\dot\Omega_k> 1$,  $\varepsilon_k$ has the same sign as $i_k$, a case that can be compared with Cassini State I. In the limit case $-\omega_f/\dot\Omega_k \rightarrow \infty$, the torque outweighs inertia, $\varepsilon_k$ tends to $0$ ($\theta_k \rightarrow i_k$), and the axis of rotation tends to align with the orbit normal, on a mode-by-mode basis.
    \item [\textit{Resonant regime:}]  When the ratio is close to $1$, $\varepsilon_k$ can reach large values compared to $i_k$, negative or positive depending on the exact value of the ratio and the sign of the inclination amplitude.  
\end{description}
CSI and CSII regimes are mutually exclusive, but do not exclude the resonant regime.

\begin{figure}[!htb]
      \begin{center}
        \hspace{0cm}
\includegraphics[width=16 cm]{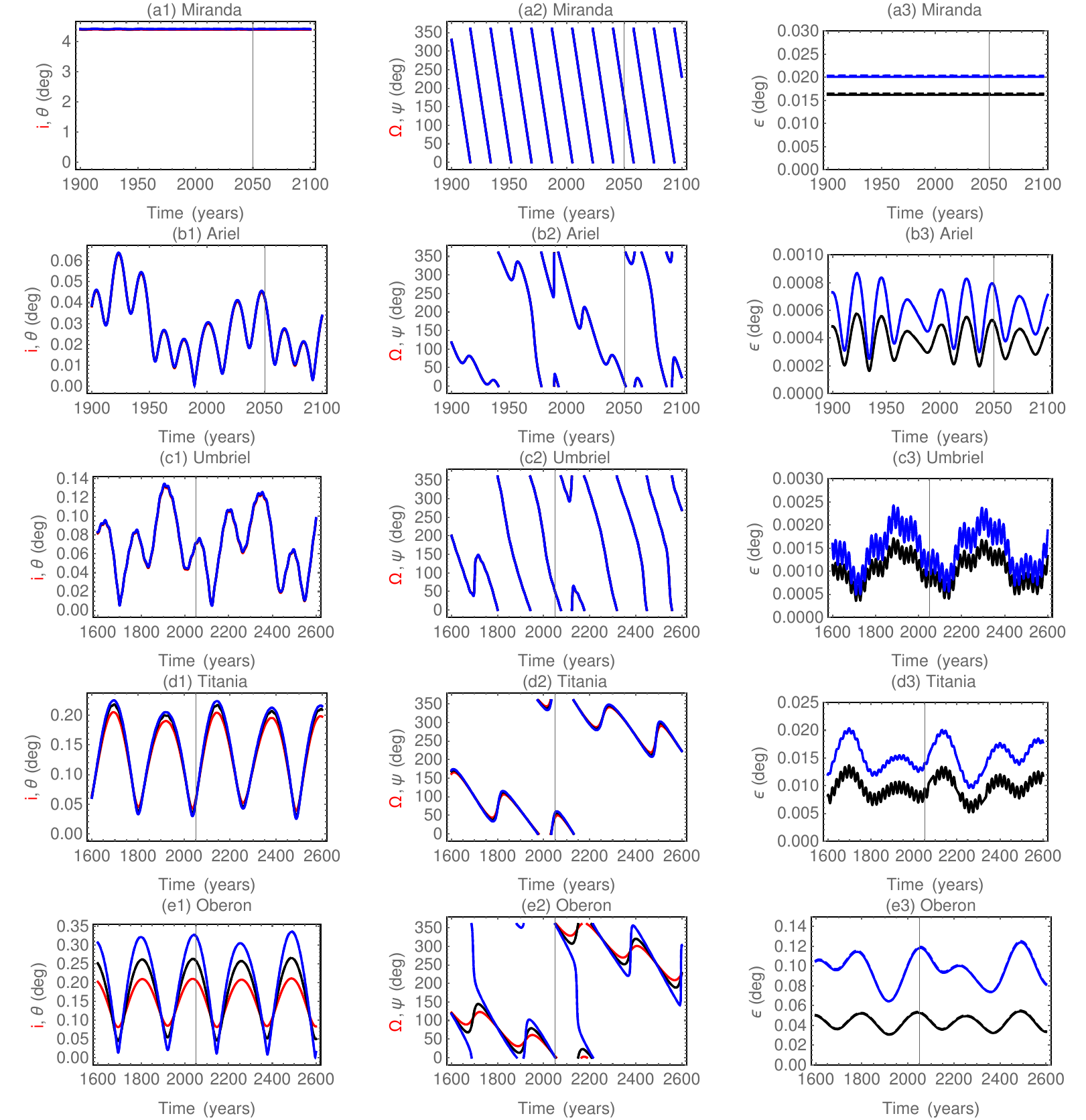}
\caption{Inertial obliquity $\theta$, node longitude $\psi$, and obliquity $\varepsilon$ angles of the satellites’ equatorial planes. $\theta$ and $\varepsilon$ are measured from the Laplace plane (here the equator of Uranus in J2000) and the satellites’ orbital planes, respectively, whereas $\psi$ is measured on the Laplace plane, see Fig.~\ref {FigAngles}. Black and blue curves correspond to the cases of homogeneous and two-layer solid interiors, respectively, see Table~\ref{TabSerieseps}. Solid and dashed lines correspond to rigid and elastic cases, respectively (their difference is small). The orbital inclination $i$ and node longitude $\Omega$ are plotted in red, for comparison. Vertical lines indicate the year 2050. }
\label{fig_obliquitime}
      \end{center}
\end{figure}

For Miranda, Ariel and Umbriel, $-\omega_f/\dot\Omega_k\gtrsim 10$ for all $k$, so that all five obliquity amplitudes fall in the CSI regime, see Fig.~\ref{reson} panels (a-c). As a result, their rotation axes are characterized by a relatively small obliquity $\varepsilon(t)$ compared to their orbital inclination $i(t)$ and $\theta(t)\simeq i(t)$, see Fig.~\ref{fig_obliquitime} panels (a1,a3,b1,b3,c1,c3). Miranda stands out as a satellite whose spin axis almost precesses at constant rate and with a constant obliquity because $i_1 \gg i_{k\neq1}$, see Fig.~\ref{fig_obliquitime} (panels a1-a3), whereas the spin precession of Ariel and Umbriel is affected by the gravitational perturbations exerted by the other satellites on their orbit, see Fig.~\ref{fig_obliquitime} (panels a-c).

For Titania, all obliquity amplitudes but $\varepsilon_1$ are in the CSI regime, see Fig.~\ref{reson} panel (d). $\varepsilon_1$ falls in the CSII regime and is resonantly amplified ($\varepsilon_1\simeq 0.001^\circ$, about $6$ times larger than $i_1=-0.00016^\circ$ since the free precession period of about $21$ years is close to the first forcing period of $18$ years). However, this amplification is not sufficient for $\varepsilon_1$ to substantially affect the total $\varepsilon(t)$. In the end, Titania is essentially in the CSI regime, with $\varepsilon(t)$ one order of magnitude smaller than inclination and dominated by $\varepsilon_3$, $\varepsilon_4$ and $\varepsilon_5$ (panels d of Fig.~\ref{fig_obliquitime}).

Oberon's first two obliquity amplitudes fall in the CSII regime, with $\varepsilon_2$ being resonantly amplified ($\varepsilon_2/i_2=-4$ for MOI $=0.4$, see Fig.~\ref{reson} panel e), but not to the point of having a significant impact on the total obliquity. Like Titania, Oberon therefore falls essentially in the CSI regime, with $\varepsilon(t)$ dominated by $\varepsilon_3$, $\varepsilon_4$ and $\varepsilon_5$. However, the free precession period (about $78$ and $120$ years for the homogeneous and two-layer cases, respectively) being relatively close to the third and fourth forcing periods (about $130$ and $200$ years), $\varepsilon_3$ and $\varepsilon_4$ are resonantly amplified (the amplification being stronger for the two-layer case, see Fig.~\ref{reson} panel f). As a result, the obliquity $\varepsilon(t)$ of Oberon can be of the same magnitude as its inclination (panels e1 and e3, Fig.~\ref{fig_obliquitime}). At certain times in the two-layer case, the node longitude $\psi(t)$ of Oberon's equator may be very different from $\Omega(t)$, the node longitude of the orbital plane, see panel (e2) of Fig.~\ref{fig_obliquitime}).

In table \ref{TabObl}, we give the values of $\varepsilon$ at time 2050 for the rigid and elastic homogeneous and two-layer cases, with a view to discussing the possibilities of interpreting a measurement of the orientation of the spin axis of a satellite in terms of its internal structure. Note that, as mentioned above, this is a truncated view of the problem, since the obliquity $\varepsilon$ is not enough to orientate the axis of rotation in space.  
If Uranus's satellites were rigid, homogeneous bodies, their obliquity at 2050 would range from around $0.0005$ deg (Ariel) to $0.05$ deg (Oberon), see Table \ref{TabObl}. Unlike for the libration amplitude, there is no general dependence of a satellite's obliquity on its distance from the planet. Even if the difference $(C-A)/M\!R^2$ tends to decrease with distance from the planet, this is not enough to ensure that obliquity also increases, because of the complexity of orbital precession.
In particular, Miranda's large inclination and precession rate implies an obliquity larger than those of Ariel, Umbriel and Titania. For a body differentiated into two layers, the reduction in $C$ compared with the homogeneous case does not compensate for the reduction in $(C-A)$ due to differentiation. Consequently, a smaller MOI means a larger obliquity. For the two-layer interiors of \cite{Hus06}, the increase in obliquity compared to the homogeneous case ranges from around $3$ m (Ariel) to around $900$ m (Oberon), see Table \ref{TabObl} (last column). These differences give an idea of the order of magnitude needed on an obliquity measurement to distinguish between a homogeneous and a differentiated solid body.
In general, elastic deformations tend to increase the obliquity, compared with the rigid case. For large icy satellites considered as solid bodies, the decrease is limited (about $8\%$ for Ganymede, about $1\%$ for Europa, see \cite{Bal16}). For all the small/medium-sized satellites of Uranus, the difference is less than about $1\%$ for both homogeneous and two-layer interiors, see Table \ref{TabObl}.

\begin{table}[h]
\footnotesize
\begin{center}
\begin{tabular}{lccccccccc}
\hline
& $2\pi/ \omega_f$ & $\varepsilon$   & $\varepsilon_{el}$ & Diff & $2\pi/ \omega_f$ & $\varepsilon$ & $\varepsilon_{el}$  & Diff  & Differentiation\\
& [years]& [deg]  &  [deg] &  &  [years] &  [deg] &  [deg] &   & effect\\
& \multicolumn{1}{l}{[} &
\multicolumn{2}{c}{MOI $=0.4$}    & \multicolumn{1}{r}{]}& \multicolumn{1}{l}{[} & \multicolumn{2}{c}{MOI $<0.4$} & \multicolumn{1}{r}{]}& [m] - [deg]\\
\hline
Miranda 	 &0.07   &0.01621 & 0.01622 &0.04$\%$ & 0.08  &0.02010 & 0.02011 &0.04$\%$ &16.00 - 0.00389 \\
Ariel 	     &0.48   &0.00052 & 0.00052 &0.45$\%$ &  0.71 &0.00078 & 0.00078 &0.17$\%$ &2.66 - 0.00026\\
Umbriel 	 &2.1    &0.00088 & 0.00089 &0.45$\%$ & 2.94  &0.00125 & 0.00125 &0.21$\%$ &3.76 - 0.00037 \\
Titania 	 &21.06  &0.00946 & 0.00951 &0.46$\%$ & 31.49  &0.01403 & 0.01408 &0.36$\%$ &62.84 - 0.00456\\
Oberon 	     &78.36  &0.05260 & 0.05336 &1.43$\%$ & 116.61  &0.11815 & 0.11923 &0.92 $\%$ &871.02  -  0.06554\\
\hline
\end{tabular}
\end{center}
\caption{Obliquity in the solid case at time 2050 (
01-JAN-2050 12:00:00 UTC). Columns 2 to 5 show the free precession period, the rigid and elastic obliquities, and their relative difference for homogeneous bodies (MOI of $0.4$). Columns 6 to 9 show the free period, rigid and elastic obliquities and their difference, for the two-layer interiors as defined in \cite{Hus06} (MOI $<0.4$). The last column shows the difference in meters at the surface between the solid obliquities of the homogeneous and two-layer interiors.} 
\label{TabObl}
\end{table} 

\newpage

\subsubsection{Ocean case}
\label{Sec332}

Even though the tidal Love number $k_2$ increases by an order of magnitude for an interior with a liquid ocean compared to the solid case, we here consider the solid layers of the Uranian satellites as rigid, with the aim of keeping the spin precession model in the ocean case as simple as possible, without losing relevance. We will use the Cassini state model for a satellite with an ocean and no polar motion of \cite{Bal19} (see their Appendix 5), which extends the model of \cite{Bal11,Bal12} by including the hydrodynamic pressure (Poincaré flow) at the interfaces between the ocean and the solid layers. We will note $\varepsilon_k^s$ and $\varepsilon_k^c$ the obliquity amplitudes of the shell and core, respectively.
The assumption of rigid solid layers is motivated by two facts. Firstly, because the ratio $k_2/k_f\lesssim10\%$, the elastic deformations have little effect on the diurnal librations. We assume that this conclusion also applies to spin precession. Secondly, we will show that, outside the resonant regime, the obliquity in the ocean case is close to that in the case without ocean. Effects of elastic deformations will be evaluated approximately with the elastic solid model, but using numerical values for $k_2$ from the ocean case. 

We have seen in Section \ref{Sec331} that the obliquity amplitudes $\varepsilon_k$ at orbital forcing frequency $\Omega_k$ of a solid satellite can be resonantly amplified if the ratio of the free frequency over the forcing frequency $\omega_f/\dot\Omega_k$ is close to $-1$. In the ocean case, the spin precession model is characterized by three free modes: the Free Precession (FP), the Free Ocean Nutation (FON), and the Free Interior Nutation (FIN), which correspond to free motions of the rotation axes of the shell, ocean, and interior with respect to the Laplace frame, respectively, see \cite{Bal19}. By denoting the free modes frequencies $\omega_{FP}$, $\omega_{FON}$ and $\omega_{FIN}$, we write the shell and core obliquity amplitudes as 
\begin{subequations}
\begin{eqnarray}
\varepsilon_k^s&=&-\frac{i_k \, \dot\Omega_k\, n_k^s}{C_s \, C_o \, C_c \, (\omega_{FP} + \dot\Omega_k)  (\omega_{FON}+\dot\Omega_k)(\omega_{FIN}+\dot\Omega_k)},\\
\varepsilon_k^c&=&-\frac{i_k \, \dot\Omega_k\, n_k^c}{C_s \, C_o \, C_c \, (\omega_{FP} + \dot\Omega_k)  (\omega_{FON}+\dot\Omega_k)(\omega_{FIN}+\dot\Omega_k)}.
\end{eqnarray}
\end{subequations}
with $n_k^s$ and $n_k^c$ defined in \cite{Bal19} (as $n_s$ and $n_i$ in their notations).  The product $(\omega_{FP} + \dot\Omega_k)  (\omega_{FON} + \dot\Omega_k)(\omega_{FIN} + \dot\Omega_k)$ in the denominator is a rearrangement of the term $(d_0 + d_1 \dot\Omega_k + d_2 \dot\Omega_k^2 + \dot\Omega_k^3)$ of Eq.~(182) of \cite{Bal19}. 
The quantities $n_s$, $n_c$, $d_0$, $d_1$ and $d_2$ depend non-trivially on the properties (density, size and flattening) of the different layers.
In these notations, positive (negative) free frequencies correspond to retrograde (prograde) motions in space, 
whereas $\omega_{FP}$, $\omega_{FON}$ and $\omega_{FIN}$ are the eigenvalues of $-K_d$, with $K_d$ the matrix of a homogeneous system, see Eq.~(177) of \cite{Bal19}.

As pointed out in \cite{Bal19}, the well-known analytical formulae for the FCN (Free core nutation) and FICN (Free Inner Core Nutation) of a biaxial Earth (e.g.~page 293 of \cite{Dehant2015}), which are good approximations to the FCN and FICN of the Moon, see \cite{Dum2016}, are by no means correct approximations to the FON and FIN of synchronous icy satellites with relatively large solid interiors. In that case, the matrix $K_d$ includes terms related to the external gravitational torques on the solid layers which are sufficiently large to affect the FON and FIN
periods. By including those "external" terms, we also allow for the existence of the FP associated with the shell, unlike the model for the FCN and FICN of the Earth, which assumes that the mantle spin axis is fixed in space. 

\begin{figure}[!htb]
      \begin{center}
        \hspace{0cm}
\includegraphics[height=3.5cm]{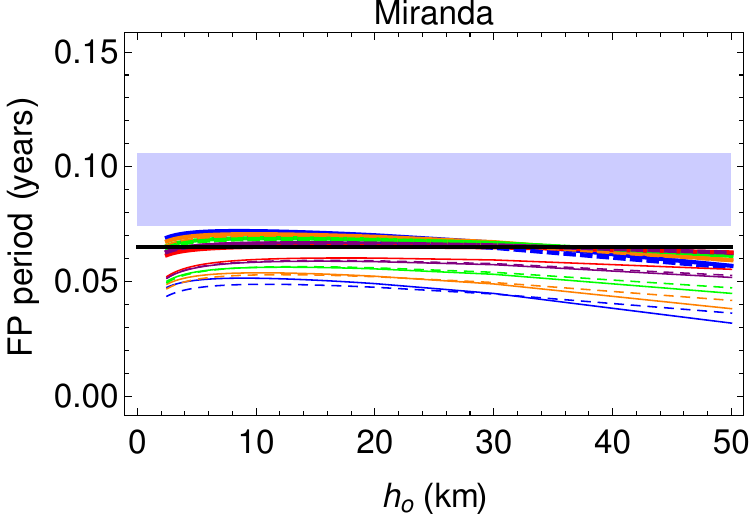}\quad
\includegraphics[height=3.5cm]{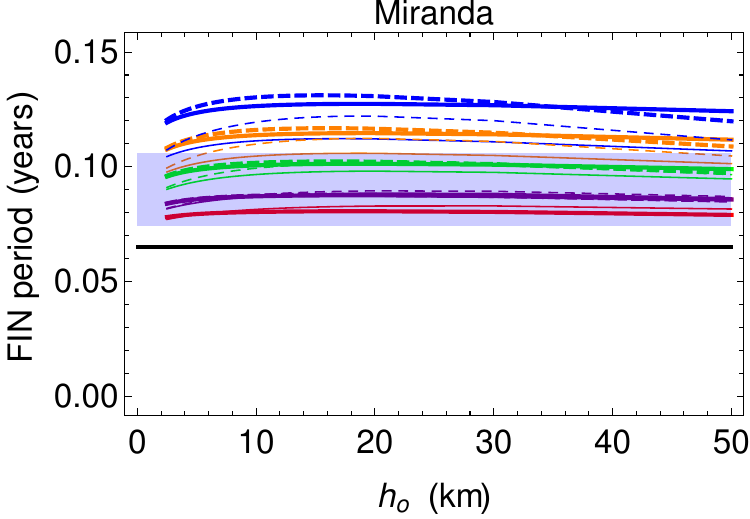}\quad
\includegraphics[height=3.5cm]{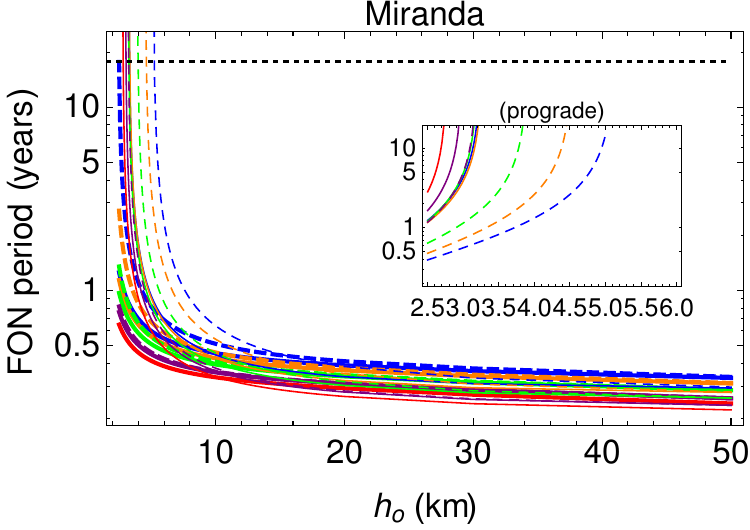}\\
\vspace{0.25cm}
\includegraphics[height=3.5cm]{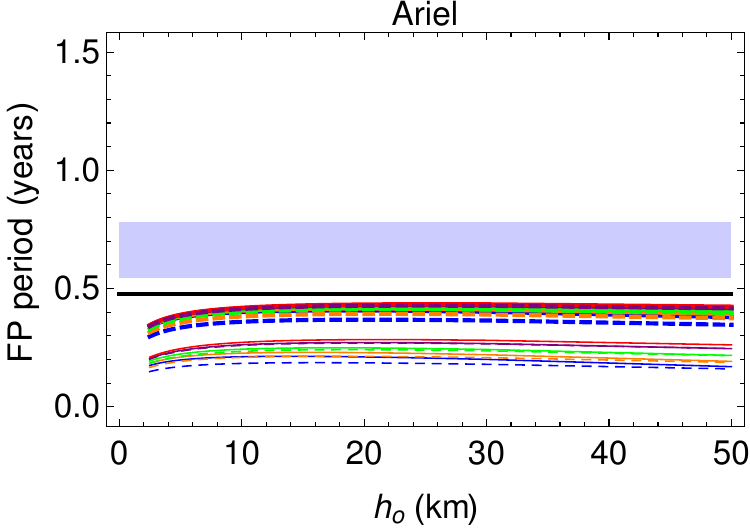}\quad
\includegraphics[height=3.5cm]{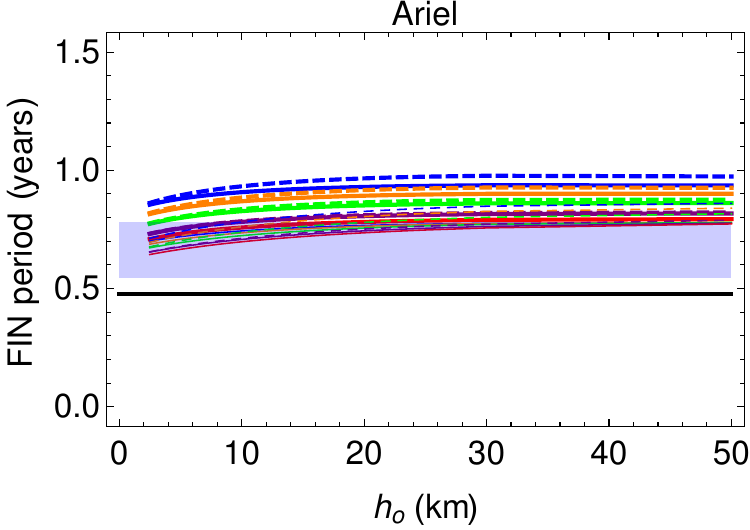}\quad
\includegraphics[height=3.5cm]{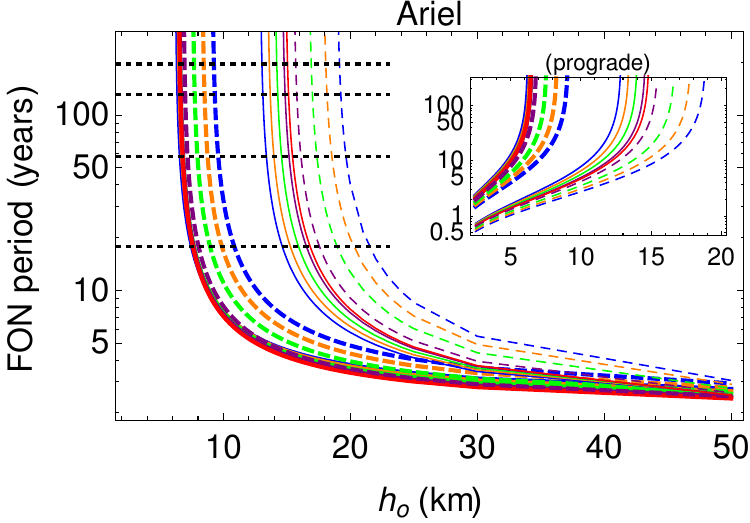}\\
\vspace{0.25cm}
\includegraphics[height=3.5cm]{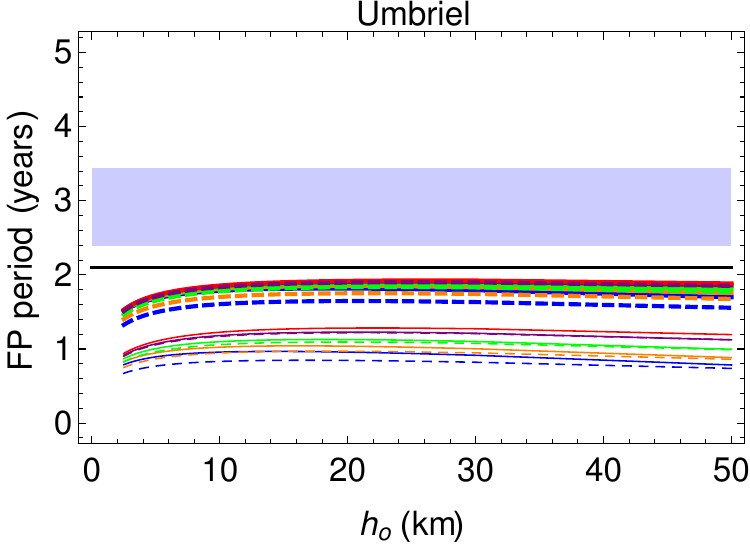}\quad
\includegraphics[height=3.5cm]{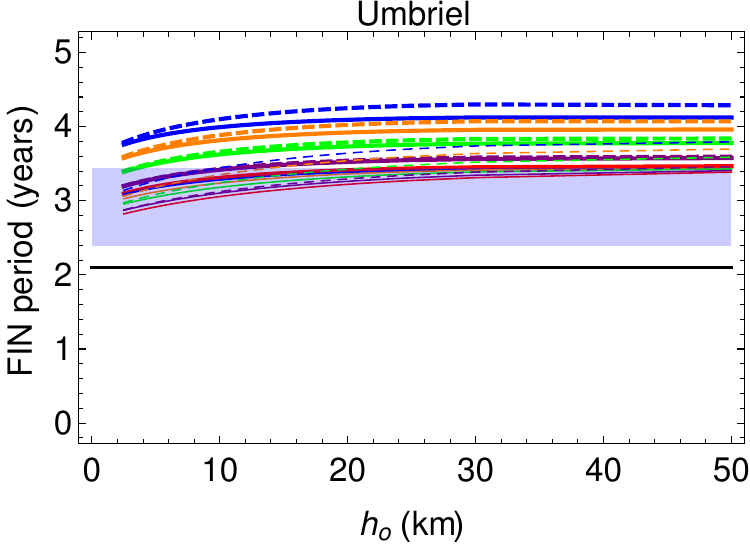}\quad
\includegraphics[height=3.5cm]{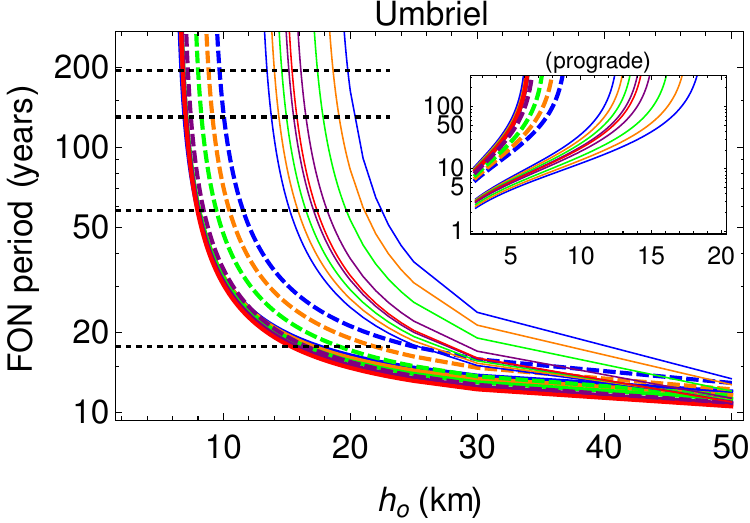}
\vspace{0.25cm}
\includegraphics[height=3.5cm]{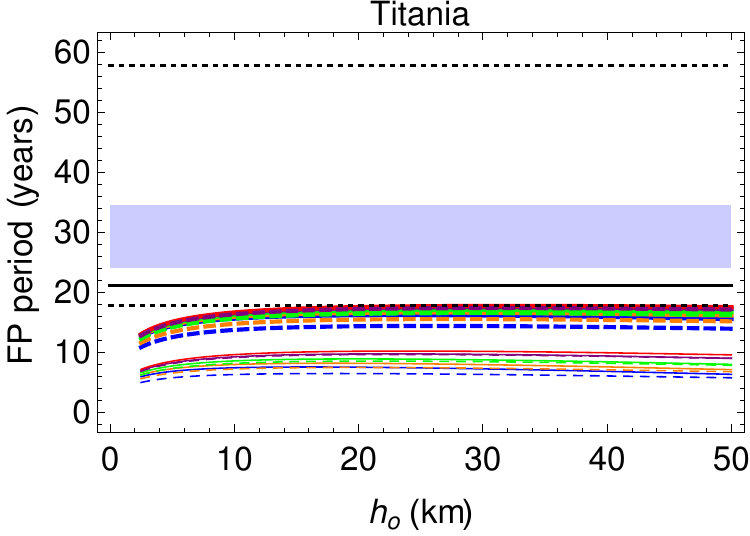}\quad
\includegraphics[height=3.5cm]{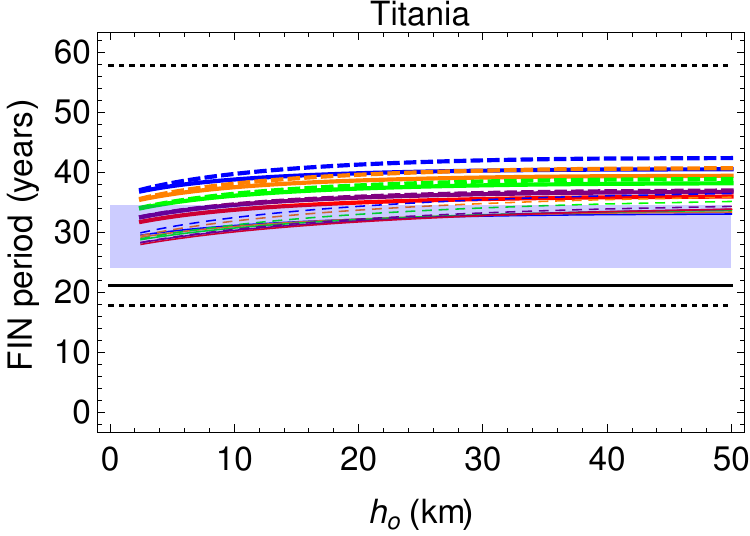}\quad
\includegraphics[height=3.5cm]{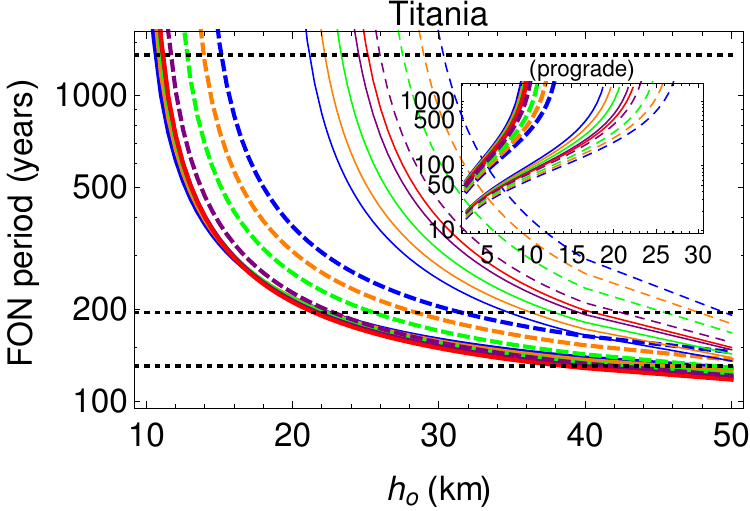}
\vspace{0.25cm}
\includegraphics[height=3.5cm]{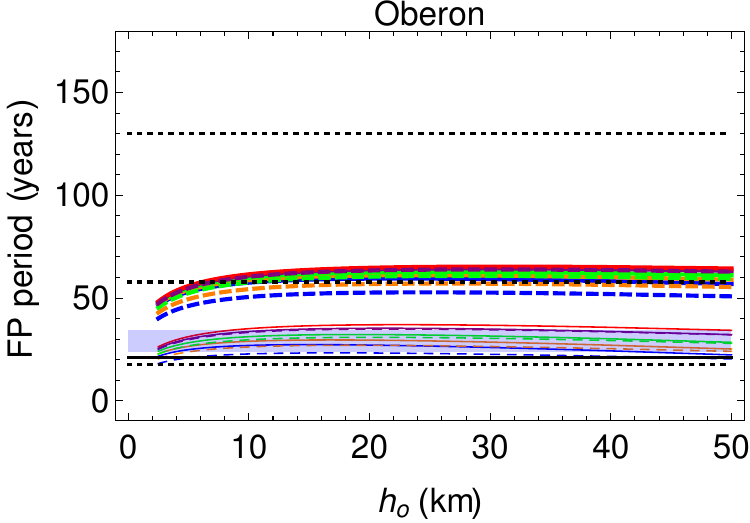}\quad
\includegraphics[height=3.5cm]{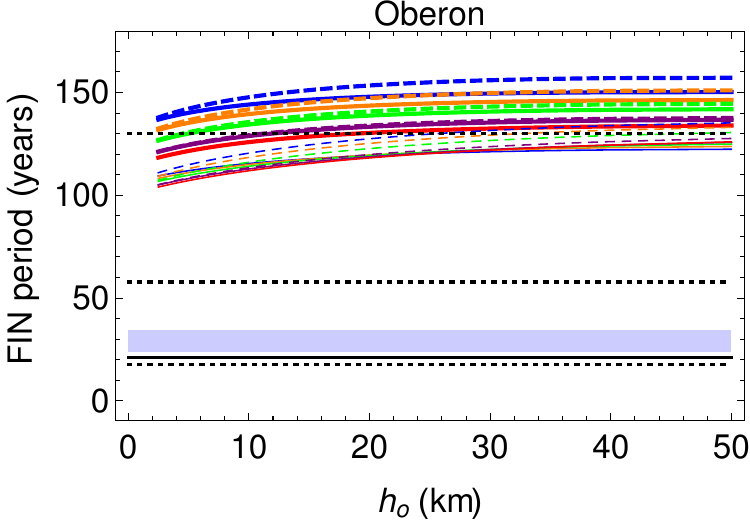}\quad
\includegraphics[height=3.5cm]{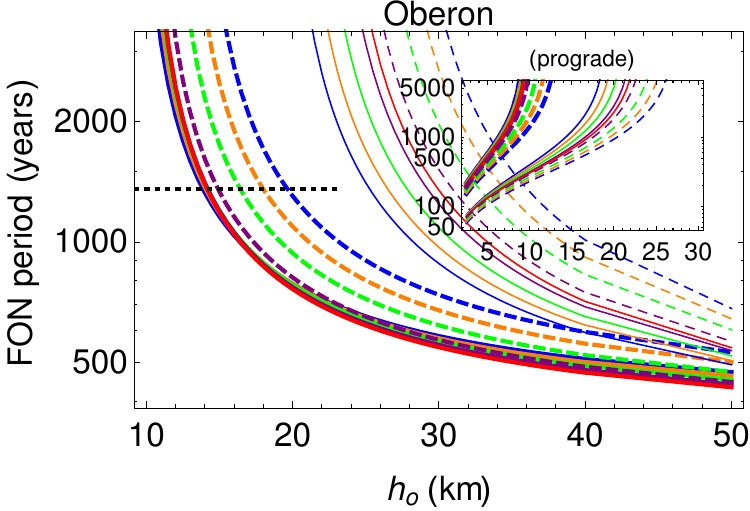}
 \caption{Periods of the FP, FIN, and FON as a function of the ocean thickness $h_o$. Solid and dashed colored lines defined as in Fig.~\ref{gsMiranda}. The FP and FIN are always retrograde (positive periods), whereas the FON can be retrograde (in most cases) or prograde (see inner frames where negative periods are displayed in absolute values). The blue region in the FP and FIN panels corresponds to the range of possible values for two-layer solid models. The black solid line corresponds to the free precession period of a single-layer homogeneous body. The horizontal black dotted lines correspond to absolute forcing periods (all retrograde). Resonances are possible when such a line is crossed.}
\label{FPFONFIN_Miranda}
      \end{center}
\end{figure}

Figure \ref{FPFONFIN_Miranda} shows the exact periods corresponding to the eigenvalues of $-K_d$. The FP and FIN periods are positive and of the same order of magnitude as the FP of the solid case, for all five Uranian satellites. They depend little on the ocean thickness. The period of the FON can be either positive or negative, and large or small. When the period of the FON tends towards $0$, the obliquity amplitude also tends towards $0$. The obliquity amplitude tends towards the same value when the period of the FON is very large, whether positive or negative.
When the FON period is positive and close to one of the forcing periods, it leads to a resonant amplification of the corresponding obliquity amplitudes. This also applies to the FP and FIN periods.
Resonant amplifications with the FON can occur for all five satellites. It only occurs for interiors with very thin oceans ($h_o\lesssim 6$ km) for Miranda. For Miranda, Ariel and Umbriel, all forcing periods can be reached by the FON. For Titania, the first two forcing periods (the shortest, about $18$ and $60$ years) cannot be reached, since the FON period does not fall below one hundred years. For Oberon, only the fifth forcing period can be reached by the FON. Note that the first forcing periods can be reached by the FP and/or the FIN for Titania and Oberon. 

In the non resonant regime, which applies mainly to Miranda in view of the above, but also possibly to other satellites depending on the thickness of the ocean, an order of magnitude analysis shows that the shell obliquity amplitudes can be roughly approximated by
\begin{equation}\label{Eqappeks}
    \varepsilon_k^s\simeq -i_k\, \dot\Omega_k \frac{(C_s+C_{ot})}{\kappa_s},\\
\end{equation}
with 
\begin{equation}
    \kappa_s =  \frac{3}{2}n (C_s-A_s+C_{ot}-A_{ot}).
\end{equation}
$(A_s,C_s)$ and $(A_{ot},C_{ot})$ are the smallest and largest principal moments of inertia of the shell and of the part of the ocean aligned with the shell (top ocean), respectively.
The expression of Eq.~(\ref{Eqappeks}) has a form similar to the solid solution in the CS1 non resonant regime ($\varepsilon_k\simeq -i_k \, \dot\Omega_k C /\kappa $ with $\kappa=3/2 n (C-A)$). Since the shell and top ocean account for most of the body inertia ($(C_s-A_s+C_{ot}-A_{ot})/(C_s+C_{ot})$ varies from about $95\%$ to $99\%$ of $(C-A)/C$), the ocean approximated solution suggests that the ocean exact solution is roughly similar to the solid solution, in the non resonant regime. 
For Miranda, Ariel, and Umbriel, the shell obliquity amplitudes can be approximated further, with an accuracy of a few tens of percent, as
\begin{equation}
\label{epsapp}
    \varepsilon_k^s\approx -\frac{5\,i_k\,\dot\Omega_k}{12 \, n}\frac{[ R^5+R_o^5(\rho_o -\rho_s)/\rho_s ]}{\alpha_s R^5}.
\end{equation}
This approximated obliquity behaves in the same way as the exact solution with respect to interior parameters. In particular note the dependence on $\alpha_s$ (or equivalently on the MOI) and on $(\rho_o-\rho_s)/\rho_s$. This approximation does not work for Titania and Oberon, and for large icy satellites (e.g.~Titan, the Galilean satellites) in general.

We describe below the obliquity amplitudes and the time-varying obliquity of the five Uranian satellites, distinguishing Miranda from the other since its precession is only slightly affected by the gravitational perturbations of the other satellites.

\subsubsection*{Miranda }
 
\begin{figure}[!htb]
      \begin{center}
        \hspace{0cm}
\includegraphics[height=5 cm]{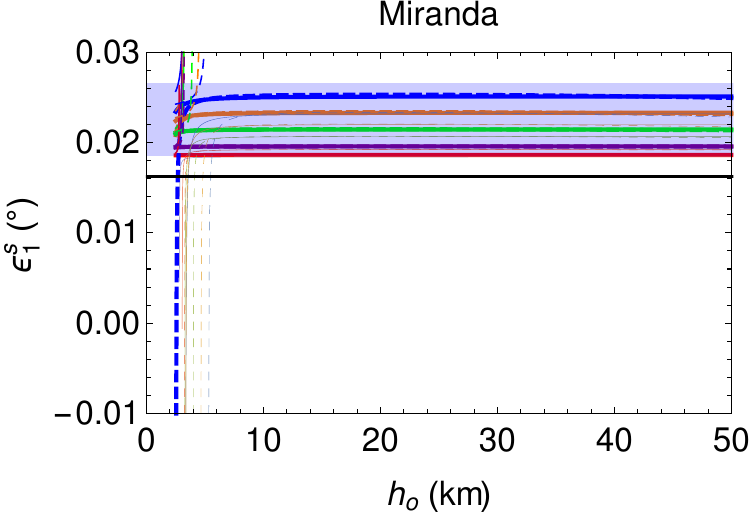}\quad
\includegraphics[height=5 cm]{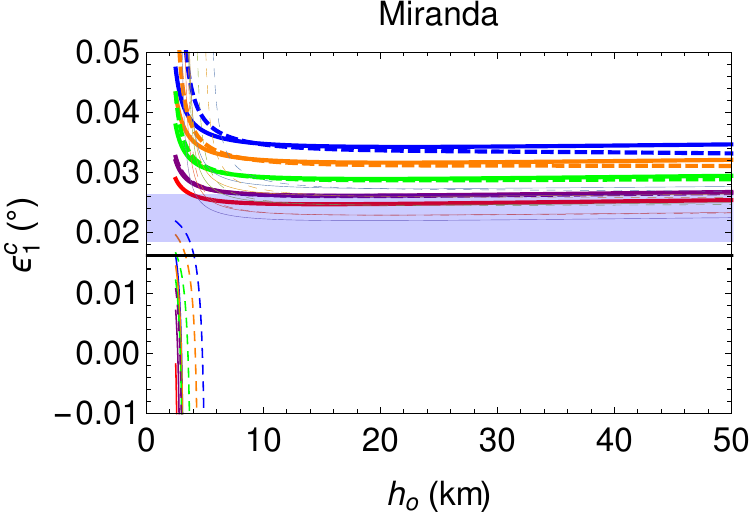}
\caption{Miranda shell ($\varepsilon_1^s$, left panel) and core ($\varepsilon_1^c$, right panel) first obliquity amplitude  as a function of the ocean thickness  $h_o$ for a selection of interiors representative of the range of interiors defined in Section \ref{Secinteriors}.  Solid and dashed lines defined as in Fig.~\ref{gsMiranda}. The blue region corresponds to the range of possible values for two-layer solid models with a MOI ranging from $0.288$ to $0.367$. The black line corresponds to the first obliquity amplitude of a single-layer homogeneous body.}
\label{epssMiranda} 
      \end{center}
\end{figure}

For Miranda, we only discuss the first shell ($\varepsilon_1^s$) and core ($\varepsilon_1^c$) obliquity amplitudes. Even if, in the theoretical framework presented here, a resonance with the FON could result in values for the other obliquity amplitudes as large as the first ones, we don't consider this to be a likely possibility. Amplification by several orders of magnitude would require a very good match between frequencies.
Moreover, under real-life conditions, such strong resonant amplifications would be counteracted by dissipative processes. The first obliquity amplitudes are therefore excellent approximations of shell and core obliquities over time ($\varepsilon_s(t)\simeq \varepsilon_1^s$ and $\varepsilon_c(t)\simeq \varepsilon_1^c$). 

We identify in Fig.~\ref{epssMiranda} the non-resonant and resonant regimes anticipated above. The non-resonant obliquities depend essentially on the total MOI, $\varepsilon_1^s$ being close to that of the solid case. When the densities of the different layers are fixed and $\rho_o=\rho_s$, $\varepsilon_1^s$ does not vary with the thickness of the ocean as can be understood from Eq.~(\ref{epsapp}) which shows no $h_s$ and no $h_o$ dependence. Since $\varepsilon_1^s\propto 1/\alpha_s$ and since $\alpha_s$ increases with $\rho_s$, $\varepsilon_s$ decreases with increasing shell density. A density contrast between the ocean and the shell allows for slightly different shell obliquities, see the second term in the numerator of Eq.~(\ref{epsapp}), but the effect is very limited or even negligible. A change in core density $\rho_c$ affects the MOI and therefore $\alpha_s$ ($\alpha_s$ increases when $\rho_c$ decreases), so a smaller core density implies a smaller shell obliquity. The obliquity of the core $\varepsilon_1^c$ is also almost independent of the ocean thickness and larger than that of the shell for a given MOI.
In the resonant regime, which corresponds to a thin ocean in the case of Miranda, the obliquities can deviate sharply from the non-resonant values and be either larger or smaller (and even negative). When the resonance is strong enough, the shell and core can precess out of phase with each other ($\varepsilon_1^c$ is negative when $\varepsilon_1^s$ is positive, and vice versa).

\begin{figure}[!htb]
      \begin{center}
        \hspace{0cm}
\includegraphics[height=5 cm]{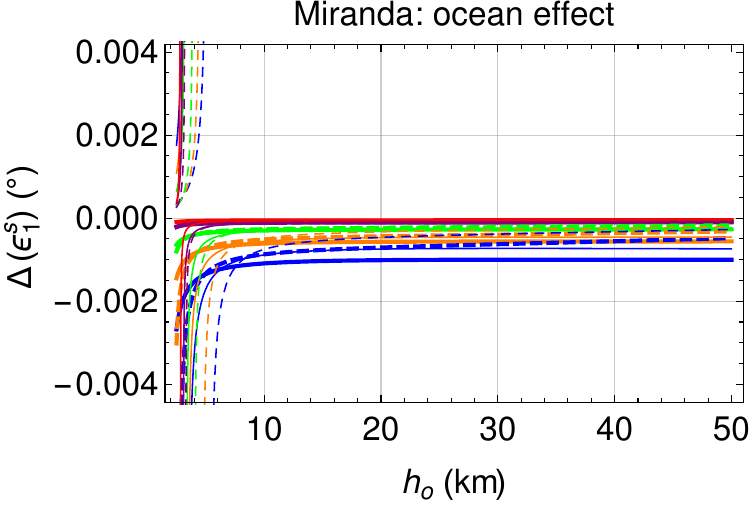}\quad
\includegraphics[height=5 cm]{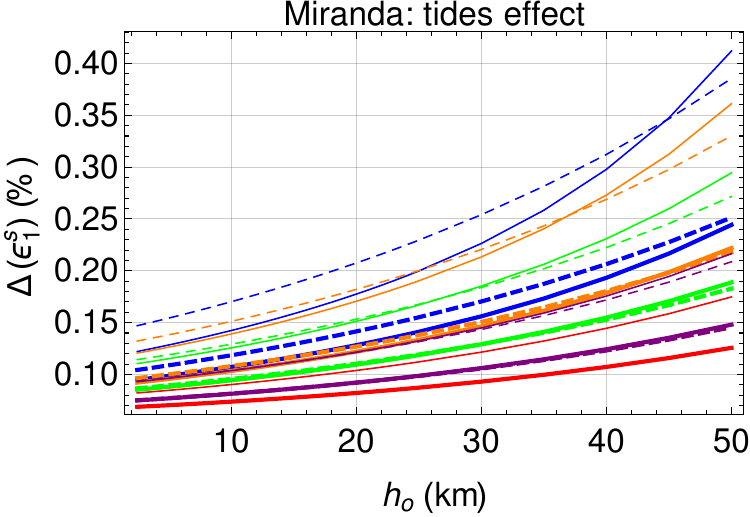}
\caption{Left: Difference in first shell obliquity amplitude $\varepsilon_1^s$ of Miranda between the case with and without ocean as a function of shell ocean thicknesses $h_o$. Solid and dashed lines defined as in Fig.~\ref{gsMiranda}. As each individual curve corresponds approximately to a given MOI (see bottom panel of Fig.~\ref{gsMiranda}), these differences correspond to the precision required to detect the ocean from a surface obliquity measurement if the MOI is determined independently (from the gravity field, for example). Right: Relative difference in first shell obliquity amplitude of Miranda between solid cases with and without periodic deformations as a function of shell thickness $h_s$, considering the tidal Love number of an interior with a global ocean. 
Solid and dashed lines defined as in Fig.~\ref{gsMiranda}}
\label{FigDeltaepssMiranda} 
      \end{center}
\end{figure}

In the non-resonant regime, and in the absence of any constraint on the MOI of the satellite, it is impossible to detect an ocean from the measurement of Miranda's obliquity, since the shell obliquity values are within the range of values expected for a solid body, see Fig.~\ref{epssMiranda}. In principle, an independent constraint on the MOI would reduce the range of values expected in the solid case and could help confirm the presence of an ocean. However, a precision of the order of $0.0005$ degrees or less would be required to achieve this, see Fig.~\ref{FigDeltaepssMiranda}. In the resonant regime, the presence of a thin ocean could be confirmed if the obliquity is significantly smaller or larger than the expected solid value. 

As explained above, we evaluate the effect of elasticity on Miranda's obliquity in the presence of an internal global ocean by comparing the outputs of the rigid and elastic solid models, using the Love number values $k_2$ obtained for the case of the ocean in the latter, see Section \ref{Seck2}. The difference is less than $0.5\%$, see Fig.~\ref{FigDeltaepssMiranda}.

\subsubsection*{Ariel, Umbriel, Titania and Oberon}

\begin{figure}[!htb]
      \begin{center}
        \hspace{0cm}
\includegraphics[height=4.5 cm]{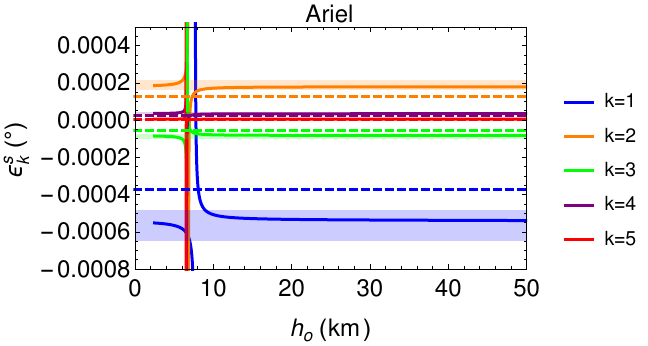}\quad
\includegraphics[height=4.5 cm]{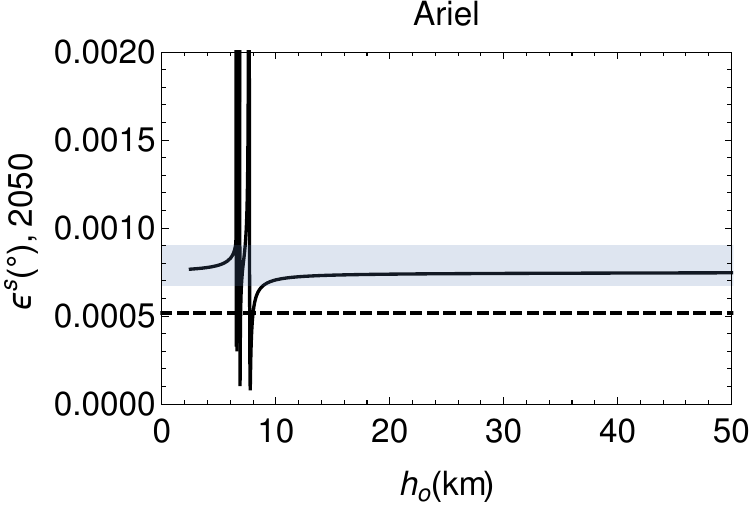}\\
\vspace{0.25cm}
\includegraphics[height=4.5 cm]{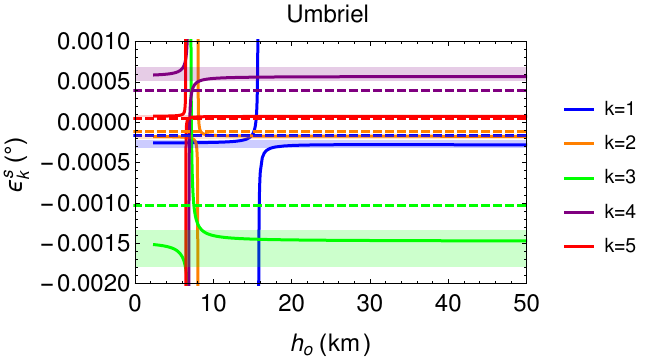}\quad
\includegraphics[height=4.5 cm]{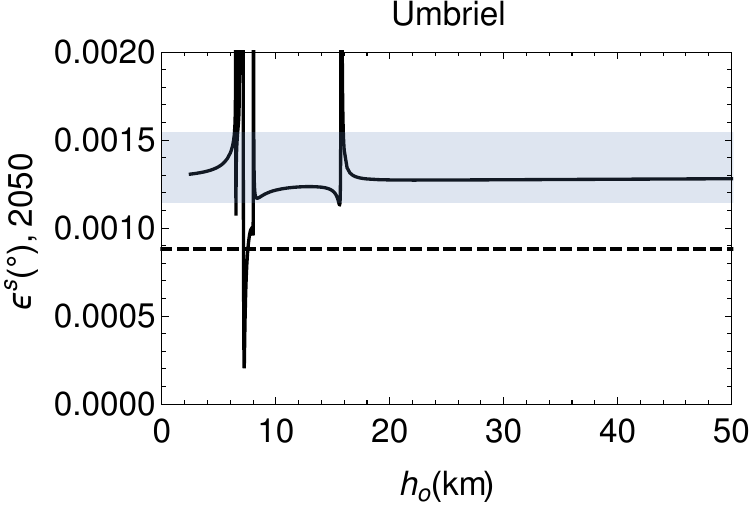}\\
\vspace{0.25cm}
\includegraphics[height=4.5 cm]{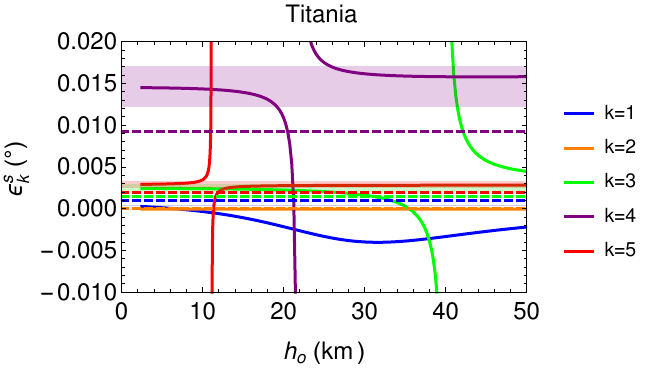}\quad
\includegraphics[height=4.5 cm]{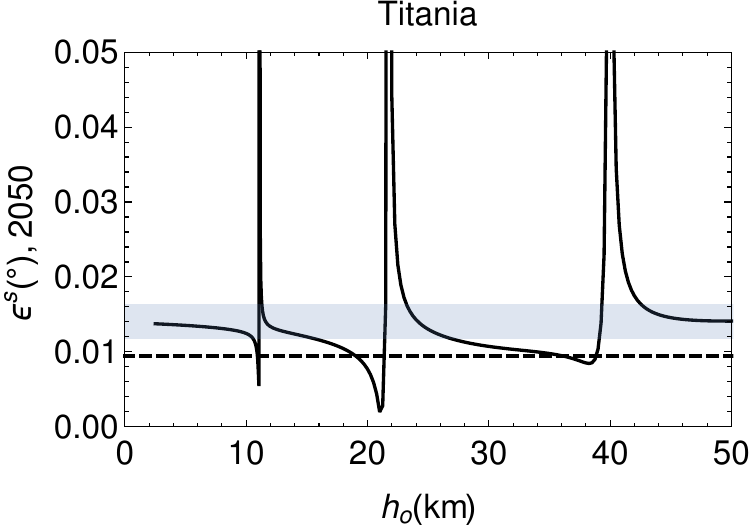}\\
\vspace{0.25cm}
\includegraphics[height=4.5 cm]{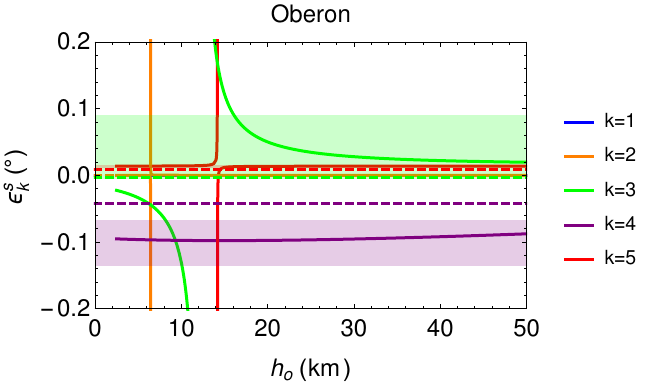}\quad
\includegraphics[height=4.5 cm]{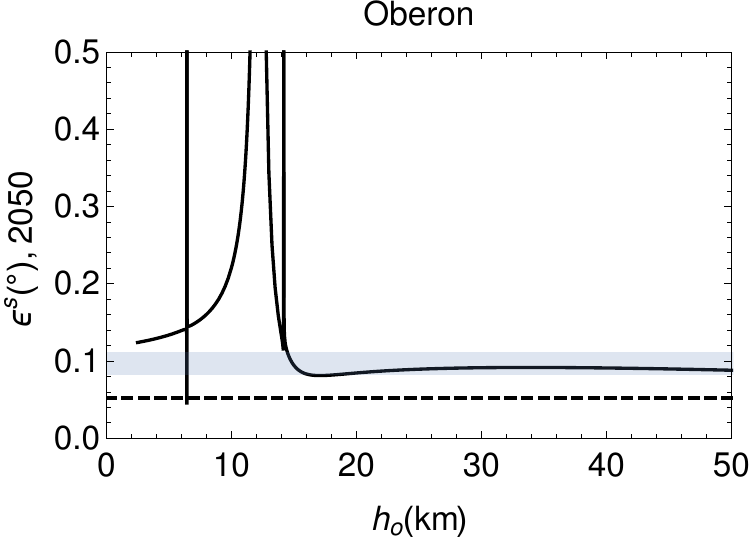}\\
\caption{Shell obliquity amplitude ($\varepsilon_k^s$, left panel) and obliquity in 2050 (right panel) as a function of the ocean thickness $h_o$ for interior models with fixed values of the density of the different layers ($\rho_s=\rho_o=1000$ kg/m$^3$ and $\rho_c=3500$ kg/m$^3$). 
The shaded areas correspond to the range of possible values for two-layer solid models with a MOI ranging from $0.288$ to $0.367$. The dashed horizontal lines correspond to the values for a single-layer homogeneous body.}
\label{epssAriel} 
      \end{center}
\end{figure}

Given the many possibilities for resonances between the eigenmodes and the forcings for Ariel, Umbriel, Titania and Oberon, and given that resonances can occur for a wide range of interior parameters, it is unlikely that an obliquity measurement will allow to constrain the latter. We are therefore not exploring the entire parameter space. 
Fig.~\ref{epssAriel} shows the obliquity amplitudes and the obliquity in 2050 (a date representative of the time of the UOP mission) for fixed values of the density of the different layers ($\rho_s=\rho_o=1000$ kg/m$^3$ and $\rho_c=3500$ kg/m$^3$), while allowing the thickness of the ocean to vary, and this for the four satellites. 

With a few exceptions, the behavior of the obliquity amplitudes is essentially identical to that of the first obliquity amplitude of Miranda, where non-resonant and resonant regimes governed by the FON can be identified. The difference with Miranda is the (larger) thickness of the ocean at which the regime changes occurs (e.g.~between $6$ and $8$ km for Ariel). 
Titania's first obliquity amplitude is also amplified by the FP for relatively thick oceans, with a maximum in absolute value around $h_o=32$ km. Oberon's second and third obliquity amplitudes can be amplified by the FP and FIN respectively, but these resonances are unlikely to play a role in real conditions. Some obliquity amplitudes are not amplified for the chosen density profiles (the second for Titania and the first and fourth for Oberon). 

The obliquity in 2050 is close to that of the solid case, except around specific values of ocean thickness where the obliquity can be very large or very small, due to resonances. When several resonances occur in a small range of ocean thicknesses, the plot can look a little chaotic. As with Miranda, detecting the ocean from a measurement of obliquity, if the satellite is in a non-resonant regime, would be a challenge. Titania and Oberon tend to have larger obliquities than Ariel and Umbriel, which combined with the possibility of resonance, would give us a real chance of detecting a subsurface ocean. The effect of elasticity on the obliquity in 2050, in the presence of an internal global ocean, and estimated as for Miranda, ranges from about $1\%$ (Ariel) to $10\%$ (Oberon).

\section{Conclusions}
\label{Sec4}

We have calculated the amplitude of the diurnal libration and the obliquity of the five largest satellites of Uranus in order to assess the potential of rotation observations to constrain their interior. Assuming hydrostatic equilibrium, we first defined ranges of three-layer interiors (shell, ocean and core) for which we also computed the tidal Love number $k_2$.

Our three-layer interior ranges are conservative extensions of \cite{Cas23}'s two-layer interior ranges, in which for instance we allow a contrast between the outer shell and ocean densities. 
For all five satellites but Miranda, we found that an accuracy of $\pm0.005$ on MOI corresponds to no better than $50$ km on the hydrosphere thickness, in agreement with \cite{Cas23}, see Section~\ref{Secinteriors}. However, we found that an accuracy of $\pm0.005$ on the MOI corresponds to an uncertainty of around $30$ km on the hydrosphere thickness of Miranda, twice the value reported in \cite{Cas23}. 
The prospects for excluding homogeneous interiors (corresponding to MOI $=0.4$) are slightly better, since an accuracy of less than half the difference between $0.4$ and the actual value would be sufficient. 
Assuming hydrostatic equilibrium, the targeted accuracies for the MOI can be translated into targeted accuracies for the gravity coefficients $J_{2}$ and $C_{22}$, using Radau's equation. 

Depending on the thickness of the ice shell, the Love number $k_2$ of a satellite with a global subsurface ocean is $2$ to $20$ times larger than that of a differentiated entirely solid body, see Section~\ref{Seck2}. The value for the homogeneous case is similar to the lowest values for the three-layer case. The tidal Love number $k_2$ can therefore be a good indicator of the existence of a subsurface ocean. For example, assuming differentiation, and depending on the thickness of the ice shell, a measurement accuracy on $k_2$ of between $0.003$ and $0.017$ would enable us to detect a global ocean in Ariel.
Such precision seems difficult to achieve when compared to Titan's $k_2$ estimated from $10$ flybys with a precision of $0.06$, \citealt{Goossens2024}.  However, depending on the geometry of the UOP spacecraft's flybys of Ariel, an accuracy of $0.010$ to $0.015$ could be achieved \citep{Filice2024}. The provisional UOP Decadal Tour includes $12$ close flybys of Ariel at $25$ km altitude ($4\%$ of the radius), \cite{Simon2021}, whereas Titan's Cassini flybys were at $>1500$ km altitude ($60\%$ of the radius).

If the Uranian satellites were two-layer solid bodies, their libration amplitude would range from about $1$ m for Oberon to about $50$ m for Miranda, see Section~\ref{Sec321}. Measurement accuracies from around $0.25$ m for Oberon to around $6$ m for Miranda would be required to distinguish homogeneous solid bodies from differentiated solid bodies, see Section~\ref{Sec322}. In the presence of an internal global ocean, we found larger libration amplitudes than in the solid differentiated case. However, the ranges for the solid and ocean cases overlap, which makes it difficult to detect an ocean using a libration measurement without an independent constraint on the MOI. If its MOI was known to a few percents, either from a precise measurement of the gravitational field or of the shape, and its shell was $150$ km thick, we would need a precision of about $5$ m on the libration estimate to detect Ariel's ocean. Constraining the thickness of the shell would require even more accurate measurements. 
As a point of reference, the precision on the libration of Mimas and Enceladus measured with Cassini's imaging system is of the order of $1$ arcmin (or $60$ m at the surface, \citealt{Taj2014,Tho16}). Assuming a state-of-the-art resolution of 2'' for narrow- and wide-angle cameras, \cite{Filice2024} find an accuracy ranging from $0.5$ to $15$ m on the libration amplitude of Ariel, depending on the geometry of the UOP flybys and the number of landmarks observed by the cameras per flyby.
Tidal deformation has the effect of reducing the amplitude of libration compared with the case where the solid layers are rigid, but the effect is limited for small and medium-sized icy satellites (less than $2\%$ for Miranda, and less than about $15\%$ for Titania and Oberon).

\begin{figure}[!htb]
      \begin{center}
        \hspace{0cm}
\includegraphics[height=5 cm]{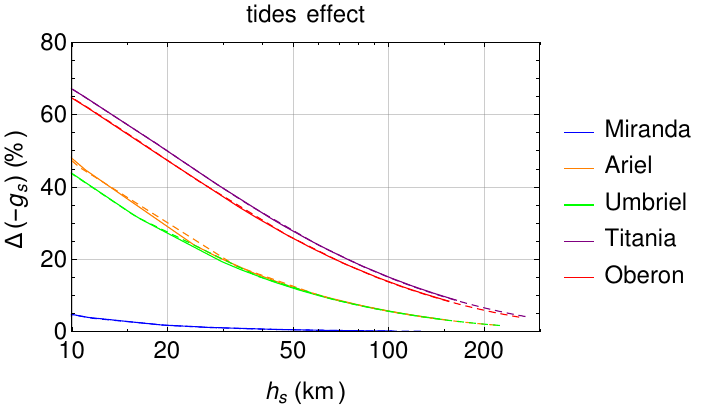}
\caption{Relative decrease in libration absolute amplitude $(-g_s)$ between ocean-bearing cases with and without periodic deformations as a function of shell thickness $h_s$. Deformations are computed with $\mu_s=3.3\times 10^9$ Pa and $\mu_c=50\times 10^9$ Pa for the shell and core rigidity, respectively (e.g.~\cite{Hus06}, see also Section \ref{Seck2}). We consider here the same values as \cite{Hem24} for layer sizes and densities. In particular, shell and ocean densities are set at $930$ kg/m$^3$ and $1030$ kg/m$^3$ respectively. Solid and dotted lines correspond to core densities of $2400$ kg/m$^3$ and $3500$ kg/m$^3$, respectively.}
 \label{FigDeltagsAll} 
\end{center}
\end{figure}

Compared with the largest values reported in \cite{Nim23,Hem24}, our libration values are relatively small. These low amplitudes are mainly explained by the fact that we do not consider here interior models with a thin shell, since \cite{Cas23} consider that if an ocean exists, it must be no more than $50$ km thick. 
If we nevertheless consider thinner shells, as \cite{Nim23,Hem24} did, we find, like them, larger values for libration than for thicker shells, just as in their Fig.~1 (with deformable solid layers) and S6 (with rigid solid layers). However, we disagree with their assertion that libration amplitudes are only slightly smaller when elastic deformations are taken into account. A comparison between the ocean cases with and without periodic deformations shows that libration amplitudes, while remaining well above the values for the entirely solid case, are reduced by around $45\%$ or $65\%$ for Ariel and Umbriel or Titania and Oberon, respectively, for a thin shell with thickness of $10$ km, see Fig.~\ref{FigDeltagsAll}. Since the Figs.~1 and S6 of \cite{Hem24} uses a logarithmic scale spread over many orders of magnitude, this reduction is more difficult to visualize. The effect of deformations is limited only for Miranda ($5\%$). These results are consistent with the $50\%$ and $90\%$ decrease in libration amplitude of Io with a magma ocean and Europa with a subsurface liquid ocean, respectively \citep{VH2020,VH2013}.

Due to their mutual gravitational interactions, the orbits and equatorial planes of the Uranian satellites do not precess at constant rates and with constant inclinations and obliquities. Only Miranda precesses almost like a body isolated from the other satellites and with  almost constant-over-time inclination and obliquity. We first applied the secular perturbation method to the URA111 ephemerides, to obtain satellite orbital precession series characterized by inclination amplitudes, see Section~\ref{AppA}. The corresponding obliquity amplitudes were then calculated using Cassini state models. Depending on the values of the forcing and free precession frequencies, the obliquity amplitudes are of the same or opposite sign to the inclination amplitudes. Resonant amplification occurs when the free and forcing frequencies are close to each other.

The obliquity of a homogeneous solid satellite is smaller than that of a two-layer solid body. For example, a measurement accuracy of around $8$ m at the surface would be needed to exclude the homogeneous case for a solid Miranda, see Section~\ref{Sec331}. 
\cite{Filice2024} find an accuracy ranging from $0.2$ to $5$ m on the obliquity of Ariel, depending on the geometry of the UOP flybys and the number of landmarks observed by the cameras per flyby, whereas an accuracy of $1.3$ m would be required.
In the presence of an internal global ocean, the Cassini state model is characterized by three free modes (FP, FON, FIN), which allow resonant amplification to occur for the 5 satellites (only for interiors with very thin oceans for Miranda), see Section~\ref{Sec332}. In the non-resonant regime, the shell obliquity values are all mostly within the range expected for a solid body, so detecting an ocean from an obliquity measurement seems difficult. The presence of an ocean could still be confirmed if it turns out that the satellite is in the resonant regime. Tidal deformation has the effect of increasing the obliquity compared with the case where the solid layers are rigid, but the effect is limited for Uranus's small and medium-sized icy satellites (less than $0.5\%$ for Miranda, and up to $10\%$ for Titania and Oberon).

The results presented here contradict a reasoning occasionally reported in the literature, where it is assumed that the ice shell of an icy satellite is mechanically decoupled from the solid interior and precesses somewhat independently with a larger obliquity than that of the solid case (e.g.~\cite{Bil08,Che14,Nim23}). This reasoning is based on an analogy with diurnal librations, for which even in a non-resonant regime, the shell librates somewhat independently from the solid interior at the short timescales of the rotation. However, in the presence of an internal liquid layer, and outside the resonant regime, the obliquity of the surface layer of a body in spin-orbit resonance is close to that of a solid body because at the long timescale of precession, the body behaves almost as a solid body (see also e.g.~\cite{Bal24} and references therein for Mercury, \cite{Coy24} for Europa). 

Our results rely on certain assumptions that, we believe, do not significantly affect our interpretation in terms of the measurement precisions required to detect an ocean or constrain an interior. For example, it is possible that the largest Uranian satellites deviate slightly from hydrostatic equilibrium. The secular model of orbital precession that we use here introduces an error in the description of orbital precession that propagates to the description of spin precession. If satellite rotation is ever measured with the UOP mission, it may be necessary to consider interior models that deviate from hydrostatic equilibrium (as was done for Enceladus, e.g.~\cite{VH2016}) and a more accurate orbital model to interpret these measurements.

In this study, we have discussed separately the potential of libration, obliquity and, to a lesser extent, tide measurements to constrain the interior of the five largest Uranian satellites. Combined with estimates of the static gravity field, tidal and rotational measurements should make it possible to characterize their interior structure (size, density, rigidity of the different layers). This is further discussed in a companion paper where we use a Markov Chain Monte Carlo (MCMC) approach to infer the interior properties of the Uranian satellites. In this subsequent study, the targeted measurements accuracies are computed, providing valuable inputs to refine and help define the UOP mission (see \cite{Filice2024}).

\section*{CRediT authorship contribution statement}

\textbf{Rose-Marie Baland:} Conceptualization, Investigation, Methodology, Writing - original draft, Writing - Review \& Editing, Visualization.
\textbf{Valerio Filice:} Conceptualization, Writing - Review \& Editing.
\textbf{Sébastien Le Maistre:} Conceptualization, Writing - Review \& Editing, Project administration.
\textbf{Antony Trinh:} Methodology, Investigation, Validation, Writing - Review \& Editing.
\textbf{Marie Yseboodt:}  Methodology, Validation, Writing - Review \& Editing, Visualization.
\textbf{Tim Van Hoolst:} Methodology, Writing - Review \& Editing, Funding acquisition.

\section*{Declaration of competing interest}

The authors declare that they have no known competing financial interests or personal relationships that could have appeared to influence the work reported in this paper.

\section*{Acknowledgments}
We would like to thank two anonymous reviewers for their comments and suggestions that helped improve our manuscript.
This work was financially supported by the Belgian PRODEX program managed by the European Space Agency in collaboration with the Belgian Federal Science Policy Office.

\appendix

\section{Secular model for the satellites orbital precession}
\label{AppA}

Four different ephemerides for the largest satellites of Uranus are currently available electronically. The semi-analytical theory of \cite{Las87}, which is nearly
forty years old, is available online via the MULTI-SAT server. It can also be extracted manually from the paper or from the associated Fortran routine (ftp://ftp.imcce.fr/pub/ephem/satel/gust86). The MULTI-SAT server also provides the numerical ephemerides of \cite{Eme13} and the "Lainey15" ephemerides, which is the most recent update of \cite{Lai08}. The numerical ephemeris URA111 of \cite{Jac14} is available from e.g.~the JPL Horizons online solar system data.
A description and comparison regarding the fitted observations, accuracy, and time span coverage of these different ephemerides can be found in \cite{Hil15}.
The choice of ephemerides for the orbit of the satellites influences the description of their orbital precession, which in turn will influence the modeling of the precession of their axis of rotation. Note that the ephemerides of \cite{Eme13} extend only until 2031, so they
are not suited for a description of the satellites orbits and rotation during the UOP mission. 

\subsection{Uranus Body Frame}
\label{sec_UBF}
The orbit of a Uranian satellite, and hence its precession, is most easily described in an non-rotating inertial reference frame associated with the Body Frame (BF) of its parent planet, defined in such a way that the rotation of the planet, the revolution of the satellite and its rotation are all prograde. The $Z-$axis of such a frame defines the direction of the planet north pole. It is customary for the $X-$axis to be chosen in the direction of the ascending node of the planet's equator on the ICRF equator. The $Y-$axis is at $90^\circ$ with respect to the $X-$axis in such a way that the three axes form a right-handed Cartesian frame. 

\begin{figure}[!htb]
      \begin{center}
        \hspace{0cm}
\includegraphics[width=12 cm]{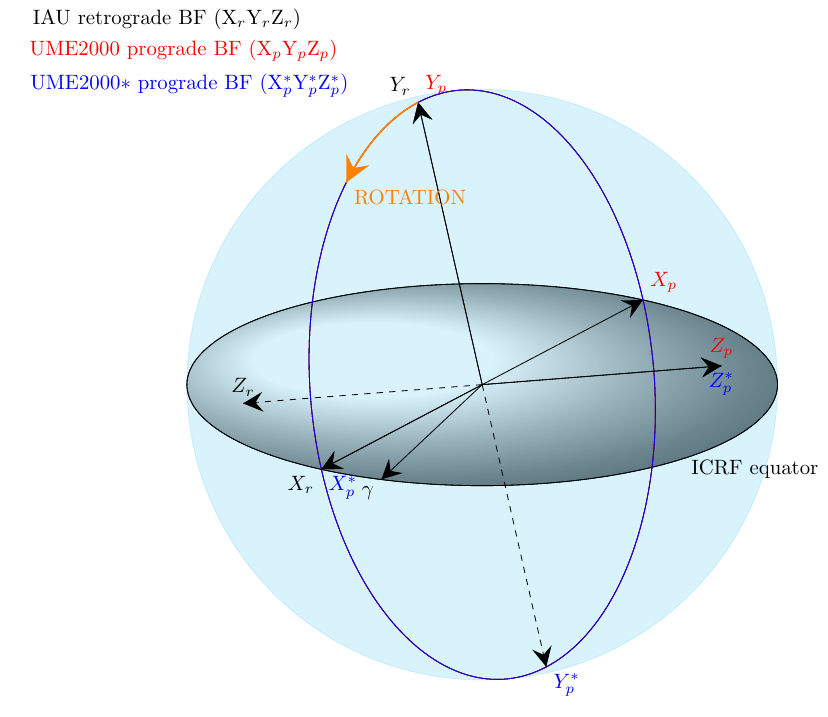}
\caption{Orientation of the IAU retrograde Body Frame ($X_r,Y_r,Z_r$), UME2000 prograde BF ($X_p,Y_p,Z_p$), and of the UME2000* prograde BF ($X_p^*,Y_p^*,Z_p^*$) of Uranus with respect to the ICRF. $\gamma$ is in the direction of the vernal point. Realistic angle values were used.  }
\label{FigBF}
      \end{center}
\end{figure}

Following \cite{Fre86}, \cite{Las87} defined such a reference frame, called \textbf{UME50}, associated with the mean equator of Uranus in 1950 and with the $X-$axis in the direction of the ascending node. They also defined the \textbf{UME50*} frame, with the $X-$axis in the direction of the descending node. We here chose J2000 as the reference epoch and refer to the "prograde Uranus BF" suited for the description of the orbit of the satellites as \textbf{UME2000} or \textbf{UME2000*} depending on the choice for the $X-$axis. Unfortunately, neither of these reference systems corresponds to the conventions of the IAU Working Group on cartographic Coordinates and Rotational elements \citep{Arc18}. The IAU defines the north pole of Uranus as the pole opposite to that of UME2000, and the $X-$axis in the same direction as the one of UME2000*, see Fig.~\ref{FigBF}. In the "IAU Uranus BF", the revolution and rotation of the satellites are retrograde. The right ascension $\alpha$ and declination $\delta$ of the UME2000 and UME2000* frames with respect to ICRFJ2000 are given in Table~\ref{URABF}, and are adapted from those of the IAU conventions. 

\begin{table}[h]
\begin{center}
\begin{tabular}{lrr}
\hline
Frame & $\alpha$ (deg) & $\delta$  (deg) \\
\hline
UME2000 "prograde" Body Frame       & +77.311 & +15.175\\
UME2000* "prograde" Body Frame      & +257.311 & +164.825 \\
IAU Uranus "retrograde" Body Frame  & +257.311 & -15.175\\
\hline
\end{tabular}
\end{center}
\caption{Right ascension $\alpha$ and declination $\delta$ of the UME2000 and UME2000* "prograde" Body Frame associated with Uranus mean equator at J2000, adapted from those of the IAU "retrograde" BF of \cite{Arc18}.} 
\label{URABF}
\end{table}

In his paper, \cite{Jac14} describes the Uranian satellites’ orbit in a prograde BF very similar to UME2000 ($\alpha=77^\circ.310$ and $\delta=15^\circ.172$). However the ephemerides provided by Jacobson to the community, under designation URA111 and accessed through the Horizons system are described in the IAU retrograde BF of \cite{Arc18}.
The MULTI-SAT server also provides the ephemerides (GUST86, Lainey and Emelyanov) in the IAU retrograde BF, though the GUST86 theory was originally described in UME50*. 
\cite{Lai08} and \cite{Eme13} chose the Uranian North Pole in accordance with the IAU conventions. \cite{Eme13} recognise its disadvantages for the description of the motion of the satellites. In the following, we will describe the orbits in the UME2000 frame. In practice, this means taking the opposite of the Horizons and MULTI-SAT outputs for the node longitudes, and the supplementary angles for the inclinations.

\subsection{Secular model for the orbital motion}

We apply the secular perturbations method presented in \cite{Der86} to the URA111 ephemerides, to obtain series for the satellites orbital precession. We use the URA111 ephemerides because \cite{Jac14} has also solved for the satellite masses and the $J_2$ and $J_4$ gravitational coefficients of Uranus, all of which are required for a secular precession model. There are no orbit-orbit resonances between the largest satellites of Uranus, so their precession is essentially due to the planet's flattening on the one hand, and mutual perturbations between the satellites on the other hand. Solar perturbations are negligible for the Uranian satellites. With this method, the orbital precession for each satellite is described as series for the following variables 
\begin{subequations}\label{pq}
\begin{eqnarray}
    p(t)&=&\sin i(t) \sin \Omega(t)=\sum_{k=1}^{5} \sin{i_k} \sin (\dot\Omega_k t +\phi_k)\\
    q(t)&=&\sin i(t) \cos\Omega(t)=\sum_{k=1}^{5} \sin{i_k} \cos(\dot\Omega_k t +\phi_k)
\end{eqnarray}    
\end{subequations}
$i(t)$ and $\Omega(t)$ are the time-variable orbital inclination and node longitude, see Fig.~\ref{FigAngles}. The frequencies $\dot\Omega_k$ are the eigenfrequencies of the $B$ matrix of Eqs.~(1-5) of \cite{Der86}. 
There are as many eigenfrequencies as there are satellites under consideration. The inclination amplitudes $i_k$ are components of the $B$ eigenvectors, scaled according the initial conditions (here the inclination and node of each satellite at J2000). The phases $\phi_k$ at the J2000 epoch are also determined by the initial conditions. The frequencies and phases are common to the five satellites, whereas the set of inclination amplitudes $i_k$ differ for each satellite, see Table \ref{TabSeries}. 
Note that the eigenfrequencies obtained here are relatively close (differences of the order of a percent) to the fundamental frequencies obtained by \cite{Gomes2024} with a frequency analysis of the orbits while our main inclination amplitudes can differ by up to $20\%$ from theirs.
For Miranda, the effect of the planet's flattening dominates the perturbations from the other satellites, so that its orbital precession is almost uniform and its inclination is almost constant and close to $i_1=4^\circ.41$. For the other satellites, which are further away from the planet, mutual perturbations cannot be neglected, and their inclination varies significantly over time.

\begin{table}[h]
\begin{center}
\begin{tabular}{llccccc}
\hline
&&   $k=1$ & $k=2$ & $k=3$ &$k=4$ &  $k=5$\\
\hline
&&&&&&\\
& Period (yr)  & -17.74 & -57.78 & -130.17 & -194.46 & -1355.97\\
&$\dot\Omega_k$   (deg/yr)        & -20.29508 & -6.23006& -2.76572& -1.85128& -0.26549 \\
&$\phi_k$ (deg)          & 100.69897 & 301.65462 & 353.53767 & 256.94581 &24.38595 \\ 
&&&&&&\\
Miranda &$i_k$  (deg)  &  4.40574 & 0.00106 & -0.0016 & 0.00119 & 0.00161 \\
Ariel   &$i_k$  (deg) & -0.01341 & 0.01508 & -0.01565 & 0.00988 & 0.01045 \\
Umbriel &$i_k$  (deg) &  -0.00123 & -0.00315 & -0.06272 & 0.03654 & 0.03352 \\
Titania & $i_k$ (deg) &  -0.00016 & -0.00005 & 0.00773 & 0.07571 & 0.12227 \\
Oberon & $i_k$  (deg) &  -0.00005 & -0.00001 & -0.00188 & -0.06283 & 0.14665 \\
\hline
\end{tabular}
\end{center}
\caption{Periods, frequencies, phases (origin of time is J2000), and amplitudes of the orbital precession of the Uranian satellites in the UME2000 frame, obtained with the secular perturbations method of \cite{Der86} applied to the URA111 ephemerides of \cite{Jac14}. $k$ is the frequency number.} 
\label{TabSeries}
\end{table}

\begin{figure}[!htb]
      \begin{center}
        \hspace{0cm}
\includegraphics[height=4.2 cm]{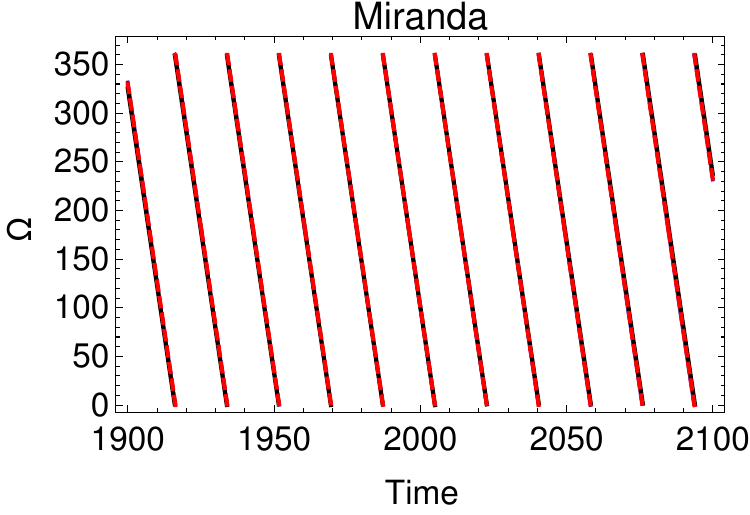}\quad
\includegraphics[height=4.2 cm]{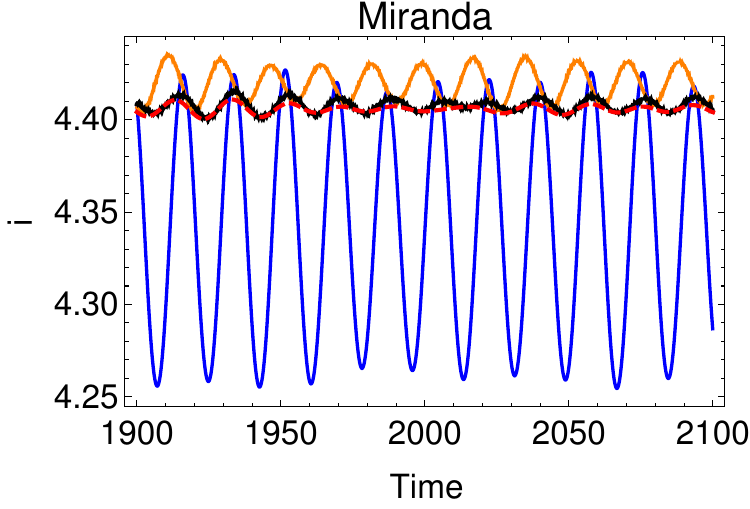}\\
\vspace{0.10 cm}
\includegraphics[height=4.2 cm]{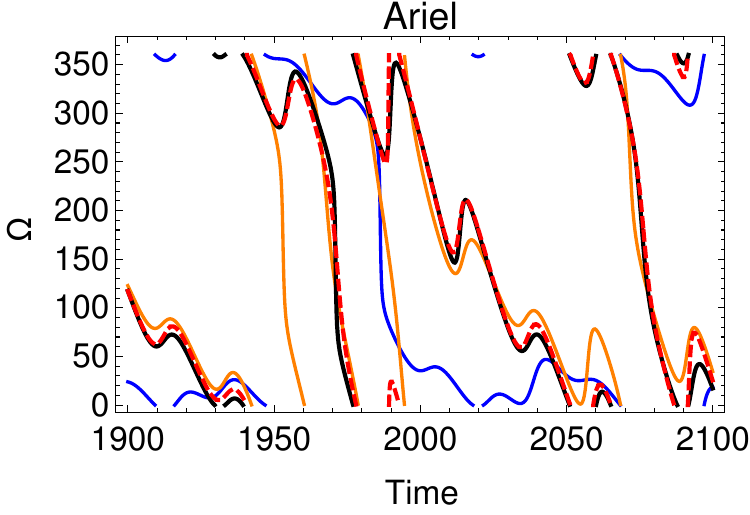}\quad
\includegraphics[height=4.2 cm]{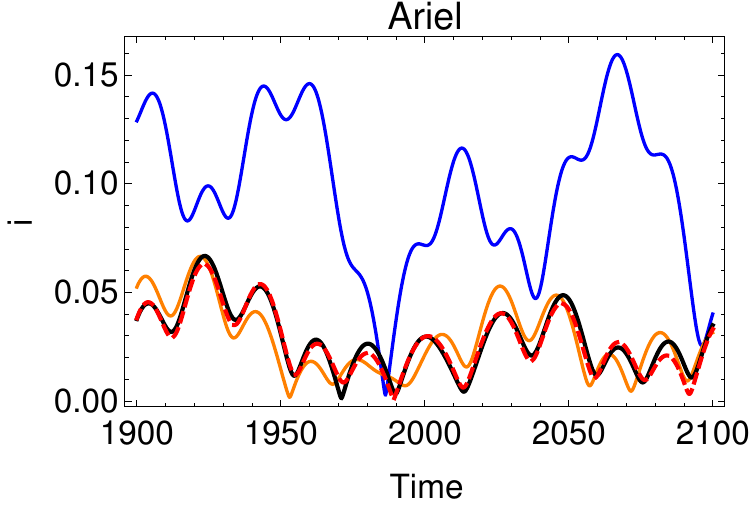}\\
\vspace{0.10 cm}
\includegraphics[height=4.2 cm]{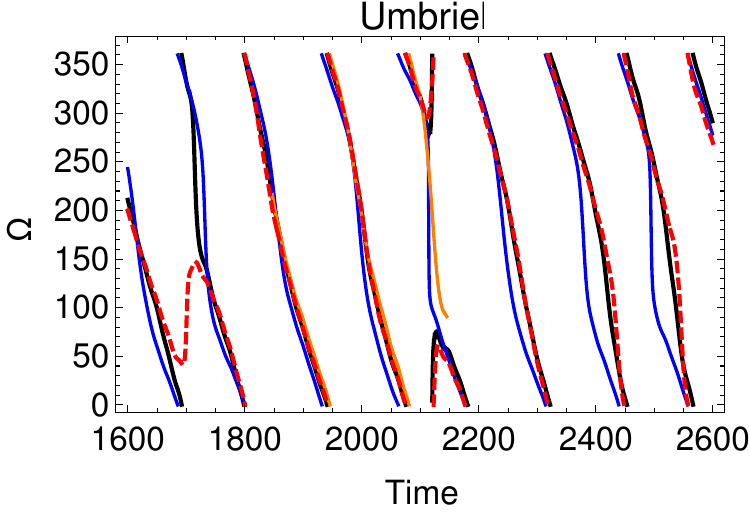}\quad
\includegraphics[height=4.2 cm]{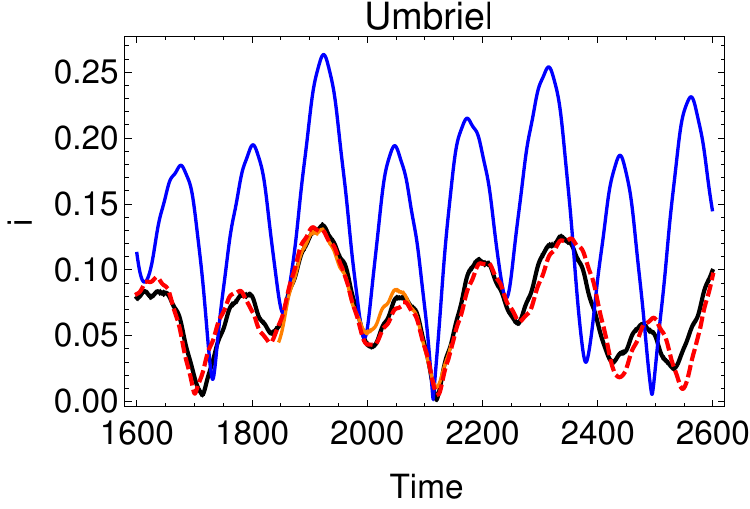}\\
\vspace{0.10 cm}
\includegraphics[height=4.2 cm]{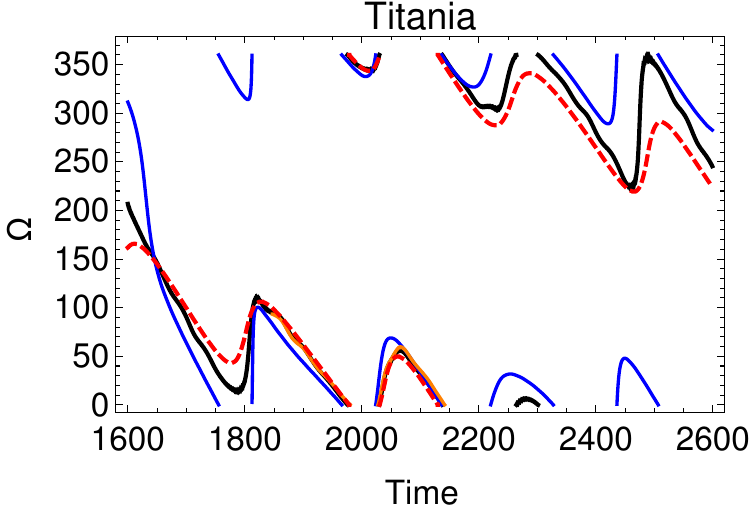}\quad
\includegraphics[height=4.2 cm]{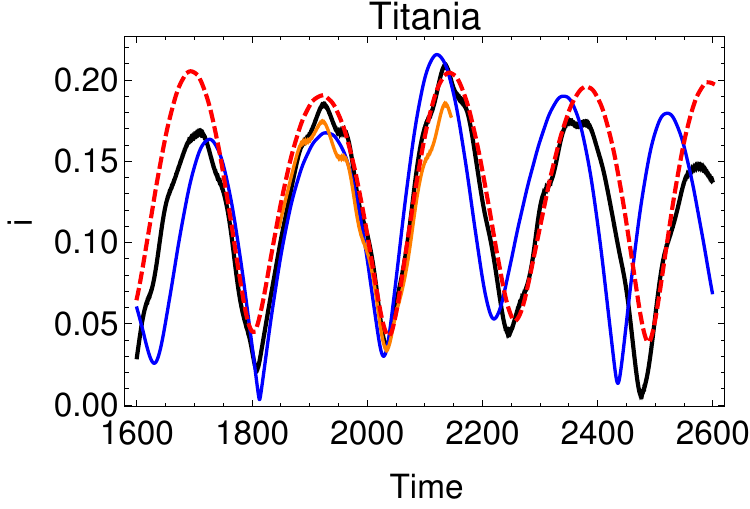}\\
\vspace{0.10 cm}
\includegraphics[height=4.2 cm]{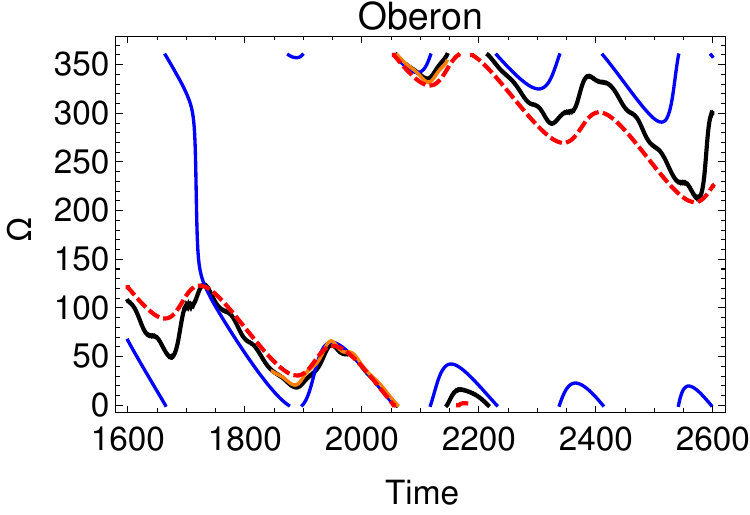}\quad
\includegraphics[height=4.2 cm]{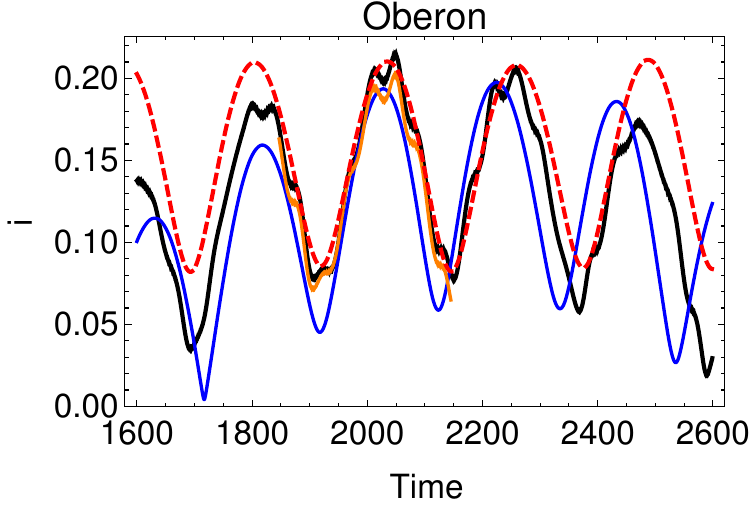}\\
\caption{Orbital node longitude and inclination of the Uranian satellites as a function of time (URA111 in black, GUST86 in blue, Lainey2015 in orange, secular model of Table~\ref{TabSeries} in dashed red).}
\label{series}
      \end{center}
\end{figure}

The series description obtained here is similar in form to that of the GUST86 analytical theory (\cite{Las87}, note the factor $1/2$ in their definition for $\zeta=\sin{i/2} \exp(\sqrt{-1}\,\Omega)$). The General Uranus Satellite Theory (GUST) is fitted to Earth-based observations from 1911 to 1986 and to data obtained during Voyager's flyby of Uranus, and provides series for the orbital elements of the satellites. They include short-period and secular terms for the description of the eccentricity and longitude of the pericenter, but only secular terms for the description of the inclination and longitude of the node. As noted in \cite{Arl06}, the GUST86 model was built from observations made mainly when the Uranian system was viewed from its pole, leading to an incorrect determination of orbit precession and inclination. 
Our series for the inclination and node longitude are in better agreement with the URA111 ephemerides than GUST86 theory, see Fig.~\ref{series}.  
However, the secular model is not ideal. In particular, it loses precision the further one moves away from the epoch chosen for the initial conditions and is not suitable for a description of shorter-period variations such as can be seen for Miranda with the URA111 ephemeris. Our intention here is not to replace the URA111 ephemeris with a secular model, but simply to obtain a description of orbital precession that meets our needs. As mentioned by \cite{Lai08}, a more precise analytical theory to replace GUST86 could be developed on the basis of all the measurements currently available, but this is beyond the scope of this study.

\newpage

\bibliographystyle{plainnat}
\bibliography{references}

\begin{thebibliography}{68}
\providecommand{\natexlab}[1]{#1}
\providecommand{\url}[1]{\texttt{#1}}
\expandafter\ifx\csname urlstyle\endcsname\relax
  \providecommand{\doi}[1]{doi: #1}\else
  \providecommand{\doi}{doi: \begingroup \urlstyle{rm}\Url}\fi

\bibitem[{Archinal} et~al.(2018){Archinal}, {Acton}, {A'Hearn}, {Conrad}, {Consolmagno}, {Duxbury}, {Hestroffer}, {Hilton}, {Kirk}, {Klioner}, {McCarthy}, {Meech}, {Oberst}, {Ping}, {Seidelmann}, {Tholen}, {Thomas}, and {Williams}]{Arc18}
B.~A. {Archinal}, C.~H. {Acton}, M.~F. {A'Hearn}, A.~{Conrad}, G.~J. {Consolmagno}, T.~{Duxbury}, D.~{Hestroffer}, J.~L. {Hilton}, R.~L. {Kirk}, S.~A. {Klioner}, D.~{McCarthy}, K.~{Meech}, J.~{Oberst}, J.~{Ping}, P.~K. {Seidelmann}, D.~J. {Tholen}, P.~C. {Thomas}, and I.~P. {Williams}.
\newblock {Report of the IAU Working Group on Cartographic Coordinates and Rotational Elements: 2015}.
\newblock \emph{Celestial Mechanics and Dynamical Astronomy}, 130\penalty0 (3):\penalty0 22, February 2018.
\newblock \doi{10.1007/s10569-017-9805-5}.

\bibitem[{Arlot} et~al.(2006){Arlot}, {Lainey}, and {Thuillot}]{Arl06}
J.~E. {Arlot}, V.~{Lainey}, and W.~{Thuillot}.
\newblock {Predictions of the mutual events of the Uranian satellites occurring in 2006-2009}.
\newblock \emph{\aap}, 456\penalty0 (3):\penalty0 1173--1179, September 2006.
\newblock \doi{10.1051/0004-6361:20065153}.

\bibitem[{Assafin} et~al.(2023){Assafin}, {Santos-Filho}, {Morgado}, {Gomes-J{\'u}nior}, {Sicardy}, {Margoti}, {Benedetti-Rossi}, {Braga-Ribas}, {Laidler}, {Camargo}, {Vieira-Martins}, {Swift}, {Dunham}, {George}, {Bardecker}, {Anderson}, {Nolthenius}, {Bender}, {Viscome}, {Oesper}, {Dunford}, {Getrost}, {Kitting}, {Green}, {Bria}, {Olsen}, {Scheck}, {Billard}, {Wasiuta}, {Tatum}, {Maley}, {Cicco}, {Gamble}, {Ceravolo}, {Ceravolo}, {Hanna}, {Smith}, {Carlson}, {Messner}, {Bean}, {Moore}, and {Venable}]{Ass23}
M.~{Assafin}, S.~{Santos-Filho}, B.~E. {Morgado}, A.~R. {Gomes-J{\'u}nior}, B.~{Sicardy}, G.~{Margoti}, G.~{Benedetti-Rossi}, F.~{Braga-Ribas}, T.~{Laidler}, J.~I.~B. {Camargo}, R.~{Vieira-Martins}, T.~{Swift}, D.~{Dunham}, T.~{George}, J.~{Bardecker}, C.~{Anderson}, R.~{Nolthenius}, K.~{Bender}, G.~{Viscome}, D.~{Oesper}, R.~{Dunford}, K.~{Getrost}, C.~{Kitting}, K.~{Green}, R.~{Bria}, A.~{Olsen}, A.~{Scheck}, B.~{Billard}, M.~E. {Wasiuta}, R.~{Tatum}, P.~{Maley}, D.~di {Cicco}, D.~{Gamble}, P.~{Ceravolo}, D.~{Ceravolo}, W.~{Hanna}, N.~{Smith}, N.~{Carlson}, S.~{Messner}, J.~{Bean}, J.~{Moore}, and R.~{Venable}.
\newblock {Kilometer-precise (UII) Umbriel physical properties from the multichord stellar occultation on 2020 September 21}.
\newblock \emph{\mnras}, 526\penalty0 (4):\penalty0 6193--6204, December 2023.
\newblock \doi{10.1093/mnras/stad3093}.

\bibitem[{Baland} et~al.(2011){Baland}, {Van Hoolst}, {Yseboodt}, and {Karatekin}]{Bal11}
R.~M. {Baland}, T.~{Van Hoolst}, M.~{Yseboodt}, and {\"O}.~{Karatekin}.
\newblock {Titan's obliquity as evidence of a subsurface ocean?}
\newblock \emph{\aap}, 530:\penalty0 A141, June 2011.
\newblock \doi{10.1051/0004-6361/201116578}.

\bibitem[Baland(2022)]{Bal24}
Rose-Marie Baland.
\newblock The obliquity of mercury: Models and interpretation.
\newblock \emph{Proceedings of the International Astronomical Union}, 18\penalty0 (S382):\penalty0 1–11, 2022.
\newblock \doi{10.1017/S1743921323004398}.

\bibitem[{Baland} and {Van Hoolst}(2010)]{Bal10}
Rose-Marie {Baland} and Tim {Van Hoolst}.
\newblock {Librations of the Galilean satellites: The influence of global internal liquid layers}.
\newblock \emph{\icarus}, 209\penalty0 (2):\penalty0 651--664, October 2010.
\newblock \doi{10.1016/j.icarus.2010.04.004}.

\bibitem[{Baland} et~al.(2012){Baland}, {Yseboodt}, and {Van Hoolst}]{Bal12}
Rose-Marie {Baland}, Marie {Yseboodt}, and Tim {Van Hoolst}.
\newblock {Obliquity of the Galilean satellites: The influence of a global internal liquid layer}.
\newblock \emph{\icarus}, 220\penalty0 (2):\penalty0 435--448, August 2012.
\newblock \doi{10.1016/j.icarus.2012.05.020}.

\bibitem[{Baland} et~al.(2016){Baland}, {Yseboodt}, and {{V}an {H}oolst}]{Bal16}
Rose-Marie {Baland}, Marie {Yseboodt}, and Tim {{V}an {H}oolst}.
\newblock {The obliquity of Enceladus}.
\newblock \emph{\icarus}, 268:\penalty0 12--31, April 2016.
\newblock \doi{10.1016/j.icarus.2015.11.039}.

\bibitem[{Baland} et~al.(2019){Baland}, {Coyette}, and {{V}an {H}oolst}]{Bal19}
Rose-Marie {Baland}, Alexis {Coyette}, and Tim {{V}an {H}oolst}.
\newblock {Coupling between the spin precession and polar motion of a synchronously rotating satellite: application to Titan}.
\newblock \emph{Celestial Mechanics and Dynamical Astronomy}, 131\penalty0 (2):\penalty0 11, February 2019.
\newblock \doi{10.1007/s10569-019-9888-2}.

\bibitem[{B{\'e}ghin} et~al.(2012){B{\'e}ghin}, {Randriamboarison}, {Hamelin}, {Karkoschka}, {Sotin}, {Whitten}, {Berthelier}, {Grard}, and {Sim{\~o}es}]{Beg12}
Christian {B{\'e}ghin}, Or{\'e}lien {Randriamboarison}, Michel {Hamelin}, Erich {Karkoschka}, Christophe {Sotin}, Robert~C. {Whitten}, Jean-Jacques {Berthelier}, R{\'e}jean {Grard}, and Fernando {Sim{\~o}es}.
\newblock {Analytic theory of Titan's Schumann resonance: Constraints on ionospheric conductivity and buried water ocean}.
\newblock \emph{\icarus}, 218\penalty0 (2):\penalty0 1028--1042, April 2012.
\newblock \doi{10.1016/j.icarus.2012.02.005}.

\bibitem[{Beuthe}(2015)]{Beu15}
Mikael {Beuthe}.
\newblock {Tides on Europa: The membrane paradigm}.
\newblock \emph{\icarus}, 248:\penalty0 109--134, March 2015.
\newblock \doi{10.1016/j.icarus.2014.10.027}.

\bibitem[{Beuthe} et~al.(2016){Beuthe}, {Rivoldini}, and {Trinh}]{Beu16}
Mikael {Beuthe}, Attilio {Rivoldini}, and Antony {Trinh}.
\newblock {Enceladus's and Dione's floating ice shells supported by minimum stress isostasy}.
\newblock \emph{\grl}, 43\penalty0 (19):\penalty0 10,088--10,096, October 2016.
\newblock \doi{10.1002/2016GL070650}.

\bibitem[{Bierson} and {Nimmo}(2022)]{Bie22}
Carver~J. {Bierson} and Francis {Nimmo}.
\newblock {A note on the possibility of subsurface oceans on the Uranian satellites}.
\newblock \emph{\icarus}, 373:\penalty0 114776, February 2022.
\newblock \doi{10.1016/j.icarus.2021.114776}.

\bibitem[{Bills}(2005)]{Bil05}
Bruce~G. {Bills}.
\newblock {Free and forced obliquities of the Galilean satellites of Jupiter}.
\newblock \emph{\icarus}, 175\penalty0 (1):\penalty0 233--247, May 2005.
\newblock \doi{10.1016/j.icarus.2004.10.028}.

\bibitem[{Bills} and {Nimmo}(2008)]{Bil08}
Bruce~G. {Bills} and Francis {Nimmo}.
\newblock {Forced obliquity and moments of inertia of Titan}.
\newblock \emph{\icarus}, 196\penalty0 (1):\penalty0 293--297, July 2008.
\newblock \doi{10.1016/j.icarus.2008.03.002}.

\bibitem[{Carr} et~al.(1998){Carr}, {Belton}, {Chapman}, {Davies}, {Geissler}, {Greenberg}, {McEwen}, {Tufts}, {Greeley}, {Sullivan}, {Head}, {Pappalardo}, {Klaasen}, {Johnson}, {Kaufman}, {Senske}, {Moore}, {Neukum}, {Schubert}, {Burns}, {Thomas}, and {Veverka}]{Car98}
M.~H. {Carr}, M.~J.~S. {Belton}, C.~R. {Chapman}, M.~E. {Davies}, P.~{Geissler}, R.~{Greenberg}, A.~S. {McEwen}, B.~R. {Tufts}, R.~{Greeley}, R.~{Sullivan}, J.~W. {Head}, R.~T. {Pappalardo}, K.~P. {Klaasen}, T.~V. {Johnson}, J.~{Kaufman}, D.~{Senske}, J.~{Moore}, G.~{Neukum}, G.~{Schubert}, J.~A. {Burns}, P.~{Thomas}, and J.~{Veverka}.
\newblock {Evidence for a subsurface ocean on Europa}.
\newblock \emph{\nat}, 391\penalty0 (6665):\penalty0 363--365, January 1998.
\newblock \doi{10.1038/34857}.

\bibitem[{Castillo-Rogez} et~al.(2023){Castillo-Rogez}, {Weiss}, {Beddingfield}, {Biersteker}, {Cartwright}, {Goode}, {Melwani Daswani}, and {Neveu}]{Cas23}
Julie {Castillo-Rogez}, Benjamin {Weiss}, Chloe {Beddingfield}, John {Biersteker}, Richard {Cartwright}, Allison {Goode}, Mohit {Melwani Daswani}, and Marc {Neveu}.
\newblock {Compositions and Interior Structures of the Large Moons of Uranus and Implications for Future Spacecraft Observations}.
\newblock \emph{Journal of Geophysical Research (Planets)}, 128\penalty0 (1):\penalty0 e2022JE007432, January 2023.
\newblock \doi{10.1029/2022JE007432}.

\bibitem[{Chen} et~al.(2014){Chen}, {Nimmo}, and {Glatzmaier}]{Che14}
E.~M.~A. {Chen}, F.~{Nimmo}, and G.~A. {Glatzmaier}.
\newblock {Tidal heating in icy satellite oceans}.
\newblock \emph{\icarus}, 229:\penalty0 11--30, February 2014.
\newblock \doi{10.1016/j.icarus.2013.10.024}.

\bibitem[{Cochrane} et~al.(2021){Cochrane}, {Vance}, {Nordheim}, {Styczinski}, {Masters}, and {Regoli}]{Cochrane2021}
C.~J. {Cochrane}, S.~D. {Vance}, T.~A. {Nordheim}, M.~J. {Styczinski}, A.~{Masters}, and L.~H. {Regoli}.
\newblock {In Search of Subsurface Oceans Within the Uranian Moons}.
\newblock \emph{Journal of Geophysical Research (Planets)}, 126\penalty0 (12):\penalty0 e06956, December 2021.
\newblock \doi{10.1029/2021JE006956}.

\bibitem[{Comstock} and {Bills}(2003)]{Com03}
Robert~L. {Comstock} and Bruce~G. {Bills}.
\newblock {A solar system survey of forced librations in longitude}.
\newblock \emph{Journal of Geophysical Research (Planets)}, 108\penalty0 (E9):\penalty0 5100, September 2003.
\newblock \doi{10.1029/2003JE002100}.

\bibitem[Coyette et~al.(2022)Coyette, Baland, and Van~Hoolst]{Coy24}
Alexis Coyette, Rose-Marie Baland, and Tim Van~Hoolst.
\newblock Revisiting the cassini states of synchronous satellites with an angular momentum approach.
\newblock \emph{Proceedings of the International Astronomical Union}, 18\penalty0 (S382):\penalty0 73–79, 2022.
\newblock \doi{10.1017/S1743921323004118}.

\bibitem[{Dehant} and {Mathews}(2015)]{Dehant2015}
V.~{Dehant} and P.~M. {Mathews}.
\newblock \emph{{Precession, Nutation and Wobble of the Earth}}.
\newblock April 2015.

\bibitem[{Dermott} and {Nicholson}(1986)]{Der86}
S.~F. {Dermott} and P.~D. {Nicholson}.
\newblock {Masses of the satellites of Uranus}.
\newblock \emph{\nat}, 319\penalty0 (6049):\penalty0 115--120, January 1986.
\newblock \doi{10.1038/319115a0}.

\bibitem[{Dumberry} and {Wieczorek}(2016)]{Dum2016}
Mathieu {Dumberry} and Mark~A. {Wieczorek}.
\newblock {The forced precession of the Moon's inner core}.
\newblock \emph{Journal of Geophysical Research (Planets)}, 121\penalty0 (7):\penalty0 1264--1292, July 2016.
\newblock \doi{10.1002/2015JE004986}.

\bibitem[Durante et~al.(2019)Durante, Hemingway, Racioppa, Iess, and Stevenson]{Durante2019}
Daniele Durante, D.J. Hemingway, P.~Racioppa, L.~Iess, and D.J. Stevenson.
\newblock Titan's gravity field and interior structure after cassini.
\newblock \emph{Icarus}, 326:\penalty0 123--132, jul 2019.
\newblock \doi{10.1016/j.icarus.2019.03.003}.

\bibitem[{Emelyanov} and {Nikonchuk}(2013)]{Eme13}
N.~V. {Emelyanov} and D.~V. {Nikonchuk}.
\newblock {Ephemerides of the main Uranian satellites}.
\newblock \emph{\mnras}, 436\penalty0 (4):\penalty0 3668--3679, December 2013.
\newblock \doi{10.1093/mnras/stt1851}.

\bibitem[{Filice} et~al.(2024){Filice}, {Cascioli}, {Le Maistre}, {Baland}, {Trinh}, {Mazarico}, and {Goossens}]{Filice2024}
V.~{Filice}, G.~{Cascioli}, S.~{Le Maistre}, R.-M. {Baland}, A.~{Trinh}, E.~{Mazarico}, and S.G. {Goossens}.
\newblock {Geophysical Investigation of the Uranian Moons using Radiometric and Landmark Tracking Data}.
\newblock \emph{The Planetary Science Journal, in revision}, 2024.

\bibitem[French et~al.(1986)French, Elliot, and Levine]{Fre86}
R.G. French, J.L. Elliot, and S.E. Levine.
\newblock Structure of the uranian rings: {II}. ring orbits and widths.
\newblock \emph{\icarus}, 67\penalty0 (1):\penalty0 134--163, 1986.
\newblock ISSN 0019-1035.
\newblock \doi{https://doi.org/10.1016/0019-1035(86)90181-8}.

\bibitem[{Gomes} and {Correia}(2024)]{Gomes2024}
S{\'e}rgio R.~A. {Gomes} and Alexandre C.~M. {Correia}.
\newblock {Dynamical evolution of the Uranian satellite system I.: From the 5/3 Ariel{\textendash}Umbriel mean motion resonance to the present}.
\newblock \emph{\icarus}, 424:\penalty0 116282, December 2024.
\newblock \doi{10.1016/j.icarus.2024.116282}.

\bibitem[Goossens et~al.(2024)Goossens, van Noort, Mateo, Mazarico, and van~der Wal]{Goossens2024}
Sander Goossens, Bob van Noort, Alfonso Mateo, Erwan Mazarico, and Wouter van~der Wal.
\newblock A low-density ocean inside {Titan} inferred from {Cassini} data.
\newblock \emph{Nature Astronomy}, pages 1--10, April 2024.
\newblock ISSN 2397-3366.
\newblock \doi{10.1038/s41550-024-02253-4}.
\newblock URL \url{https://www.nature.com/articles/s41550-024-02253-4}.

\bibitem[{Hemingway} and {Nimmo}(2024)]{Hem24}
D.~J. {Hemingway} and F.~{Nimmo}.
\newblock {Looking for Subsurface Oceans Within the Moons of Uranus Using Librations and Gravity}.
\newblock \emph{\grl}, 51\penalty0 (18):\penalty0 e2024GL110409, September 2024.
\newblock \doi{10.1029/2024GL110409}.

\bibitem[{Henrard} and {Schwanen}(2004)]{Hen04}
Jacques {Henrard} and Gabriel {Schwanen}.
\newblock {Rotation of Synchronous Satellites: Application to the Galilean Satellites}.
\newblock \emph{Celestial Mechanics and Dynamical Astronomy}, 89\penalty0 (2):\penalty0 181--200, March 2004.
\newblock \doi{10.1023/B:CELE.0000034515.57763.33}.

\bibitem[{Hilton}(2015)]{Hil15}
J.~L. {Hilton}.
\newblock {Desirability of Upgrading the Ephemerides of the Uranian Satellites Used in the Publication of The Astronomical Almanac}.
\newblock Technical report, Astronomical Applications Department, Technical Note 2015-03, 2015.

\bibitem[{Hussmann} et~al.(2006){Hussmann}, {Sohl}, and {Spohn}]{Hus06}
Hauke {Hussmann}, Frank {Sohl}, and Tilman {Spohn}.
\newblock {Subsurface oceans and deep interiors of medium-sized outer planet satellites and large trans-neptunian objects}.
\newblock \emph{\icarus}, 185\penalty0 (1):\penalty0 258--273, November 2006.
\newblock \doi{10.1016/j.icarus.2006.06.005}.

\bibitem[{Iess} et~al.(2012){Iess}, {Jacobson}, {Ducci}, {Stevenson}, {Lunine}, {Armstrong}, {Asmar}, {Racioppa}, {Rappaport}, and {Tortora}]{Ies12}
Luciano {Iess}, Robert~A. {Jacobson}, Marco {Ducci}, David~J. {Stevenson}, Jonathan~I. {Lunine}, John~W. {Armstrong}, Sami~W. {Asmar}, Paolo {Racioppa}, Nicole~J. {Rappaport}, and Paolo {Tortora}.
\newblock {The Tides of Titan}.
\newblock \emph{Science}, 337\penalty0 (6093):\penalty0 457, July 2012.
\newblock \doi{10.1126/science.1219631}.

\bibitem[{Jacobson}(2014)]{Jac14}
R.~A. {Jacobson}.
\newblock {The Orbits of the Uranian Satellites and Rings, the Gravity Field of the Uranian System, and the Orientation of the Pole of Uranus}.
\newblock \emph{\aj}, 148\penalty0 (5):\penalty0 76, November 2014.
\newblock \doi{10.1088/0004-6256/148/5/76}.

\bibitem[{Jara-Oru{\'e}} and {Vermeersen}(2014)]{Jar14}
Hermes~M. {Jara-Oru{\'e}} and Bert L.~A. {Vermeersen}.
\newblock {The forced libration of Europa{\textquoteright}s deformable shell and its dependence on interior parameters}.
\newblock \emph{\icarus}, 229:\penalty0 31--44, February 2014.
\newblock \doi{10.1016/j.icarus.2013.10.027}.

\bibitem[{Khurana} et~al.(1998){Khurana}, {Kivelson}, {Stevenson}, {Schubert}, {Russell}, {Walker}, and {Polanskey}]{Khu98}
K.~K. {Khurana}, M.~G. {Kivelson}, D.~J. {Stevenson}, G.~{Schubert}, C.~T. {Russell}, R.~J. {Walker}, and C.~{Polanskey}.
\newblock {Induced magnetic fields as evidence for subsurface oceans in Europa and Callisto}.
\newblock \emph{\nat}, 395\penalty0 (6704):\penalty0 777--780, October 1998.
\newblock \doi{10.1038/27394}.

\bibitem[{Kivelson} et~al.(2002){Kivelson}, {Khurana}, and {Volwerk}]{Kiv02}
M.~G. {Kivelson}, K.~K. {Khurana}, and M.~{Volwerk}.
\newblock {The Permanent and Inductive Magnetic Moments of Ganymede}.
\newblock \emph{\icarus}, 157\penalty0 (2):\penalty0 507--522, June 2002.
\newblock \doi{10.1006/icar.2002.6834}.

\bibitem[{Lainey}(2008)]{Lai08}
V.~{Lainey}.
\newblock {A new dynamical model for the Uranian satellites}.
\newblock \emph{\planss}, 56\penalty0 (14):\penalty0 1766--1772, November 2008.
\newblock \doi{10.1016/j.pss.2008.02.015}.

\bibitem[{Lainey} et~al.(2024){Lainey}, {Rambaux}, {Tobie}, {Cooper}, {Zhang}, {Noyelles}, and {Bailli{\'e}}]{Lai24}
V.~{Lainey}, N.~{Rambaux}, G.~{Tobie}, N.~{Cooper}, Q.~{Zhang}, B.~{Noyelles}, and K.~{Bailli{\'e}}.
\newblock {A recently formed ocean inside Saturn's moon Mimas}.
\newblock \emph{\nat}, 626\penalty0 (7998):\penalty0 280--282, February 2024.
\newblock \doi{10.1038/s41586-023-06975-9}.

\bibitem[{Laskar} and {Jacobson}(1987)]{Las87}
J.~{Laskar} and R.~A. {Jacobson}.
\newblock {GUST86 - an analytical ephemeris of the Uranian satellites}.
\newblock \emph{\aap}, 188\penalty0 (1):\penalty0 212--224, December 1987.

\bibitem[{Melchior}(1973)]{Mel73}
Paul {Melchior}.
\newblock \emph{Physique et dynamique planétaire, géodynamique. Volume 4. Chapitre 1: le freinage séculaire de la rotation de la Terre}.
\newblock Vander, Louvain, Belgique. 257 pp., 1973.

\bibitem[{National Academies of Sciences, Engineering, and Medicine}(2023)]{NAP26522}
{National Academies of Sciences, Engineering, and Medicine}.
\newblock \emph{Origins, Worlds, and Life: A Decadal Strategy for Planetary Science and Astrobiology 2023-2032}.
\newblock The National Academies Press, Washington, DC, 2023.
\newblock ISBN 978-0-309-47578-5.
\newblock \doi{10.17226/26522}.
\newblock URL \url{https://nap.nationalacademies.org/catalog/26522/origins-worlds-and-life-a-decadal-strategy-for-planetary-science}.

\bibitem[{Nimmo}(2023)]{Nim23}
F.~{Nimmo}.
\newblock {Searching for Uranian Oceans: an Interdisciplinary Perspective}.
\newblock In \emph{LPI Contributions}, volume 2808 of \emph{LPI Contributions}, page 8009, July 2023.

\bibitem[{Nimmo} and {Pappalardo}(2016)]{Nimmo2016}
F.~{Nimmo} and R.~T. {Pappalardo}.
\newblock {Ocean worlds in the outer solar system}.
\newblock \emph{Journal of Geophysical Research (Planets)}, 121\penalty0 (8):\penalty0 1378--1399, August 2016.
\newblock \doi{10.1002/2016JE005081}.

\bibitem[{Park} et~al.(2024){Park}, {Mastrodemos}, {Jacobson}, {Berne}, {Vaughan}, {Hemingway}, {Leonard}, {Castillo-Rogez}, {Cockell}, {Keane}, {Konopliv}, {Nimmo}, {Riedel}, {Simons}, and {Vance}]{Park24}
R.~S. {Park}, N.~{Mastrodemos}, R.~A. {Jacobson}, A.~{Berne}, A.~T. {Vaughan}, D.~J. {Hemingway}, E.~J. {Leonard}, J.~C. {Castillo-Rogez}, C.~S. {Cockell}, J.~T. {Keane}, A.~S. {Konopliv}, F.~{Nimmo}, J.~E. {Riedel}, M.~{Simons}, and S.~{Vance}.
\newblock {The Global Shape, Gravity Field, and Libration of Enceladus}.
\newblock \emph{Journal of Geophysical Research (Planets)}, 129\penalty0 (1):\penalty0 e2023JE008054, January 2024.
\newblock \doi{10.1029/2023JE008054}.

\bibitem[Park et~al.(2020)Park, Riedel, Ermakov, Roa, Castillo-Rogez, Davies, McEwen, and Watkins]{Park2020}
Ryan~S. Park, Joseph~E. Riedel, Anton~I. Ermakov, Javier Roa, Julie Castillo-Rogez, Ashley~G. Davies, Alfred~S. McEwen, and Michael~M. Watkins.
\newblock Advanced pointing imaging camera ({APIC}) for planetary science and mission opportunities.
\newblock \emph{Planetary and Space Science}, 194:\penalty0 105095, dec 2020.
\newblock \doi{10.1016/j.pss.2020.105095}.

\bibitem[{Porco} et~al.(2006){Porco}, {Helfenstein}, {Thomas}, {Ingersoll}, {Wisdom}, {West}, {Neukum}, {Denk}, {Wagner}, {Roatsch}, {Kieffer}, {Turtle}, {McEwen}, {Johnson}, {Rathbun}, {Veverka}, {Wilson}, {Perry}, {Spitale}, {Brahic}, {Burns}, {Del Genio}, {Dones}, {Murray}, and {Squyres}]{Por06}
C.~C. {Porco}, P.~{Helfenstein}, P.~C. {Thomas}, A.~P. {Ingersoll}, J.~{Wisdom}, R.~{West}, G.~{Neukum}, T.~{Denk}, R.~{Wagner}, T.~{Roatsch}, S.~{Kieffer}, E.~{Turtle}, A.~{McEwen}, T.~V. {Johnson}, J.~{Rathbun}, J.~{Veverka}, D.~{Wilson}, J.~{Perry}, J.~{Spitale}, A.~{Brahic}, J.~A. {Burns}, A.~D. {Del Genio}, L.~{Dones}, C.~D. {Murray}, and S.~{Squyres}.
\newblock {Cassini Observes the Active South Pole of Enceladus}.
\newblock \emph{Science}, 311:\penalty0 1393--1401, 2006.

\bibitem[{Postberg} et~al.(2011){Postberg}, {Schmidt}, {Hillier}, {Kempf}, and {Srama}]{Pos11}
F.~{Postberg}, J.~{Schmidt}, J.~{Hillier}, S.~{Kempf}, and R.~{Srama}.
\newblock {A salt-water reservoir as the source of a compositionally stratified plume on Enceladus}.
\newblock \emph{\nat}, 474\penalty0 (7353):\penalty0 620--622, June 2011.
\newblock \doi{10.1038/nature10175}.

\bibitem[{Sabadini} et~al.(2016){Sabadini}, {Vermeersen}, and {Cambiotti}]{Sab16}
Roberto {Sabadini}, Bert {Vermeersen}, and Gabriele {Cambiotti}.
\newblock \emph{{Global Dynamics of the Earth: Applications of Viscoelastic Relaxation Theory to Solid-Earth and Planetary Geophysics}}.
\newblock 2016.
\newblock \doi{10.1007/978-94-017-7552-6}.

\bibitem[{Saur} et~al.(2015){Saur}, {Duling}, {Roth}, {Jia}, {Strobel}, {Feldman}, {Christensen}, {Retherford}, {McGrath}, {Musacchio}, {Wennmacher}, {Neubauer}, {Simon}, and {Hartkorn}]{Sau15}
Joachim {Saur}, Stefan {Duling}, Lorenz {Roth}, Xianzhe {Jia}, Darrell~F. {Strobel}, Paul~D. {Feldman}, Ulrich~R. {Christensen}, Kurt~D. {Retherford}, Melissa~A. {McGrath}, Fabrizio {Musacchio}, Alexandre {Wennmacher}, Fritz~M. {Neubauer}, Sven {Simon}, and Oliver {Hartkorn}.
\newblock {The search for a subsurface ocean in Ganymede with Hubble Space Telescope observations of its auroral ovals}.
\newblock \emph{Journal of Geophysical Research (Space Physics)}, 120\penalty0 (3):\penalty0 1715--1737, March 2015.
\newblock \doi{10.1002/2014JA020778}.

\bibitem[{Schubert} et~al.(2004){Schubert}, {Anderson}, {Spohn}, and {McKinnon}]{Sch04}
G.~{Schubert}, J.~D. {Anderson}, T.~{Spohn}, and W.~B. {McKinnon}.
\newblock \emph{{Interior composition, structure and dynamics of the Galilean satellites}}, pages 281--306.
\newblock 2004.

\bibitem[Simon et~al.(2021)Simon, Nimmo, and Anderson]{Simon2021}
Amy Simon, Francis Nimmo, and Richard~C. Anderson.
\newblock Uranus orbiter and probe: Journey to an ice giant system. mission concept study report for the planetary science and astrobiology decadal survey 2023–2032.
\newblock Technical report, Johns Hopkins Applied Physics Laboratory, 2021.
\newblock URL \url{https://drive.google.com/file/d/1TxDt_qU6H2j2fYGqcDUTJQioSJ2W_KnN/view?usp=drive_link}.

\bibitem[{Smith} et~al.(1986){Smith}, {Soderblom}, {Beebe}, {Bliss}, {Boyce}, {Brahic}, {Briggs}, {Brown}, {Collins}, {Cook}, {Croft}, {Cuzzi}, {Danielson}, {Davies}, {Dowling}, {Godfrey}, {Hansen}, {Harris}, {Hunt}, {Ingersoll}, {Johnson}, {Krauss}, {Masursky}, {Morrison}, {Owen}, {Plescia}, {Pollack}, {Porco}, {Rages}, {Sagan}, {Shoemaker}, {Sromovsky}, {Stoker}, {Strom}, {Suomi}, {Synnott}, {Terrile}, {Thomas}, {Thompson}, and {Veverka}]{Smi86}
B.~A. {Smith}, L.~A. {Soderblom}, R.~{Beebe}, D.~{Bliss}, J.~M. {Boyce}, A.~{Brahic}, G.~A. {Briggs}, R.~H. {Brown}, S.~A. {Collins}, A.~F. {Cook}, S.~K. {Croft}, J.~N. {Cuzzi}, G.~E. {Danielson}, M.~E. {Davies}, T.~E. {Dowling}, D.~{Godfrey}, C.~J. {Hansen}, C.~{Harris}, G.~E. {Hunt}, A.~P. {Ingersoll}, T.~V. {Johnson}, R.~J. {Krauss}, H.~{Masursky}, D.~{Morrison}, T.~{Owen}, J.~B. {Plescia}, J.~B. {Pollack}, C.~C. {Porco}, K.~{Rages}, C.~{Sagan}, E.~M. {Shoemaker}, L.~A. {Sromovsky}, C.~{Stoker}, R.~G. {Strom}, V.~E. {Suomi}, S.~P. {Synnott}, R.~J. {Terrile}, P.~{Thomas}, W.~R. {Thompson}, and J.~{Veverka}.
\newblock {Voyager 2 in the Uranian System: Imaging Science Results}.
\newblock \emph{Science}, 233\penalty0 (4759):\penalty0 43--64, July 1986.
\newblock \doi{10.1126/science.233.4759.43}.

\bibitem[{Spencer} et~al.(2006){Spencer}, {Pearl}, {Segura}, {Flasar}, {Mamoutkine}, {Romani}, {Buratti}, {Hendrix}, {Spilker}, and {Lopes}]{Spe06}
J.~R. {Spencer}, J.~C. {Pearl}, M.~{Segura}, F.~M. {Flasar}, A.~{Mamoutkine}, P.~{Romani}, B.~J. {Buratti}, A.~R. {Hendrix}, L.~J. {Spilker}, and R.~M.~C. {Lopes}.
\newblock {Cassini Encounters Enceladus: Background and the Discovery of a South Polar Hot Spot}.
\newblock \emph{Science}, 311\penalty0 (5766):\penalty0 1401--1405, March 2006.
\newblock \doi{10.1126/science.1121661}.

\bibitem[{Spencer} et~al.(2009){Spencer}, {Barr}, {Esposito}, {Helfenstein}, {Ingersoll}, {Jaumann}, {McKay}, {Nimmo}, and {Waite}]{Spe09}
John~R. {Spencer}, Amy~C. {Barr}, Larry~W. {Esposito}, Paul {Helfenstein}, Andrew~P. {Ingersoll}, Ralf {Jaumann}, Christopher~P. {McKay}, Francis {Nimmo}, and J.~Hunter {Waite}.
\newblock {Enceladus: An Active Cryovolcanic Satellite}.
\newblock In Michele~K. {Dougherty}, Larry~W. {Esposito}, and Stamatios~M. {Krimigis}, editors, \emph{Saturn from Cassini-Huygens}, page 683. 2009.
\newblock \doi{10.1007/978-1-4020-9217-6_21}.

\bibitem[{Stiles} et~al.(2008){Stiles}, {Kirk}, {Lorenz}, {Hensley}, {Lee}, {Ostro}, {Allison}, {Callahan}, {Gim}, {Iess}, {Perci del Marmo}, {Hamilton}, {Johnson}, {West}, and {Cassini RADAR Team}]{Sti08}
Bryan~W. {Stiles}, Randolph~L. {Kirk}, Ralph~D. {Lorenz}, Scott {Hensley}, Ella {Lee}, Steven~J. {Ostro}, Michael~D. {Allison}, Philip~S. {Callahan}, Yonggyu {Gim}, Luciano {Iess}, Paolo {Perci del Marmo}, Gary {Hamilton}, William T.~K. {Johnson}, Richard~D. {West}, and {Cassini RADAR Team}.
\newblock {Determining Titan's Spin State from Cassini RADAR Images}.
\newblock \emph{\aj}, 135\penalty0 (5):\penalty0 1669--1680, May 2008.
\newblock \doi{10.1088/0004-6256/135/5/1669}.

\bibitem[{Tajeddine} et~al.(2014){Tajeddine}, {Rambaux}, {Lainey}, {Charnoz}, {Richard}, {Rivoldini}, and {Noyelles}]{Taj2014}
R.~{Tajeddine}, N.~{Rambaux}, V.~{Lainey}, S.~{Charnoz}, A.~{Richard}, A.~{Rivoldini}, and B.~{Noyelles}.
\newblock {Constraints on Mimas{\textquoteright} interior from Cassini ISS libration measurements}.
\newblock \emph{Science}, 346\penalty0 (6207):\penalty0 322--324, October 2014.
\newblock \doi{10.1126/science.1255299}.

\bibitem[{Thomas}(1988)]{Tho88}
P.~C. {Thomas}.
\newblock {Radii, shapes, and topography of the satellites of Uranus from limb coordinates}.
\newblock \emph{\icarus}, 73\penalty0 (3):\penalty0 427--441, March 1988.
\newblock \doi{10.1016/0019-1035(88)90054-1}.

\bibitem[{Thomas} et~al.(2016){Thomas}, {Tajeddine}, {Tiscareno}, {Burns}, {Joseph}, {Loredo}, {Helfenstein}, and {Porco}]{Tho16}
P.~C. {Thomas}, R.~{Tajeddine}, M.~S. {Tiscareno}, J.~A. {Burns}, J.~{Joseph}, T.~J. {Loredo}, P.~{Helfenstein}, and C.~{Porco}.
\newblock {Enceladus's measured physical libration requires a global subsurface ocean}.
\newblock \emph{\icarus}, 264:\penalty0 37--47, January 2016.
\newblock \doi{10.1016/j.icarus.2015.08.037}.

\bibitem[{Van Hoolst} et~al.(2008){Van Hoolst}, {Rambaux}, {Karatekin}, {Dehant}, and {Rivoldini}]{VH2008}
T.~{Van Hoolst}, N.~{Rambaux}, {\"O}.~{Karatekin}, V.~{Dehant}, and A.~{Rivoldini}.
\newblock {The librations, shape, and icy shell of Europa}.
\newblock \emph{Icarus}, 195:\penalty0 386--399, May 2008.
\newblock \doi{10.1016/j.icarus.2007.12.011}.

\bibitem[{Van Hoolst} et~al.(2013){Van Hoolst}, {Baland}, and {Trinh}]{VH2013}
Tim {Van Hoolst}, Rose-Marie {Baland}, and Antony {Trinh}.
\newblock {On the librations and tides of large icy satellites}.
\newblock \emph{\icarus}, 226\penalty0 (1):\penalty0 299--315, September 2013.
\newblock \doi{10.1016/j.icarus.2013.05.036}.

\bibitem[{Van Hoolst} et~al.(2016){Van Hoolst}, {Baland}, and {Trinh}]{VH2016}
Tim {Van Hoolst}, Rose-Marie {Baland}, and Antony {Trinh}.
\newblock {The diurnal libration and interior structure of Enceladus}.
\newblock \emph{\icarus}, 277:\penalty0 311--318, October 2016.
\newblock \doi{10.1016/j.icarus.2016.05.025}.

\bibitem[{Van Hoolst} et~al.(2020){Van Hoolst}, {Baland}, {Trinh}, {Yseboodt}, and {Nimmo}]{VH2020}
Tim {Van Hoolst}, Rose-Marie {Baland}, Antony {Trinh}, Marie {Yseboodt}, and Francis {Nimmo}.
\newblock {The Librations, Tides, and Interior Structure of Io}.
\newblock \emph{Journal of Geophysical Research (Planets)}, 125\penalty0 (8):\penalty0 e06473, August 2020.
\newblock \doi{10.1029/2020JE006473}.

\bibitem[{Wahr} et~al.(2006){Wahr}, {Zuber}, {Smith}, and {Lunine}]{Wah06}
J.~M. {Wahr}, M.~T. {Zuber}, D.~E. {Smith}, and J.~I. {Lunine}.
\newblock {Tides on Europa, and the thickness of Europa's icy shell}.
\newblock \emph{Journal of Geophysical Research (Planets)}, 111\penalty0 (E10):\penalty0 E12005, 2006.
\newblock \doi{10.1029/2006JE002729}.

\bibitem[{Wu} et~al.(2001){Wu}, {Bar-Sever}, {Folkner}, {Williams}, and {Zumberge}]{Wu2001}
X.~{Wu}, Y.~E. {Bar-Sever}, W.~M. {Folkner}, J.~G. {Williams}, and J.~F. {Zumberge}.
\newblock {Probing Europa's hidden ocean from tidal effects on orbital dynamics}.
\newblock \emph{Geophysical Research Letters}, 28:\penalty0 2245--2248, 2001.
\newblock \doi{10.1029/2000GL012814}.

\bibitem[{Zannoni} et~al.(2020){Zannoni}, {Hemingway}, {Gomez Casajus}, and {Tortora}]{Zan20}
Marco {Zannoni}, Douglas {Hemingway}, Luis {Gomez Casajus}, and Paolo {Tortora}.
\newblock {The gravity field and interior structure of Dione}.
\newblock \emph{\icarus}, 345:\penalty0 113713, July 2020.
\newblock \doi{10.1016/j.icarus.2020.113713}.

\end{thebibliography}

\end{document}